\documentclass[manuscript]{rmaa}

\usepackage{psfrag,color}
\usepackage{afterpage,longtable,lscape}
\usepackage{graphicx,epsfig,xcolor,subfigure}
\usepackage{paralist}
\usepackage{natbib}
\usepackage{rotating}
\usepackage{subfigure}
\usepackage{multirow}
\usepackage{tabularx,calc}
\usepackage{paralist}
\usepackage{psfrag,color}
\usepackage{array}
\usepackage{url}
\usepackage{makeidx}

\title{Optical Spectroscopic Atlas of the MOJAVE/2cm AGN Sample\mbox{\footnote[1]{}}}

\author{Janet Torrealba\altaffilmark{2,3}, Vahram Chavushyan\altaffilmark{2},  Irene Cruz-Gonz\'alez\altaffilmark{3}, Tigran G. Arshakian\altaffilmark{4,5}, Emanuele Bertone\altaffilmark{2}, Daniel Rosa-Gonz\'alez\altaffilmark{2} }

\altaffiltext{1}{These data were acquired at Observatorio Astron\'omico Nacional in  San Pedro M\'artir (OAN--SPM), M\'exico and at Observatorio Astron\'omico Guillermo Haro, in Cananea, Sonora (OAGH)}
\altaffiltext{2}{Instituto Nacional de Astrof\'isica, \'Optica y Electr\'onica, M\'exico}
\altaffiltext{3}{Instituto de Astronom\'ia, UNAM, M\'exico}
\altaffiltext{4}{Max-Planck/Institut f\"ur Radioastronomie, Bonn, Germany}
\altaffiltext{5}{Byurakan Astrophysical Observatory, Byurakan 378433,  Armenia and Isaac Newton Institute of Chile, Armenian Branch}

\shortauthor{Torrealba et al.}

\shorttitle{Spectroscopic Atlas of MOJAVE/2cm}

\fulladdresses{\item T.~G. Arshakian: Max-Planck/Institut f\"ur Radioastronomie, Auf dem H\"ugel 69, 53121 Bonn, Germany (tigar@mpifr-bonn.mpg.de)
\item E.~Bertone, V.~Chavushyan, D.~Rosa-Gonz\'alez and J.~Torrealba: Instituto Nacional de Astrof\'isica \'Optica y Electr\'onica, Apartado Postal 51 y 216, 72000 Puebla, Pue, M\'exico (ebertone,vahram,danrosa and cjanet@inaoep.mx)
\item  I.~Cruz-Gonz\'alez: Instituto de Astronom\'ia, Universidad Nacional Aut\'onoma de M\'exico, Ciudad Universitaria, Apartado Postal 70--264,  04510, M\'exico D.F., M\'exico
(irene@astroscu.unam.mx)}

\begin{document}

\abstract {We present an optical spectroscopic atlas at intermediate resolution (8--15\,\AA) for 123 core-dominated radio-loud active galactic nuclei with relativistic jets, drawn from the  MOJAVE/2cm sample at 15\,GHz. It is the first time that spectroscopic and photometric parameters for a large sample of such type of AGN are presented. The atlas includes spectral parameters for the emission lines H$\beta$, [\ion{O}{3}]$\lambda$5007, \ion{Mg}{2}$\lambda$2798 and/or \ion{C}{4}$\lambda$1549 and corresponding data for the continuum, as well as the luminosities and equivalent widths of the \ion{Fe}{II} UV/optical. It also contains the homogeneous photometric information in the B-band for 242 sources of the sample, with a distribution peak  at $B_J$=18.0 and a magnitude interval of 11.1$\,\leq\,B_J\,\leq\,$23.7.}

\resumen{Presentamos un atlas espectrosc\'opico \'optico de resoluci\'on intermedia (8--15\,\AA) para 123 n\'ucleos activos de galaxias compactos con la presencia de chorros superlum\'inicos, tomados de la muestra limitada en densidad de flujo a 15\,GHz MOJAVE/2cm. Es la primera vez que se presentan los par\'ametros espectrosc\'opicos y fotom\'etricos  para una muestra tan grande de este tipo de AGN. El atlas incluye los par\'ametros espectrales para las l\'ineas de emisi\'on H$\beta$, [\ion{O}{3}]$\lambda$5007, \ion{Mg}{2}$\lambda$2798 y/\'o \ion{C}{4}$\lambda$1549, junto con los datos para la emisi\'on del continuo correspondiente. Se presentan adem\'as las luminosidades y el ancho equivalente del  \ion{Fe}{II} UV/\'optico. Contiene tambi\'en la informaci\'on fotom\'etrica homog\'enea en la banda B para 242 fuentes de la muestra, con un pico en la distribuci\'on de $B_J$=18.0 y un intervalo en magnitud de 11.1$\,\leq\,B_J\,\leq\,$23.7.}

\addkeyword{atlases\,---\,galaxies: active}
\addkeyword{galaxies:\,nuclei}
\addkeyword{quasars:\,emission lines}
\addkeyword{techniques:\,spectroscopic}

\maketitle

\section{Introduction}

In the current paradigm of active galactic nuclei (AGN), the large amount of energy released by the AGN is generated in the very small region, the central engine, which is thought to be powered by accretion of matter onto a central black hole. This active central engine also ejects bipolar, highly collimated, relativistic outflows (jets). AGN in which one of the jets is oriented towards the line of sight is strongly beamed due to relativistic effects. This type of sources poses several fundamental questions such as: how the central engine is related to the pc-scale jets; what is the contribution of the jet emission to the total continuum emission and line emission; how are related the continuum emission region, broad/narrow emission-line region with different properties of the relativistic jet. Despite all the advances in AGN research, a global analysis of the physics involved at all spatial scales, from sub-parsec to kpc, is needed to tackle above issues.

Recently, \citet{arshakian08,arshakian09} and \citet{tavares09}, using the long-term optical spectral and radio (VLBI) monitoring data of radio galaxies 3C 390.3 and 3C 120 found a link between the variable optical emission and the kinematics of the sub-parsec-scale jet. They have been able to localize the region of a variable optical emission in the innermost sub-pc scale region of the jet, and showed that very long-term variations ($\sim$ 10 yr) of optical continuum emission are correlated with the radio emission from the base of the jet located just above the disk, while the optical long-term variations (1--2 yr) follow the radio flares from the stationary component in the jet with a time delay of about one yr. Using the flux-limited complete sample of core-dominated AGN  \citep[MOJAVE-1;][]{lister09}, \citet{arshakian10} found interesting correlations among properties of parsec-scale jets, using the photometric data presented in this atlas. They reported a significant positive correlation between optical nuclear emission and total radio emission at 15 GHz for 99 quasars. Radio emission originates in the unresolved core, at milliarcsecond scales suggesting that both radio and optical emission are beamed and originated in the innermost part of the sub-parsec-scale jet in quasars. For BL Lacs, the optical continuum emission correlates with the radio emission of the jet. These results are confirmed for a larger sample of 233 core-dominated AGN \citep{torrealba11}. 

Furthermore, spectroscopic parameters of the broad- and narrow-line profiles provide a direct clue about the physics, kinematics, and structure of the central engine of AGN. Diverse studies have  shown the existence of an anticorrelation between the prominence of the radio nucleus and the width of broad emission line,  giving a clue on the gas distribution in the broad line region (BLR) in these sources \citep[H$\alpha$, H$\beta$, \ion{Mg}{2}, and others lines; ][]{hough02,vestergaard00,WB86}. The results of \citet{rokaki03} confirmed the correlation between the jet viewing angles and the broad emission line equivalent widths of 19 superluminal quasars, which suggests a flattened structure for the line-emitting material.

Motivation of the present paper is to perform a robust statistical analysis and to study in detail the physical link between the diverse emission regions in radio-loud AGN from sub-pc to kpc scales. For this purpose it is necessary to compile a well defined sample of compact AGN with superluminal jets and a collection of good quality spectral parameters  (signal--to--noise $\sim30$). Here, we gathered the spectroscopic and photometric information for a specific subsample of 250 compact AGN (selected from the MOJAVE sample) for which the jet parameters are well characterized \citep{kovalev05}.

We present an optical spectroscopic atlas of 123 AGN that represent the $\sim$50\% of the total sample, supplemented with the photometric information for 97\%  of the same sample (242 sources). The contents of the paper are as follows: the sample is described in \S~\ref{sec:sample}, followed by the photometric data and calibration for estimating the optical luminosities at 5100 \AA\, (\S~\ref{sec:L5100});  details of optical spectroscopic observations are presented in \S~\ref{sec:observ}, followed by data reduction and calibration procedures in \S~\ref{sec:data}; the detailed procedure to extract the principal parameters of the continuum emission and various emission lines (H$\beta$, [\ion{O}{3}]$\lambda$5007, \ion{Mg}{2}$\lambda$2798, \ion{C}{4}$\lambda$1549, and \ion{Fe}{II} UV/optical) are described in \S~\ref{sec:processing}, followed by the results in \S~\ref{sec:results}, and  finally, the appendices contain the general properties of the sample in~\ref{secap:mojave}, spectral atlas in~\ref{sec:atlas}, and  emission line parameters and continuum emission in~\ref{secap:parameters}.

Throughout the paper a flat cosmology model is used with $\Omega_{m}=0.3$ ($\Omega_{\Lambda}+\Omega_{m}=1$) and $H_0=70$ km\,s$^{-1}$\,Mpc$^{-1}$.

\section{MOJAVE/2cm AGN sample}
\label{sec:sample}

We use the sample of 250 compact extragalactic sources compiled by \citet{kovalev05}. The sources were observed with the Very Large Baseline Array (VLBA) at 2\,cm (15\,GHz) and present radio jets on parsec scales.  
The sample is composed from: \textit{i)} the flux-density-limited complete sample MOJAVE-1 of 135 sources \citep[Monitoring of Jets in AGN with VLBA Experiments;][]{lister09},  hereon M1; \textit{ii)} the extension MOJAVE-2 of 53 AGN with special characteristics (e.g. high luminosities, special kinematics, $\gamma$-ray sources); and \textit{iii)} 62 sources from the VLBA 2~cm monitoring survey \citep[][]{K98,K04,Z02}. We refer to this sample as MOJAVE/2cm.  
 
Since 1994, the VLBA 2\,cm survey and MOJAVE program have monitored compact radio sources to study their jets structure with unprecedented resolution and  sensitivity \citep[see][and references within]{lister09}. Most of the sources in our sample have flat radio spectra \citep[$\alpha >-$ 0.5, $F\sim\nu^{+\alpha}$, for $\nu\,>$ 500 MHz;][]{kov99,kov00}, and their total flux density\footnote{Total flux density was obtained in the period 1994--2003, often originally estimated by extrapolation from lower frequency data \citep{kovalev05}.} at 2\,cm is greater than 1.5 Jy for sources with $\delta >$ 0$^\circ$ and $>2$\,Jy for sources with $-20^\circ<\delta<0^{\circ}$.

\begin{table*}[!t]\centering
  \setlength{\tabnotewidth}{1.0\columnwidth}
  \tablecols{6}
  
  \setlength{\tabcolsep}{2.0\tabcolsep}
  \caption{Spectroscopic classification of the MOJAVE/2cm AGN} \label{tab1:clasificacion}
 \begin{tabular}{lccccc}
    \toprule
    
\multicolumn{1}{c}{ {Sample}} &
   \multicolumn{1}{c}{ {$\#$}} &
   \multicolumn{1}{c}{ {Quasars}} &   
   \multicolumn{1}{c}{ {BL Lac}} &
   \multicolumn{1}{c}{ {RG}} &
   \multicolumn{1}{c}{ {No ID}} \\

    \midrule
MOJAVE-1\tabnotemark{a} & 135    & 101       & 22      & 8        & 4 \\

MOJAVE-2\tabnotemark{b}& 53      & 35       & 10      & 8        &\nodata\\

2cm\tabnotemark{c} &  62         & 52       & 4       & 4        & 2 \\ 
MOJAVE/2cm & 250  & 188      & 36      & 20	   & 6\\
    \bottomrule
    \tabnotetext{a}{M1: \citet{lister05}, redshift 0.004\,$\leq\,z\,\leq$\,3.408\, and 
magnitude 11.16\,$\leq\,B\,\leq\,$22.10.}
    \tabnotetext{b}{M2: Currently monitored (\url{http://www.physics.purdue.edu/MOJAVE/}), 
redshift 0.017$\leq\,z\,\leq$ 3.280 and magnitude  13.14\,$\leq\,B\,\leq\,$20.92.}
    \tabnotetext{c}{2\,cm: \citet[][]{K98,Z02,K04}, redshift 0.055$\leq\,z\,\leq$ 3.787  and magnitude 12.15\,$\leq\,B\,\leq\,$23.40.}
  \end{tabular}
\end{table*}
%%%%%%%%%%%%%%%%%%%%%%%%%%%%%%%%%%
\begin{figure}\centering

\includegraphics[width=\textwidth]{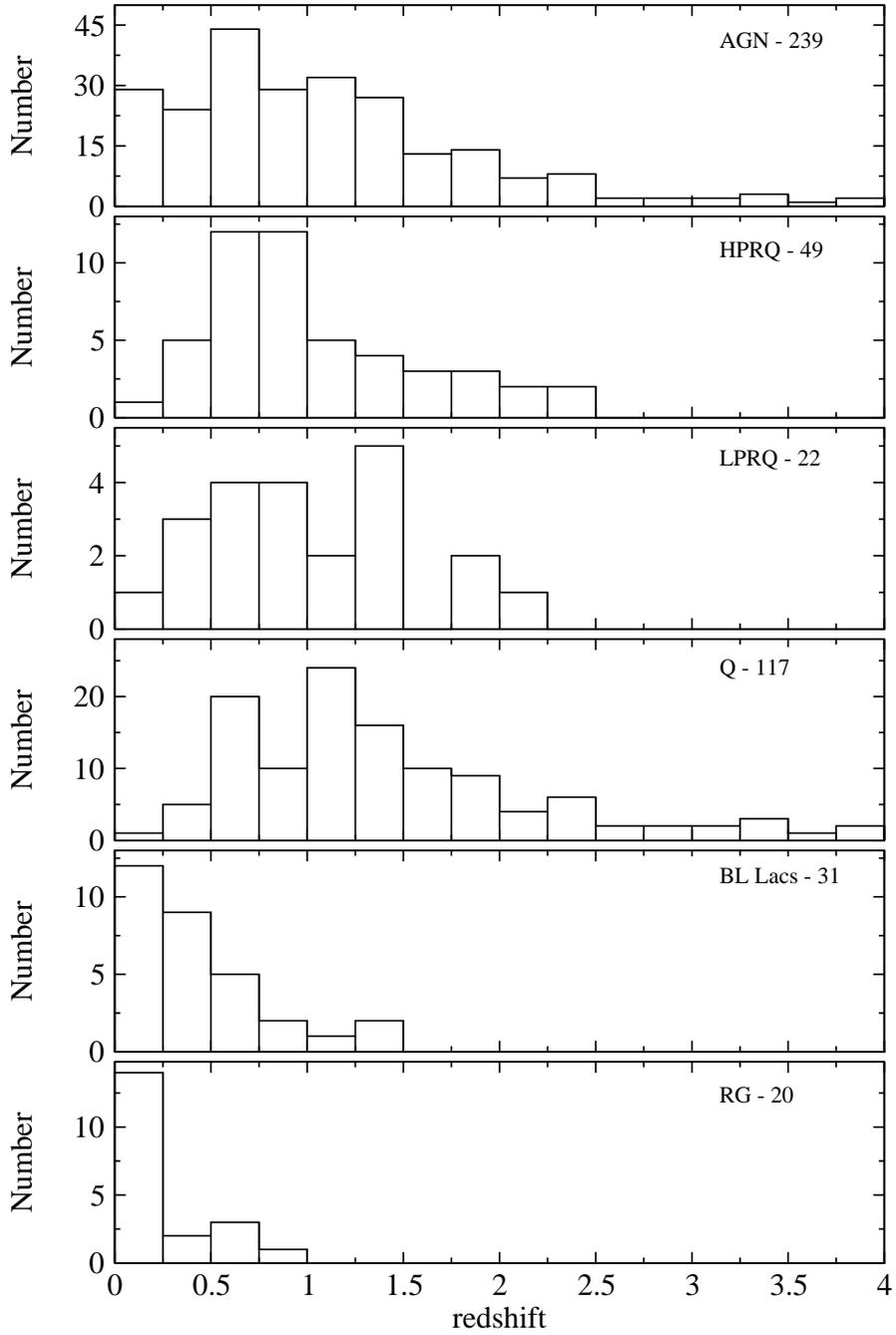}
\caption{Distribution of redshifts of the MOJAVE/2cm AGN is shown in the top panel. Other panels show the distributions of redshifts for different spectral types of AGN.\label{fig:1}}
  
\end{figure}

The summary of spectroscopic classification from \citet{VV03} AGN catalog is presented in Table~\ref{tab1:clasificacion}. The full sample consists of 188 quasars, 36 BL~Lacs, 20 radio galaxies (RG), and 6 sources with no optical identification (see Table~\ref{tabA0:sample} Appendix~\ref{secap:mojave}). Among quasars there are 49 high polarized quasars (optical polarization $P>3$\%, HPRQ), 22 low polarized quasars ($P<3$\%, LPRQ), and  117 quasars without optical polarimetry information (Q).

The distributions of redshifts of 239 MOJAVE/2cm AGN and individual types of AGN are shown in the different panels of Figure~\ref{fig:1}. The AGN redshift range  is 0.004$\,\leq\,z\,\leq\,$3.8 and the mean redshift is 1.1.
%%%%%%%%%%%%%%%%%%%%%%%%%%%%%%%%%%
\begin{figure}\centering
 
\includegraphics[width=\textwidth]{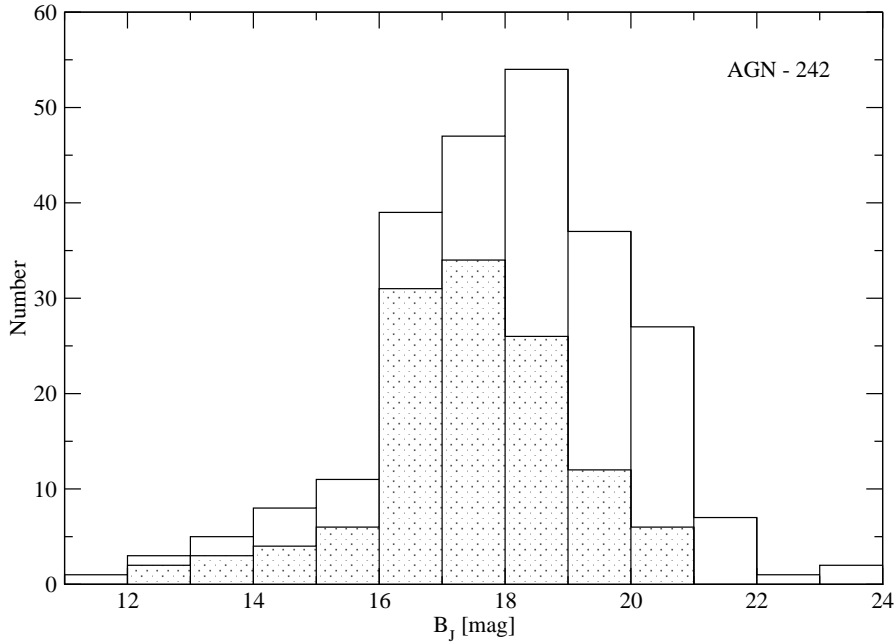}
\caption{Distribution of apparent B-magnitudes of the MOJAVE/2cm AGN. The dotted histogram represents the $\sim$50\% of the sources that have spectrophotometric data.\label{fig:2}}
  
\end{figure}
%%%%%%%%%%%%%%%%%%%%%%%%%%%%%%%%%%

The apparent B-magnitude in the Johnson's photometric system, $B_J$, for 242 MOJAVE/2\,cm sources is shown in Figure~\ref{fig:2}, where the dotted area shows sources with spectrophotometric data, which represents $\sim$50\% of the sample. The remaining sources are too faint for spectroscopy with 2m-class telescopes. The peak of the distribution is at $B_J=18.0$ and 11.1$\,\leq B_J \leq\,$23.7.

\section{Optical luminosities}
\label{sec:L5100} 

In this work we estimate the optical luminosity at 5100\AA\,($\lambda\,L_{5100}$) for 233 sources of MOJAVE/2cm compact radio sources drawn for the photometric data available for each AGN. We report as well the spectroscopical parameters of the continuum emission and diverse emission lines, measured directly from the data of 123 AGN of the same sample depending on the redshift of the source. The procedure to obtain  $\lambda\,L_{5100}$ is described below, while the treatment of the spectroscopical data is explained in section \S~\ref{sec:processing}.

The luminosity at 5100\AA\ is estimated using the following expression \citep[e.g.,][]{marziani03a}:

\begin{equation}
\lambda L_{5100}=
3.137\times10^{35-0.4(M_{\rm {B}}-A_{\rm B})}\,
\mathrm{erg\,s^{-1}},
\end{equation}
where $A_{\rm B}$ is the galactic extinction in the B-band taken from the NASA Extragalactic Database, and $M_{\rm B}$ is the absolute magnitude given 
by \citet{SG83}:
\begin{eqnarray}
M_\mathrm{B}= {B_{J}} - 5
\,\mathrm{log\,}d_{L}+2.5\,(1+\alpha_{op})\,\log\,(1+z)+ \nonumber \\
5\,\log\,(h)-42.386.
\end{eqnarray}
In this equation  $h = H_{0}/100$\,km\,s$^{-1}$\,Mpc$^{-1}$ = 0.7; the term $2.5\,(1+\alpha_{op})\,\log\,(1+z)$ reflects the effect of the redshift on measurements through a fixed color band. We adopted an optical spectral index $\alpha_{op} = -0.5$ ($S_{\nu}\propto\nu^{\alpha}$) typical value for radio-loud objects \citep[see~Table~2][]{B01}. Finally, $z$ is the redshift, and $d_{L}$ is the luminosity distance for the flat cosmology model given by
\begin{equation}
\label{eq:luminosity_distance}
d_{L}=\frac{c}{H_{0}}(1+z)\left[\eta(1,\Omega_{\rm m})-\eta\left(\frac{1}{1+z},\Omega_{\rm m}\right)\right],
\end{equation}
where $\Omega_{\rm m}$=0.3 and $\eta$ is a function of $z$ and $\Omega_{\rm m}$ \citep[see][]{Pen99}.

The details on the calculation of ${B_{J}}$ and the procedure to correct by the host galaxy contribution is described in \citet{arshakian10}.  

\begin{figure} \centering
  \includegraphics[width=\textwidth]{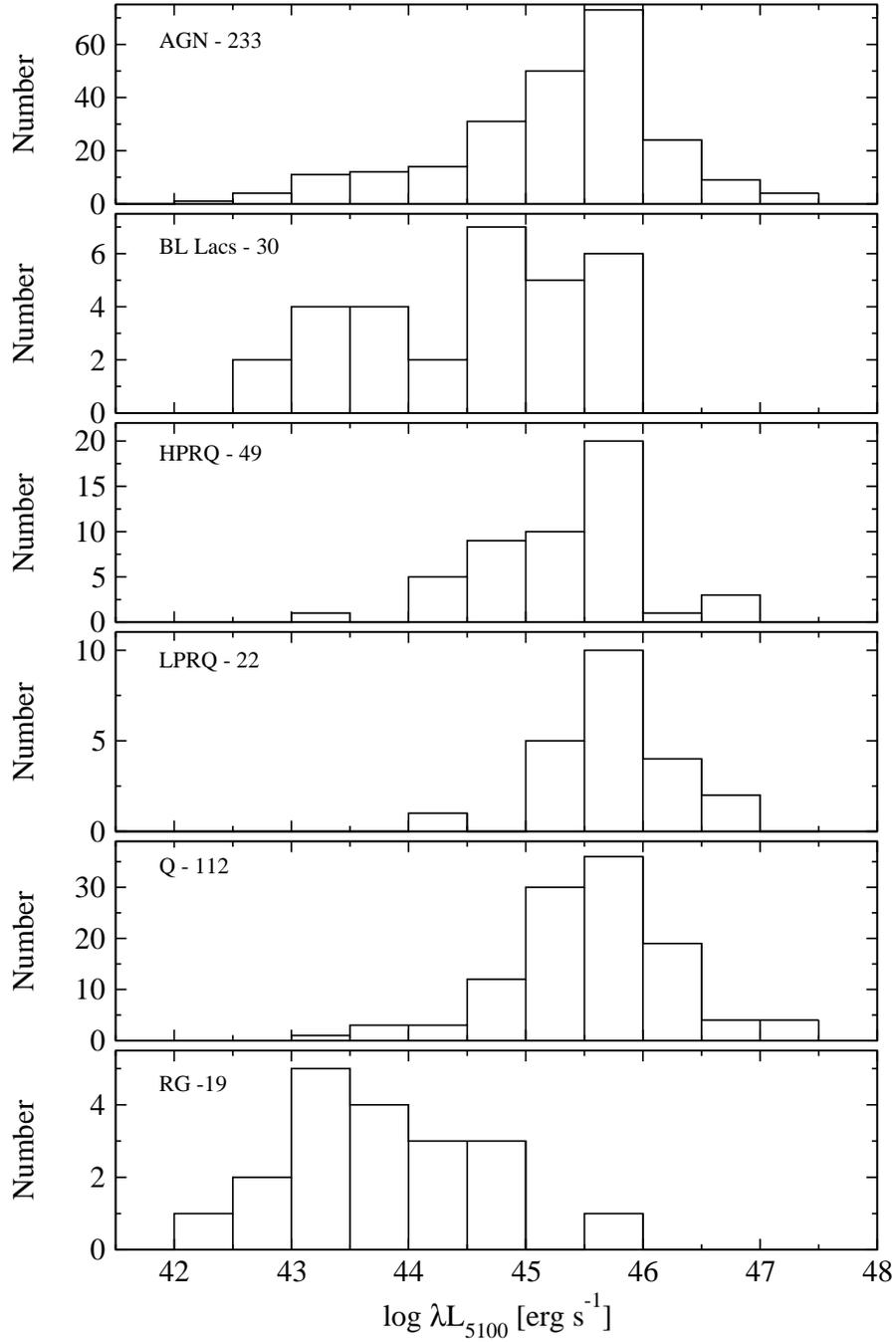}
  \caption{The distribution of corrected photometric optical luminosities at 5100\,\AA\, for the sample of 233 AGN for the MOJAVE/2cm sample (top panel), 30 BL Lacs, 49 HPRQ, 22 LPRQ, 112 quasars, 19 radio galaxies, and 1 unidentified source.
  \label{fig:3}}
\end{figure}

The distribution of $\lambda\,L_{5100}$ covers 5 orders of magnitude in luminosity with a range between $2\,\times\,10^{42}\,\rm erg\,s^{-1}$ and $1\,\times\,10^{47}\,\rm erg\,s^{-1}$,  with an average value of $9\,\times\,10^{45}\,\rm erg\,s^{-1}$; as is shown in Figure~\ref{fig:3} for the different AGN types in our sample. The radio galaxies are the sources with weaker luminosities as compared with the quasars and BL Lacs. The distributions of the optical nuclear luminosities for HPRQ and LPRQ show slightly different ranges: $\lambda\,L_{5100,\rm\,HPRQ}$ = $(10^{43}-10^{46})\,\rm erg\,s^{-1}$ and  $\lambda\,L_{5100,\rm\,LPRQ}$ = $(10^{44}-10^{46})\,\rm erg\,s^{-1}$. The Kolmogorov-Smirnov test (\textit{KS}) shows that the null hypothesis that these distributions are drawn for the same parent population cannot be rejected at a confidence level 93\% ($P_{KS}=0.072$), for M1 sample the confidence level is 81\%, ($P_{KS}=0.189$).    It is well known that HPRQ fall within the group of blazars and are known to exhibit a different behavior regarding LPRQ, and their differences cannot be explained with a model based solely on orientation effects. For example, \citet{lister00}  found that the cores of the LPRQ tend to be less luminous at 15\,GHz than the HPRQ. Besides, they reported that the jet components in the LPRQ have magnetic fields parallel to the jet, while those of HPRQ are perpendicular to the flow of the jet. We agree that high-- and --low polarized quasars are intrinsically different sources, but more observational data (polarization studies) are needed to understand the physical nature of the differences showed in these two types of quasars.

%%%%%%%%%%%%%%%%%%%%%%%%%%%%%%%%%%
%%%%%%%%%%%%%%%%%%%%%%%%%%%%%%%%%% 

\section{Spectroscopic Observations and Supplementary Data}
\label{sec:observ}
Optical spectra of the bright fraction (\textit{B} $<$ 18) of AGN of the sample were obtained in several observing runs from 2003 to 2006 with two optical telescopes in Mexico: the 2.1m telescope of the Observatorio Astron\'omico Nacional in San Pedro M\'artir, Baja California (OAN--SPM), and the 2.1m telescope of the Observatorio Astron\'omico Guillermo Haro, in Cananea, Sonora (OAGH). From these telescopes we obtained a total of 146 good quality spectra:  110  from OAN--SPM and 36 from OAGH.

The OAN--SPM spectra were obtained with a Boller \& Chivens spectrograph, using a 2.5 \arcsec\, slit, a CCD SITe3 (1024 $\times$ 1024 pixels of 24 \micron $\times$ 24 \micron), a plate scale of 40 \arcsec/mm), a 300 l/mm grating and blaze angle of $5^{\circ}50'$ for the 4000--8000 \AA\, spectral region, obtaining an effective  spectral resolution of $\sim$ 8--10\AA. 

The OAGH spectra were obtained with a Boller \& Chivens spectrograph, using a 2.5 \arcsec\, slit, a CCD Tektronix TK1024 (1024 $\times$ 1024 pixels of 24 $\mu$m $\times$ 24 $\mu$m) and a plate scale of 8.18 \arcsec/mm, a 150 l/mm grating and blaze angle of   $3^{\circ}00'$ for the 4000--7100\,\AA\, spectral region, obtaining an effective spectral resolution of $\sim 10-15$\,\AA.

We also searched for spectroscopic data in the archives of the Sloan Digital Sky Survey (\textit{SDSS}), the Hubble Space Telescope (\textit{HST}), and in the AGN sample of \citet{M03}\footnote{\url{http://web.pd.astro.it/marziani/data.html}}, hereafter M03. 
The database was also supplemented by spectra kindly provided by C.~R. Lawrence \citep[][hereafter L96]{L96}. In total, we found 114 additional spectra: 23 from \textit{SDSS}, 40 from \textit{HST}, 36 from L96, and 15 from M03.

Overall, 260 spectra of 123 AGN were analyzed from both our spectroscopic observations and the supplementary data. From these 123 AGN, 73 sources belong to the flux-limited sample MOJAVE-1 (4 BL Lacs, 7 RG, 24 HPRQ, 17 LPRQ, and 21 Q). For this sub-sample we have a total of 104 spectra: 37 from OAN-SPM, 18 from OAGH, 17 from SDSS, 13 from HST,  10 from L96, and 9 from M03.

The typical integration times of our observations were 3600\,s  splitted in three exposures per source for an object  with $B \sim 18$. The spectra for \textit{HST}, \textit{SDSS}, L96, and M03 involved typical total integration times of 1000 sec, 3800, 3100, and 10,000 sec, respectively.

Table~\ref{tab2:configuracion} summarizes the instrumental setup and spectroscopic characteristics of the different sets of optical spectra. Data is organized as follows: Col. (1) is the observing site, Col. (2)  is the telescope aperture, Col. (3) is the spectrograph, Col. (4)  is the dispersion in \AA\,mm$^{-1}$, Col. (5)  is the slit width in \arcsec, Col. (6) is the spectral resolution in \AA\,at FWHM measured on the instrumental profile, and Col. (7)  is the number of spectra. Observational details of the spectra obtained from L96 are presented in their Tables~1 and 2, and  from  M03 is presented in their Table~1. The data of L96 has a spectral resolution in the range between 6--18 \AA, and for the data of M03 the spectral resolution is in the range between 3--8 \AA.

\begin{table*}[!t]\centering
  \setlength{\tabnotewidth}{1.0\columnwidth}
  \tablecols{7}
  
\caption{Synopsis of optical spectroscopy} \label{tab2:configuracion} 

 \begin{tabular}{lcccccc}
    \toprule
%This is the header for the first page of the table...
   \multicolumn{1}{c}{ {}} &
   \multicolumn{1}{c}{ {}} &
   \multicolumn{1}{c}{ {}} &   
   \multicolumn{1}{c}{ {}} &
   \multicolumn{1}{c}{ {Slit}} &
   \multicolumn{1}{c}{ {Spectr.}} &
   \multicolumn{1}{c}{ {}} \\

   \multicolumn{1}{c}{ {}} &
   \multicolumn{1}{c}{ {Tel.}} &
   \multicolumn{1}{c}{ {}} &   
   \multicolumn{1}{c}{ {Grating}} &
   \multicolumn{1}{c}{ {Width}} &
   \multicolumn{1}{c}{ {Resol.}} &
   \multicolumn{1}{c}{ {\#}} \\

\multicolumn{1}{c}{ {Obs.}} &
\multicolumn{1}{c}{ {Apert.}} &
\multicolumn{1}{c}{{Spectr.}} &
\multicolumn{1}{c}{{(l mm$^{-1}$)}} &
\multicolumn{1}{c}{{($\arcsec$)}} &
\multicolumn{1}{c}{{(\AA)}} &
\multicolumn{1}{c}{{spectra\tablenotemark{a}}} \\
 \midrule
OAN/SPM 	 & 2.1m       & B\&Ch & 300    & 2.5 & 8-10  & 110 (37)\\
OAGH	  	 & 2.1m       & B\&Ch & 150    & 2.5 & 10-15  & 36 (18)\\
\textit{SDSS}	  	 & 2.5m       &	MOS\tablenotemark{b} & 420  &\nodata & $\sim3$   & 23 (17) \\
\textit{HST}	  	 & 2.4m       & FOS\tablenotemark{c}   & G270H  & 2.0  & $\sim2$ & 40 (13)\\
         		 
\bottomrule
\tabnotetext{a}{The numbers in parentheses refer to data acquired from the sample MOJAVE-1.}
\tabnotetext{b}{MOS: multi-object fiber spectrographs with two channels red (420 l\,mm$^{-1}$) and  blue (640 l\,mm$^{-1}$). For specifications see \url{http://www.jhu.edu/~sdss/Spectrographs/OptLayout.html.}}
\tabnotetext{c}{Spectral resolution for the \textit{HST-FOS} data varies depending on the grating and slit width: e.g. G270H-- 0.25$\times$2.0$\arcsec$--1.92 \AA, G400H --4.3$\arcsec$--2.88\AA.}
  \end{tabular}
\end{table*}

\section{Data reduction and calibration}
\label{sec:data}
The data reduction was performed with the standard procedure using IRAF\footnote{IRAF is the Image Reduction and Analysis Facility made available to the astronomical community by the National Optical Astronomy Observatories, which are operated by AURA, Inc., under contract with the U.S. National Science Foundation. It is available at \url{http://iraf.noao.edu/}} routines for a long-slit spectroscopy, i.e., bias subtraction, flat-fielding, cosmic-ray removal, and sky subtraction, to produce the final spectra. We also observed  standard stars for flux calibration. Wavelength calibration was achieved via observations of  HeAr lamp at OAGH and HeAr/CuHeNeAr lamp at OAN--SPM. The arc spectra was obtained after an exposure (if single) or between exposures (if two or more consecutive exposures were taken), with the telescope still pointing towards the target. This calibration was accomplished by fitting a polynomial of suitable order to the pixel wavelength correlation with an uncertainty of $\sim$ 0.5 \AA\,rms for all cases; this was checked using the positions of background night sky lines. Flux calibration was performed using the cataloged spectrophotometric standards stars \citep{massey88,Ok90}. Usually, one or two stars at similar airmass of the principal target were observed on the same observing night. Atmospheric extinction correction was applied using the extinction curve for OAN--SPM\footnote{Determined by \citet{SP01}; available at \url{http://www.astrossp.unam.mx/indexspm.html}} and OAGH\footnote{\url{http://www.inaoep.mx/~astrofi/cananea/oagh-sky.html\#Extinction}}. Objects with more than one observation had their spectra stacked together to increase the signal-to-noise ratio (S/N). The average S/N ratio achieved was approximately 20, 30, and 10 in the continuum near 5100 \AA, 3000 \AA, and 1350 \AA, respectively.

\section{Analysis of spectra}
\label{sec:processing}

The general data processing of spectra involved several steps.
Once the spectra were flux calibrated, they were shifted to the rest frame of the source with the available redshift data. Then the iron contribution was subtracted and the local continuum emission was fitted with a power law.  Finally, the continuum and line parameters were measured. Each of these procedures are described in more detail in the following sections.

\subsection{Continuum and iron contamination subtraction}
\label{sec:sustraccion}

Since we are interested in studying the profiles of the emission lines  (H$\beta$,  \ion{Mg}{2}$\lambda$2798, and \ion{C}{4}$\lambda$1549) it is crucial to subtract the 
\ion{Fe}{2} emission in both optical and ultraviolet spectral regions. 
%%%%%%%%%%%%%%%%%%%%%%%%%%%%%%%%%%%%%%%%%%%%%%%%%%%%%%%%%%%%%%%%%%%%%%%%%%%%%%%%%%%%%%%%%%%%%%%%
\begin{figure*}\centering
\includegraphics[width=\linewidth]{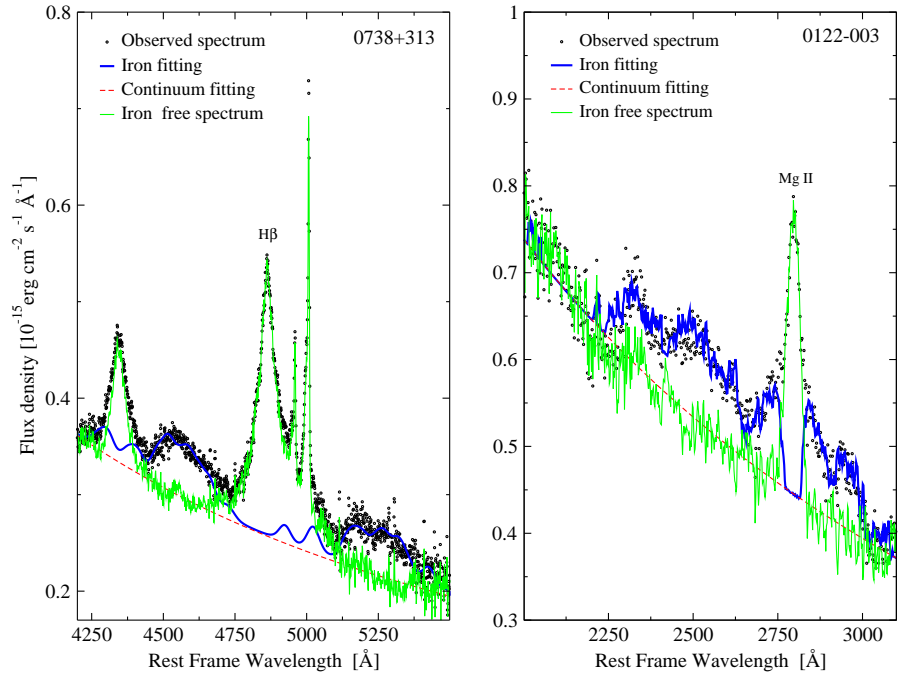}
\caption{Examples of the subtraction of \ion{Fe}{2} emission: \textit{left panel} in the H$\beta$ region for quasar 0738$+$313 and \textit{right panel} in the \ion{Mg}{ii} region for quasar 0122$-$003. It shows the observed spectrum in rest frame (circles); the \ion{Fe}{ii} fitting (\textit{blue solid line}); the power-law continuum fitting (\textit{red dashed line)}; the spectrum after  \ion{Fe}{ii} subtraction (\textit{green solid line}).  The  \ion{Mg}{2}  emission line  profile drastically changes  after removing  the iron contribution.\label{fig:4}} 
\end{figure*}

First, the continuum emission of the rest frame spectra was fitted by a power-law ($a\,\lambda^b + c$) in appropriately selected windows and subtracted using the Levenberg-Marquardt\footnote{We used the IDL routine  MPFIT from \url{http://cow.physics.wisc.edu/~craigm/idl/fitting.html}} least-squares minimal routine. Then, the resulting spectra is compared with the \ion{Fe}{2} templates in selected spectral regions, which can be modified for the corresponding  observed spectral resolution and broadening. The best fit of iron emission is subtracted from the spectra. 

The  following  \ion{Fe}{2}  templates of the NLS1 I Zw 1 galaxy ($z\,=\,$0.0611; FWHM of $\simeq\,$900 km\,s$^{-1}$)  were used:  (a) the template of \citet{VC04} based on spectra from the 4.2~m William Hershel and the 3.9~m Anglo-Australian telescopes for the optical band (from 3575--7530\, \AA); and (b) the template of \citet{VW01}, based on spectra from the \textit{HST-FOS}, for the UV-band (from 1250--3090\,\AA). These are the more accurate templates available in the literature. The intrinsic narrow lines of this source and its rich iron spectrum make the templates particularly suitable for use with AGN spectra. 

Possible continuum windows for each iron region of interest are presented in Table~\ref{tab3:Fe}. An example of the iron subtraction is illustrated in Figure~\ref{fig:4} for two sources in the H$\beta$ and \ion{Mg}{2} regions.

\begin{table*}[!t]\centering
 \setlength{\tabnotewidth}{0.8\columnwidth}
  \tablecols{3}
   \caption{Selected continuum and \ion{Fe}{2} windows} \label{tab3:Fe}
\begin{tabular}{ccc}
    \toprule
        & \multicolumn{2}{c}{Rest wavelength interval (\AA)}\\
    \cline{2-3}
    
    \multicolumn{1}{c}{ {Emission Line}} &
    \multicolumn{1}{c}{ {Continuum}} &
     \multicolumn{1}{c}{ {\ion{Fe}{2}}} \\

\midrule
H$\beta$\tablenotemark{a}	& 4210--4230 &  4400--4750   \\
$\lambda$4861               & 5080--5100 &  5150--5500  \\
                             &            &               \\
\ion{Mg}{2}\tablenotemark{a} & 2220--2230 & 2300--2650  \\
$\lambda$2798                & 3010--3040 & 2900--3090    \\
                             &            &               \\
\ion{C}{4}\tablenotemark{b}  & 1265--1290 &      \nodata       \\
$\lambda$1549                & 1340--1375 &       \nodata   \\
                             & 1425--1470 &    \nodata     \\
                             & 1680--1705 &    \nodata     \\
                             & 1950--2050 &    \nodata     \\
\bottomrule
    \tabnotetext{a}{Continuum windows from \citet{K02}.}
    \tabnotetext{b}{Continuum windows from \citet{VP06}.}
\end{tabular}
\end{table*}

\subsection{Continuum flux}
\label{subsec:continuum}
We obtained the mean value of the continuum emission flux density $f_{\lambda}$ centered at the given wavelength in an interval of $\pm\,50\,$\AA\, from the iron free spectrum of each AGN. In this manner, we estimated the flux density  $f_{5100}$ in the interval 5050 \AA$\,<\,\lambda\,<\,$5150 \AA, $f_{3000}$  in 2950 \AA$<\,\lambda\,<$3050 \AA, and   $f_{1350}$ in  1300 \AA$<\,\lambda\,<$1400 \AA. The associated error in the continuum flux density is $\sim$10~\% and depends on the S/N ratio of individual spectrum. In few particular cases, where the spectral interval was not enough to get a mean value, we estimated $f_{\lambda}$ by using the power-law  $f_{\lambda}$=$a\,\lambda^b + c$ that fits the continuum as described in \S~\ref{sec:sustraccion} above.

\subsection{Emission-line parameters}
The emission lines were fitted with Gaussian profiles \citep{V80}, assuming a broad and narrow components, which characterize the kinematics of the broad and narrow line regions (BLR and NLR, respectively). The minimal least-square fitting provides the central position of the Gaussian $\lambda_{c}$, full width at half maximum (FWHM), equivalent width (EW), and line flux.

Four Gaussian components were used for the spectral region of H$\beta$: two for H$\beta$ line (narrow and broad) and two for forbidden lines [\ion{O}{3}]\,$\lambda\lambda$4959, 5007. The narrow component (NC) of H$\beta$ (H$\beta_{\mathrm{NC}}$) was modelled and subtracted using the Gaussian profile fitted to the  [\ion{O}{3}] line,  assuming that the [\ion{O}{iii}] line ratio 
[\ion{O}{3}]$\lambda$5007/[\ion{O}{3}]$\lambda$4959 is  2.96 \citep{Os89}.  In this way, the   H$\beta_{\mathrm{NC}}$ FWHM was determined by the [\ion{O}{3}] FWHM.  The remaining parameters were set free in the fitting. In particular cases, where the narrow component position was not evident, it was fixed to the central wavelength of the H$\beta$ line. The fitting was done in the spectral interval 4700--5200~\AA.

For the  \ion{Mg}{2} and \ion{C}{4} lines, the restriction used was that the narrow component FWHM$\,\leq\,$2000\,km~s$^{-1}$. This condition was used by  \citet{MD04}, because there is no narrow line in the spectral range that could be modelled to subtract the narrow component  from the total line profile.  As in H$\beta$, if necessary, the position of the Gaussian NC was fixed in the fitting. Fitting of the \ion{Mg}{2} and \ion{C}{4} lines were performed in the spectral ranges 2600--3000\,\AA\, and 1450--1650\,\AA, respectively.  The local continuum in each case was also fitted in the spectral ranges previously mentioned. 

\begin{figure}[ht]
\centering
\includegraphics[width=\textwidth]{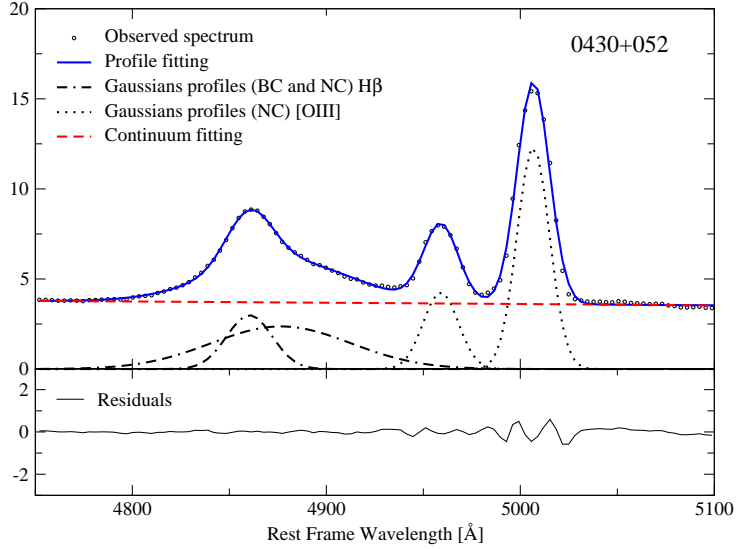}
\caption{Fitting of the H$\beta$\,$\lambda$4861 and [\ion{O}{3}] lines  profiles of the AGN~0430$+$052. The abscissa corresponds to the rest wavelength in \AA\, while the ordinate is the  flux density in units of 10$^{-15}$\,erg\,s$^{-1}$\,cm$^{-2}$\,\AA$^{-1}$. The upper panel shows the observed spectrum (\textit{empty circles}), the profile fitting (\textit{solid line}) comprises by the fitting for 4 Gaussians: 2 for  H$\beta$ (\textit{dotted lines and points}) and  2 for [\ion{O}{3}]\,$\lambda\lambda$4959,5007 (\textit{dotted lines}), where the intensity ratio was fixed to 2.96; and  the power-law local continuum fitting (\textit{dashed line}). The width of the narrow component (NC) of  H$\beta$ was fixed using the line [\ion{O}{3}] $\lambda$5007. In the lower panel are shown the residuals of the fit, with a mean value of 1.9~\% compared to the observed spectra.\label{fig:5}} 
\end{figure}

\begin{figure}[ht]\centering
\includegraphics[width=\textwidth]{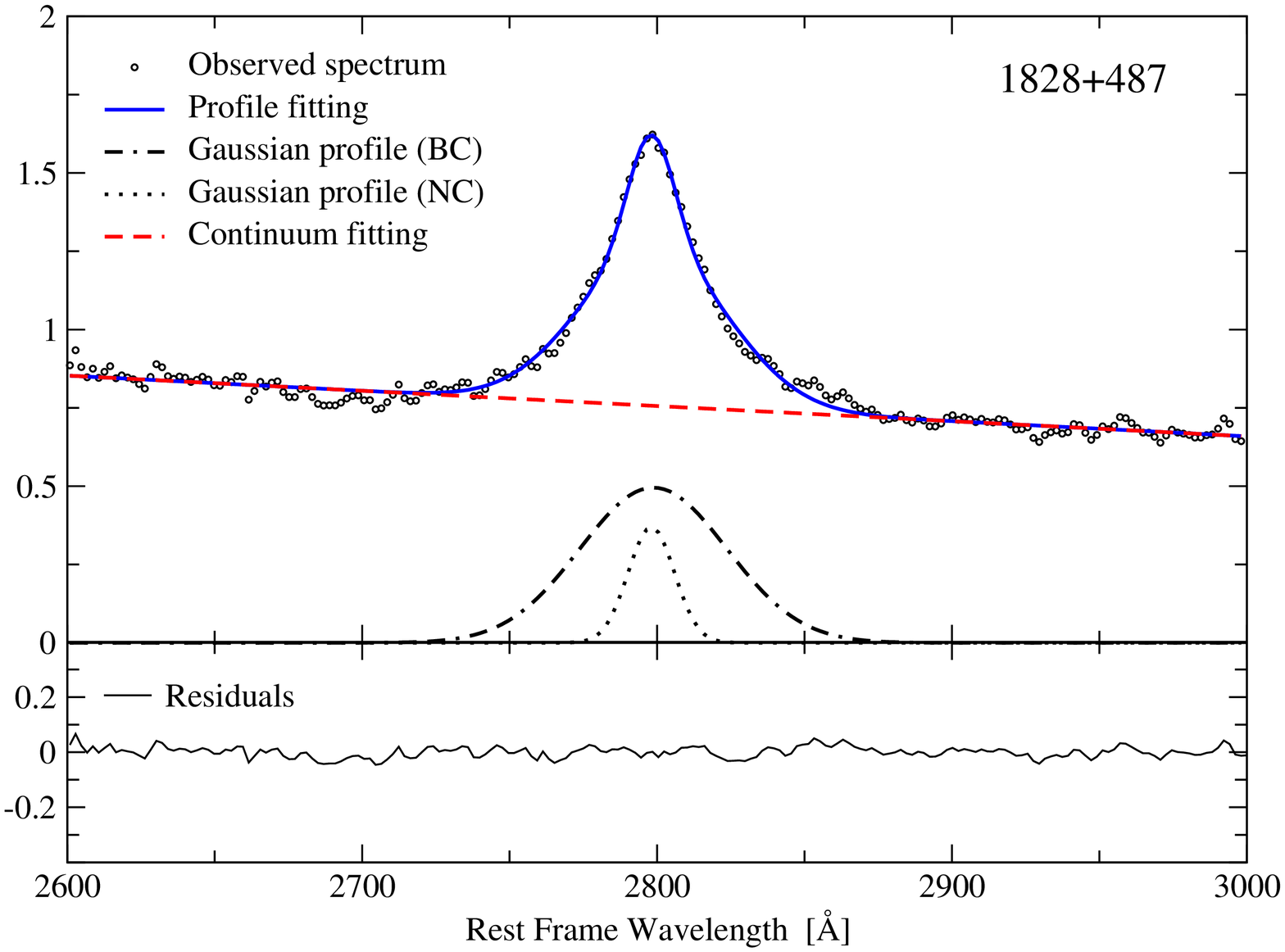}
\caption{Fitting of the \ion{Mg}{2}$\lambda$2798 line profile for the AGN~1828$+$487. The abscissa corresponds to the rest wavelength in \AA\, while the ordinate is the  flux density in units of 10$^{-15}$\,erg\,s$^{-1}$\,cm$^{-2}$\,\AA$^{-1}$.  The upper panel shows the profile fitting (\textit{solid line}) comprises by the fitting with 2 Gaussians curves and the power-law local continuum emission (\textit{dashed line}). The Gaussian functions represent the broad component (BC, \textit{dotted line and points}) and the narrow component (NC, \textit{dotted line})  of the line.  The width of the NC was restricted to  a FWHM\,$<$~2000\,km\,s$^{-1}$. In the lower panel are shown the residuals of the fit, with a mean value of 2.4~\% compared to the observed spectra.\label{fig:6}}
\end{figure}

\begin{figure}[ht]\centering
\includegraphics[width=\textwidth]{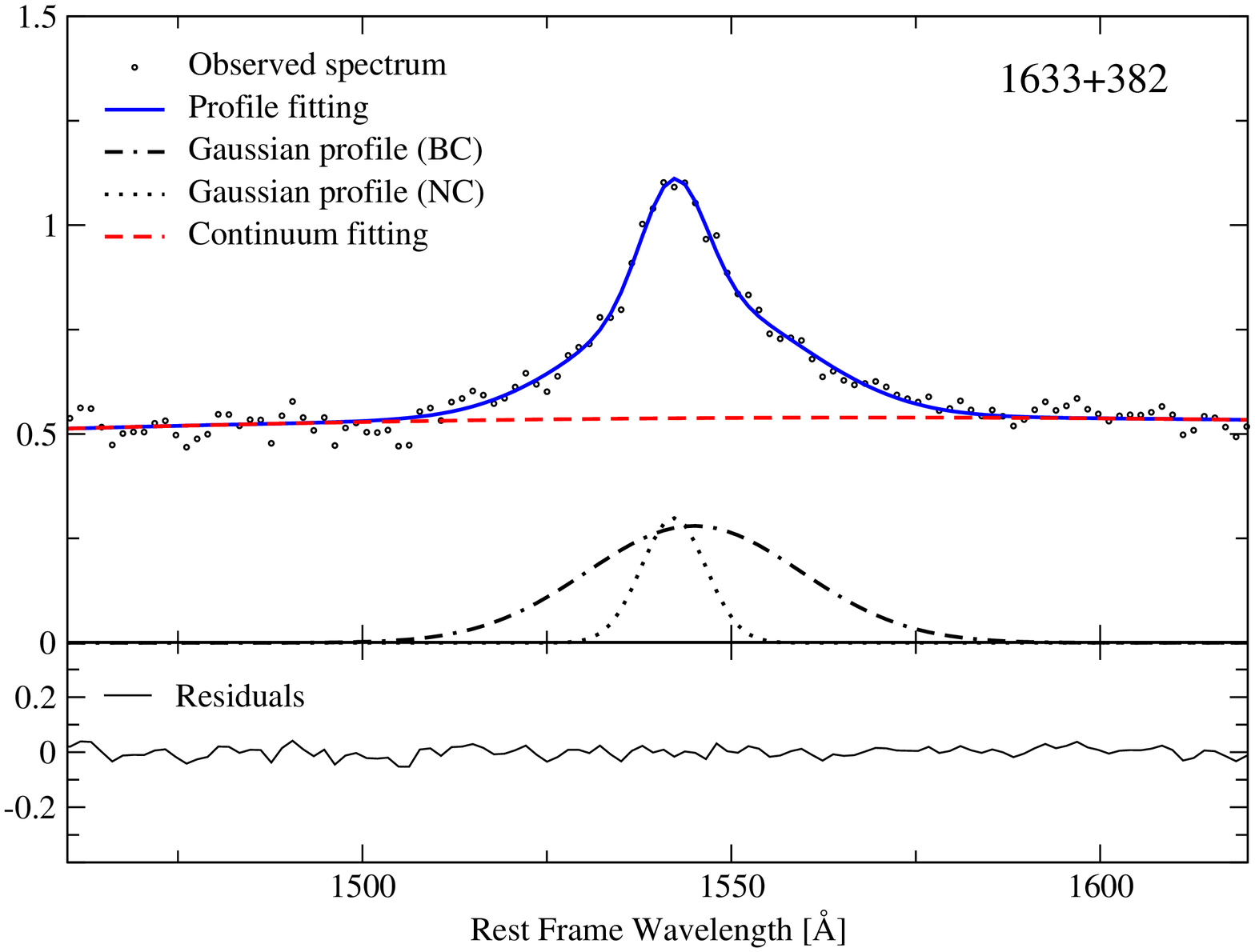}
\caption{Fitting of the \ion{C}{4} $\lambda$1549 profile for the AGN~1633$+$382. The symbols are the same as in Figure~\ref{fig:5}. Also the restriction for the NC was applied to the fit. In the lower panel, the residuals of the fit are shown, with a mean value of 3.4~\% compared to the observed spectra.\label{fig:7}}
\end{figure}

The obtained residuals in the fitting compared to the observed spectra were always less than 5~\% in the three subsamples. In Figures~\ref{fig:5}, \ref{fig:6}, and \ref{fig:7} an example of the fitting is shown for the lines H$\beta$, \ion{Mg}{2}, and \ion{C}{4}, respectively.

Once the FWHM of the lines was obtained, an instrumental resolution correction was applied by means of the equation:
\begin{equation}
\mathrm{FWHM}_{corrected}=(\mathrm{FWHM}_{observed}^2-\mathrm{FWHM}_{instrumental}^2)^{1/2}
\end{equation}
%%%%%%%%%%%%%%%%%%%%%%%%%%%%%%%%%%%%%%%%%%%%%%%%%%%%%%%%%%%%%%%%%%%%%%%%%%%%%%%%%%%%%%%%%%%%

The mean value of the instrumental resolution for the spectra obtained in OAN-SPM and OAGH was $\sim\,$8.5\,\AA\, and $\sim\,$15\,\AA, respectively.  For the \textit{SDSS} spectra a mean value of $\sim\,$3\,\AA\, was used for the instrumental resolution, and  2\,\AA\, was used for the \textit{HST-FOS} spectra. The instrumental resolution for the spectra of L96  was obtained from their Table 2, and for M03 spectra it was extracted from their Table~1.

%%%%%%%%%%%%%%%%%%%%%%%%%%%%%%%%%%%%%%%%%%%%%%%%%%%%%%%%%%%%%%%%%%%%%%%%%%%%%%%%%%%%%%%
%fig:ajuste_hb
%%%%%%%%%%%%%%%%%%%%%%%%%%%%%%%%%%%%
%fig:ajuste_mgii
%%%%%%%%%%%%%%%%%%%%%%%%%%%%%%%%%%%%

\subsection{Continuum and emission line luminosities.}\label{sec:luminosidades}

Once we had the continuum and emission line fluxes, the corresponding luminosities were calculated. First, the rest frame flux f$_{\lambda}$ was corrected for local reddening, applying the following expression: 
\begin{equation}
 \label{ec:flujo_cor} f_{c}= f_{\lambda}\,10^{({0.4}\,A_{V}\,K_{\lambda})},
\end{equation}
where  $f_{c}$ is the reddening corrected flux represented by a density flux in the case of continuum emission and integrated flux in the case of line emission. 
The reddening  $A_{V}$ was calculated assuming  $A_{V}$=$A_{B}/$1.32, where $A_B$ is the $B$-band extinction and  $K_{\lambda}$ is estimated from the Galaxy extinction curve of \citet{H83}. The $A_{B}$ values were obtained from the NED\footnote{NASA/IPAC Extragalactic Database.}, where dust extinction maps of \citet{SFD98} are used, which are based on the infrared difuse emission data from IRAS/DIRBE. The typical extinction for the H$\beta$, \ion{Mg}{2}, and \ion{C}{4}  subsamples is 0.566 mag, 0.342 mag, and 0.547 mag, respectively. The values of $A_{B}$ used are given in Table~\ref{tabA0:sample}  for each AGN.

The luminosity was calculated for the corrected fluxes with the usual relation $L=4\pi\,{d_{L}}^{2}f_{c}$, where $d_{L}$ is the luminosity distance of the object given by  equation~(\ref{eq:luminosity_distance}).

\subsection{Measurement of \ion{Fe}{2} emission} 
\label{subsec:FeII}

The identification of iron in the spectra of AGN dates from 40 years ago \citep[e.g.,][]{GS64}. \citet{J87} studied the physical conditions required to explain the observed \ion{Fe}{2} in AGN spectra.
Iron emission measurements in the optical region have shown that this emission (\ion{Fe}{2}$_{opt}$) is a distinctive parameter to separate type I from type II AGN \citep{M03}. In the ultraviolet \citet{W80} showed spectra of intermediate redshift quasars ($z\sim1$) with prominent \ion{Fe}{2} emission.

The \ion{Fe}{2} emission lines can be split into various wavelength bands. In the ultraviolet approximately 2000--3000\,\AA\, and 3000--3400\,\AA; and in the optical $\sim$4500--4700\,\AA\, and $\sim$5000--5400\,\AA\, \citep[e.g.][]{O77,P78a,J87}.

In the optical, the strongest feature, designated as   \ion{Fe}{2}\,$\lambda4570$ arises from lines in the range 4470--4670\,\AA\,  \citep[see][]{O77,P78a,J87}, while other lines contribute to the prominent \ion{Fe}{2} bands denoted as  \ion{Fe}{2}\,$\lambda5190$ and  \ion{Fe}{2}\,$\lambda5320$. The AGN redshifts and the instrumental setup used in our sample favours the detection of the \ion{Fe}{2}\,$\lambda4570$ multiplet, which is extensively studied in the literature \citep[e.g.][]{collin00,J87,M03,Z06,zhang07}

The equivalent width of the \ion{Fe}{2}\,$\lambda4570$ emission, denoted as EW(\ion{Fe}{2} $\lambda4570$), was calculated
via the usual definition EW(\ion{Fe}{2}\,$\lambda4570$)= $f_{FeII}/f_c$, where $f_{FeII}$ is the \ion{Fe}{2} total flux in the spectral range 4450--4600\,\AA, and $f_c$ is the continuum flux intensity measured in the interval 4520--4620\,\AA.  The mean value for the EW(\ion{Fe}{2} $\lambda4570$) is 18.2\,\AA\, with a standard deviation of 16.6\,\AA\,(c.f. Table~\ref{tab5:average_lines_c}).

In the ultraviolet at $\lambda\,<$ 3000\,\AA, the different regions involve the wings of the \ion{Mg}{2} $\lambda$2798 emission line. The blue part, in the range 2100--2800\,\AA, is formed by  \ion{Fe}{2}\,$\lambda$2100, \ion{Fe}{2}\,$\lambda$2500, and \ion{Fe}{2}\,$\lambda$2750 \citep{G81} and is denoted as \ion{Fe}{2} $\lambda2490$.  The first spectral region is also known as \ion{Fe}{2}\,$\lambda$2050, the second corresponds to  the range 2300--2600\,\AA, and the third spectral region was named by \citet{W80}, and partially includes  the \ion{Mg}{2}\,$\lambda$2798 line. The red part is called \ion{Fe}{2}\,$\lambda$2950 and \ion{Fe}{2}\,$\lambda$3200. The instrumental setup used and the AGN redshifts limit our study to the blue part of the spectrum. The same procedure used for deriving the \ion{Fe}{2} $\lambda4570$ parameters was used to estimated the \ion{Fe}{2} $\lambda2490$ parameters. The mean value estimated for EW(\ion{Fe}{2} $\lambda2490$) was 54.2\,\AA\, with a standard deviation of 25.8\,\AA.

\section{Results}
\label{sec:results}
\subsection{Photometric Data for MOJAVE/2cm AGN Sample}

The MOJAVE/2cm sample of 250 AGN is presented in the Appendix~\ref{secap:mojave},  together with the homogeneous photometric information at the B-band. Table~\ref{tabA0:sample} is organized as follows: 
Col. (1) is the source name,  Col. (2)--(3) are the J2000.0 coordinates,  Col. (4) is the spectroscopic classification from \citet{VV03} (B = BL Lac, G = radio galaxy, H = HPRQ, high polarized radio quasar, L = LPRQ, low polarized radio quasars, Q = Quasar with no polarization information), Col. (5) is the redshift, Col. (6) is the B-band extinction taken from NED, Col. (7) is the apparent B magnitude in Johnson's photometric system, Col. (8) is the absolute B-magnitude, Col. (9) is the MOJAVE monitoring identification (M1, M2 or 2cm; see Table~\ref{tab1:clasificacion}), and Col. (10) is the type of the radio spectra between 0.6 GHz and 22 GHz: $\alpha \ga$−0.7 F = flat, $\alpha <$−0.7 S = steep, C = compact steep spectra, G1 = gigahertz-peaked spectrum with one-sided VLBI jets, G2 = gigahertz-peaked spectrum with two-sided VLBI jets.

\subsection{Spectral Atlas for MOJAVE/2cm AGN Sample}
\label{sec:spectral}
The Spectral Atlas is presented in Appendix~\ref{sec:atlas}, it shows 142 spectra corresponding to 123 AGN of the MOJAVE/2cm sample in the observed frame with the redshift information for each source (see Figure~\ref{fig:atlas1}). The spectra are ordered by right ascension. If multiple spectra are available for a single source they are presented from blue to red  wavelengths. 

We present the continuum emission and/or line parameters for 41 sources in the  H$\beta$ region, 78 in the  \ion{Mg}{2} region, and 35 in the   \ion{C}{4} region. Also,  there are 14 sources with information available for both H$\beta$ and \ion{Mg}{2} regions, 12 with \ion{Mg}{2}, and \ion{C}{4}, and 5 with  H$\beta$, \ion{Mg}{2} and  \ion{C}{4}.  The information for the MOJAVE-1 sample that are included in the Atlas are as follows: 28 sources in the  H$\beta$ region, 46 in the  \ion{Mg}{2} region, and 23 in the \ion{C}{4} region.  Table~\ref{tab4:summary} summarizes the data presented for each AGN type.  Col. (1) is the sample for which we present the spectroscopic data, Col. (2) is the AGN type, Col. (3)--(5) show the different spectral regions of the continuum emission and/or line parameters for which we report data.

\begin{table*}[!t]\centering
  \setlength{\tabnotewidth}{1.0\columnwidth}
  \tablecols{6}
  % Stretch the space between table columns 
  \setlength{\tabcolsep}{2.0\tabcolsep}
  \caption{Summary of the spectroscopic data} \label{tab4:summary}
 \begin{tabular}{lcccc}
    \toprule
    
    \multicolumn{1}{c}{ {}} &
%\multicolumn{1}{c}{ {}} &
\multicolumn{1}{c}{ {}} &
\multicolumn{3}{c}{ {Spectral Region}} \\ \cline{3-5}

\multicolumn{1}{c}{ {Sample}} &
%\multicolumn{1}{c}{ {}} &
\multicolumn{1}{c}{ {Spectral Type}} &
\multicolumn{1}{c}{ {H$\beta$}} &
\multicolumn{1}{c}{ {\ion{Mg}{2}}} &
\multicolumn{1}{c}{ {\ion{C}{4}}} \\

\multicolumn{1}{c}{(1)} &
%\multicolumn{1}{c}{ {}} &
   \multicolumn{1}{c}{(2)} &
   \multicolumn{1}{c}{(3)} &
   \multicolumn{1}{c}{(4)} &
   \multicolumn{1}{c}{(5)} \\
    \midrule

MOJAVE/2cm        & BL    & 6  & \nodata & \nodata\\
                  & RG    & 12 & 2       & 1 \\
                  & HPRQ  & 8  & 19      & 9 \\
                  & LPRQ  & 7  & 18      & 6 \\
                  & Q     & 8  & 39      & 19 \\ \hline
             &    Total   & 41 & 78      & 35 \\[0.5ex] \hline
           
MOJAVE-1          & BL    & 4  & \nodata & \nodata\\
                  & RG    & 7  & 1       & \nodata \\
                  & HPRQ  & 7  & 16      & 9 \\
                 & LPRQ  & 6  & 14      & 6 \\
                  & Q     & 4  & 15      & 8 \\ \hline
                  & Total   & 28 & 46      & 23 \\
           \bottomrule

 \end{tabular}
\end{table*}

\subsection{Tabular data}
The results of measurements for each parameter of the emission lines and continuum emission for different regions are presented in the Tables~\ref{tabA1:hb1} to \ref{tabA5:CIV_1} of the Appendix~\ref{secap:parameters}. The format of these tables is explained in what follows.

Table~\ref{tabA1:hb1} presents the emission line parameters of the region around H$\beta$ for 41  AGN. Col. (1) is the source name, Col. (2) is the spectrum reference, Col. (3)--(6) are the parameters of the H$\beta$ broad component: FWHM, EW, total flux, and luminosity, respectively, Col. (7)--(10) are the parameters of the [\ion{O}{3}] $\lambda$5007 line: FWHM, EW, flux and luminosity, respectively. Nine of 41 AGN have only narrow emission lines with FWHM H$\beta\lesssim$ 1000 km\, s$^{-1}$: 0238$-$084, 0316$+$413, 0710$+$439, 0745$+$241, 0831$+$557, 1155$+$251, 1345$+$125, 1642$+$690, and 1957$+$405.

Table~\ref{tabA2:hb2} presents the parameters obtained for the emission continuum 5100\,\AA\, and the \ion{Fe}{2} $\lambda4570$: Col. (1) is the source same, Col. (2)--(3) are the flux density and luminosity at 5100\,\AA, respectively, Col. (4) is the spectral index of the local continuum, Col. (5)--(7) are the total flux, luminosity, and EW of the \ion{Fe}{2} $\lambda4570$, respectively.

Table~\ref{tabA3:MgII_1} presents the continuum emission at 3000\,\AA, and parameters  of the \ion{Mg}{2}\,$\lambda$2798 emission line for 78  AGN: Col. (1) is the source name, Col. (2) is the spectrum reference, Col. (3)--(6) are the parameters of the \ion{Mg}{2} broad component: FWHM, EW, total flux, and luminosity, respectively, Col. (7)--(8) are the flux density and luminosity at 3000\,\AA\,, respectively, and Col. (9) is the spectral index of the local continuum at 3000\,\AA.

The emission line parameters of the \ion{Fe}{2} $\lambda2490$ are presented in Table~\ref{tabA4:MgII_2}: Col. (1) is the source name, Col. (2)--(4) are the total flux, luminosity, and EW, respectively, and Col. (5) is the comment to  \ion{Fe}{2} $\lambda2490$  measurements.

The continuum emission at 1350\,\AA\, and parameters of the \ion{C}{4}\,$\lambda$1549 emission line for 35 AGN are presented in Table~\ref{tabA5:CIV_1}: Col. (1) is the source name, Col. (2) is the spectrum reference, Col. (3)--(6) are the parameters of the \ion{C}{4} broad component: FWHM, EW, total flux, and luminosity, respectively, Col. (7)--(8) are the flux density and luminosity at 1350\,\AA, and Col. (9) is the spectral index of the local continuum at 1350\,\AA.

\subsection{Descriptive statistics for different spectral regions}
\label{sec:statistics}
We have measured  the emission line parameters: FWHM, EW, flux, and luminosity, and we obtained the continuum emission flux density and luminosity using our Spectral Atlas of 123 AGN from the MOJAVE/2cm sample. In Table~\ref{tab5:average_lines_c} we present the descriptive statistics for each spectral parameter.  Columns are as follows: Col. (1) is the parameter, Col. (2) is the spectral type, Col. (3) is the number of sources, Col. (4) is the average value of the parameter, Col.(5) is the standard deviation of the data, and Col. (6) and (7) are the minimum and maximum values of the parameter, respectively. The numbers in parentheses refer to descriptive statistics for the flux-limited sample MOJAVE-1 (M1).

The statistical parameters of the H$\beta$ region are presented for both radio galaxies and quasars because the data for each spectral type are scarce (4 Galaxies, 7 HPRQ,  7 LPRQ, and 6 Q). Note that the radio galaxy 3C 390.3 was excluded from the analysis because of large FWHM H$\beta\,\rm (BC)\sim\,12,000\,km\,s^{-1}$, as well the BL Lacs and the narrow line objects. The \ion{C}{4} region (9 HPRQ, 6 LPRQ, and 19 Q) was treated similarly. For the \ion{Mg}{2} region the statistical analysis  was performed for different types of quasars (19 HPRQ, 18 LPRQ, and 39 Q). In this case, two radio galaxies were excluded: 0007$+$106 and 3C 390.3.

The broad component of the emission line \ion{C}{4} has a larger FWHM \ion{C}{4}\,(BC)\,=\,6498$\,\pm\,$1515\, km\,s$^{-1}$ value, between H$\beta$ and \ion{Mg}{2}. A caveat here is that the data of each emission line corresponds to a different AGN, and only few sources have data for more than one emission line, as was mentioned in \S~\ref{sec:spectral}.

The continuum luminosity at $\lambda\,L_{1350}$ is  larger by a factor of $\sim5$ and $\sim30$ than the continuum luminosities  $\lambda\,L_{3000}$ and $\lambda\,L_{5100}$, respectively.

The FWHM of \ion{Mg}{2} is quite similar for different types of quasars (HPRQ, LPRQ, and Q). This parameter has an average value of FWHM \ion{Mg}{2}$\rm\,(BC)\,=\,$5108$\,\pm\,$946\,km\,s$^{-1}$.  In contrast,  the EW of \ion{Mg}{2}$\rm\,(BC)\,=\,$42$\,\pm\,$29\,\AA\, shows a small difference between types. This parameter for quasars with no polarimetry information (Q) is smaller by 36\% and  25\% than those for HPRQ and LPRQ.  The \textit{KS}-test shows that the EW of the \ion{Mg}{2} distributions of the HPRQ and LPRQ are different at a confidence level of 95.1\% ($P_{KS}=0.049$), also the  distributions of the HPRQ and Q are different at a confidence level of 97.8\% ($P_{KS}=0.011$).

Finally, the \textit{KS}-test between the EW \ion{Fe}{2}\,$\lambda$2490 distributions of the HPRQ and LPRQ showed no significant difference  ($P_{KS}=0.314$). 

Further analysis related to the emission lines parameters and the properties of the pc-scale jets  will be presented in a forthcoming paper. 
\newpage
\begin{scriptsize}
\begin{longtable}{lcccccc}

\caption{Mean values of parameters for continuum and line emission}
 \label{tab5:average_lines_c} \\

%This is the header for the first page of the table...
\hline \hline \\[-2ex]
   \multicolumn{1}{c}{Parameter} &
   \multicolumn{1}{c}{Type} &
   \multicolumn{1}{c}{Number} &
   \multicolumn{1}{c}{Average} &
   \multicolumn{1}{c}{$\sigma$} &
   \multicolumn{1}{c}{Min.} &
   \multicolumn{1}{c}{Max.} \\
   
   \multicolumn{1}{c}{(1)} &
   \multicolumn{1}{c}{(2)} &
   \multicolumn{1}{c}{(3)} &
   \multicolumn{1}{c}{(4)} &
   \multicolumn{1}{c}{(5)} &
   \multicolumn{1}{c}{(6)} &
   \multicolumn{1}{c}{(7)}    
\\[0.5ex] \hline
 %  \\[-1.8ex]
\endfirsthead
 \hline \hline
\endlastfoot
%Now the data...
FWHM H$\beta\rm\, (BC)$\,[km\,s$^{-1}$]
	&	All	&	24	&	4055	&	1331	&	1670	&	8600	\\
	&		&	(20)	&	(3957)	&	(856)	&	(2650)	&	(5750)	\\
EW H$\beta\rm\, (BC)$\,[\AA]
	&	All	&	24	&	56	&	24	&	13	&	125	\\
	&		&	(20)	&	(57)	&	(26)	&	(13)	&	(125)	\\
$L_{\rm H\beta\rm\, (BC)}$\,[10$^{42}$ erg\,s$^{-1}$]
	&	All	&	24	&	40	&	50	&	1	&	182	\\
	&		&	(20)	&	(37)	&	(44)	&	(1)	&	(182)	\\\cline{1-7}

EW \ion{Fe}{2}\,$\lambda$4570\, [\AA]
	&	All	&	22	&	18	&	17	&	2	&	69	\\
	&		&	(17)	&	(19)	&	(17)	&	(2)	&	(69)	\\
$L_{4570}$\,[10$^{41}$ erg\,s$^{-1}$]
	&	All	&	22	&	96	&	96	&	2	&	342	\\
	&		&	(17)	&	(98)	&	(99)	&	(2)	&	(342)	\\\cline{1-7}

FWHM [\ion{O}{3}]\,$\lambda$5007\,[km\,s$^{-1}$]
	&	All	&	35	&	767	&	382	&	360	&	1716	\\
	&		&	(24)	&	(731)	&	(328)	&	(360)	&	(1394)	\\
EW [\ion{O}{3}]\,$\lambda$5007\,[\AA]
	&	All	&	35	&	39	&	62	&	0.2	&	380	\\
	&		&	(24)	&	(43)	&	(75)	&	(5)	&	(380)	\\
 $L_{[\ion{O}{3}]}$\,[10$^{42}$\,erg\,s$^{-1}$]	&	All	&	35	&	14	&	33	&	0.004	&	191	\\
	&		&	(24)	&	(10)	&	(13)	&	(0.004)	&	(50)	\\\cline{1-7}

$\lambda\,L_{5100}$\,[10$^{44}$\,erg\,s$^{-1}$]
	&	All	&	41	&	31	&	42	&	0.01	&	181	\\
	&		&	(28)	&	(34)	&	(43)	&	(0.006)	&	(181)	\\\cline{1-7}

FWHM \ion{Mg}{2}$\rm\, (BC)$\,[km\,s$^{-1}$] 
	&	All	&	76	&	5108	&	946	&	2445	&	8752	\\
	&		&	(45)	&	(5174)	&	(663)	&	(3735)	&	(7495)	\\
	&	HPRQ	&	19	&	5118	&	960	&	2565	&	7495	\\
	&		&	(16)	&	(5272)	&	(789)	&	(4245)	&	(7495)	\\
	&	LPRQ	&	18	&	5151	&	1063	&	3735	&	8752	\\
	&		&	(14)	&	(4991)	&	(599)	&	(3735)	&	(5583)	\\
	&	Q	&	39	&	5083	&	908	&	2445	&	6849	\\
	&		&	(15)	&	(5239)	&	(578)	&	(4287)	&	(6266)	\\\cline{1-7}

EW \ion{Mg}{2}$\rm\, (BC)$\,[\AA]
	&	All	&	76	&	36	&	24	&	6	&	177	\\
	&		&	(45)	&	(33)	&	(27)	&	(6)	&	(177)	\\
	&	HPRQ	&	19	&	27	&	18	&	6	&	74	\\
	&		&	(16)	&	(29)	&	(19)	&	(6)	&	(74)	\\
	&	LPRQ	&	18	&	31	&	12	&	15	&	53	\\
	&		&	(14)	&	(30)	&	(11)	&	(15)	&	(53)	\\
	&	Q	&	39	&	42	&	29	&	11	&	177	\\
	&		&	(15)	&	(40)	&	(41)	&	(13)	&	(177)	\\\cline{1-7}

$L_{\ion{Mg}{2} \rm (BC)}$\,[10$^{42}$\,erg\,s$^{-1}$]
	&	All	&	76	&	177	&	695	&	7	&	6091	\\
	&		&	(45)	&	(229)	&	(899)	&	(11)	&	(6091)	\\
	&	HPRQ	&	19	&	53	&	34	&	7	&	126	\\
	&		&	(16)	&	(50)	&	(30)	&	(11)	&	(126)	\\
	&	LPRQ	&	18	&	476	&	1407	&	35	&	6091	\\
	&		&	(14)	&	(587)	&	(1591)	&	(47)	&	(6091)	\\
	&	Q	&	39	&	99	&	97	&	12	&	438	\\
	&		&	(15)	&	(85)	&	(60)	&	(17)	&	(196)	\\\cline{1-7}

EW \ion{Fe}{2}\,$\lambda$2490\,[\AA]
	&	All	&	67	&	54	&	26	&	6	&	128	\\
	&		&	(40)	&	(50)	&	(26)	&	(6)	&	(128)	\\
	&	HPRQ	&	18	&	41	&	21	&	6	&	79	\\
	&		&	(15)	&	(39)	&	(22)	&	(6)	&	(79)	\\
	&	LPRQ	&	16	&	47	&	17	&	18	&	74	\\
	&		&	(12)	&	(47)	&	(17)	&	(18)	&	(67)	\\
	&	Q	&	33	&	65	&	27	&	21	&	128	\\
	&		&	(13)	&	(64)	&	(33)	&	(22)	&	(128)	\\\cline{1-7}

$L_{2490}$\,[10$^{42}$\,erg\,s$^{-1}$]	&	All	&	67	&	1614	&	10,772	&	31	&	88,402	\\
	&		&	(40)	&	(2456)	&	(13,941)	&	(40)	&	(88,402)	\\
	&	HPRQ	&	18	&	137	&	101	&	40	&	451	\\
	&		&	(15)	&	(118)	&	(68)	&	(40)	&	(275)	\\
	&	LPRQ	&	16	&	5863	&	22014	&	36	&	88402	\\
	&		&	(12)	&	(7760)	&	(25,400)	&	(100)	&	(88,402)	\\
	&	Q	&	33	&	359	&	535	&	31	&	2760	\\
	&		&	(13)	&	(257)	&	(153)	&	(46)	&	(521)	\\\cline{1-7}

$\lambda\,L_{3000}$\,[10$^{44}$\,erg\,s$^{-1}$]
	&	All	&	76	&	204	&	936	&	8	&	8205	\\
	&		&	(45)	&	(287)	&	(1212)	&	(8)	&	(8205)	\\
	&	HPRQ	&	19	&	88	&	70	&	13	&	237	\\
	&		&	(16)	&	(84)	&	(70)	&	(13)	&	(237)	\\
	&	LPRQ	&	18	&	592	&	1905	&	23	&	8205	\\
	&		&	(14)	&	(735)	&	(2155)	&	(33)	&	(8205)	\\
	&	Q	&	39	&	82	&	90	&	8	&	386	\\
	&		&	(15)	&	(84)	&	(73)	&	(8)	&	(223)	\\\cline{1-7}

FWHM \ion{C}{4}$\rm\, (BC)$\,[km\,s$^{-1}$]
	&	All	&	34	&	6498	&	1515	&	2818	&	9150	\\
	&		&	(23)	&	(6640)	&	(1500)	&	(3445)	&	(9150)	\\
EW \ion{C}{4}$\rm\, (BC)$\,[\AA]
	&	All	&	34	&	29	&	17	&	10	&	87	\\
	&		&	(23)	&	(29)	&	(17)	&	(10)	&	(87)	\\
$L_{\ion{C}{4}\rm\, (BC)}$[10$^{42}$\,erg\,s$^{-1}$]
	&	All	&	34	&	817	&	2063	&	20	&	11,345	\\
	&		&	(23)	&	(857)	&	(2314)	&	(27)	&	(11,345)	\\
$\lambda\,L_{1350}$\,[10$^{44}$\,erg\,s$^{-1}$]
	&	All	&	34	&	983	&	3363	&	5	&	19,566	\\
	&		&	(23)	&	(1199)	&	(4043)	&	(24)	&	(19,566)	\\
\end{longtable}
\end{scriptsize}

\section{Summary}

For the first time an optical spectroscopic atlas with intermediate resolution data of the bright part of the MOJAVE/2cm sample, comprised by core-dominated AGN at 15~GHz, is presented.

The  parameters obtained  from the spectra of 123 sources, such as FWHM, EW, fluxes, and luminosities of various  emission lines (H$\beta$, [\ion{O}{3}] $\lambda$5007, \ion{Mg}{2} $\lambda$2798, and/or \ion{C}{4} $\lambda$1549), and their corresponding continuum emission are presented together with the descriptive statistics of these parameters. The luminosities and equivalent widths of the \ion{Fe}{II} $\lambda$4570 and  \ion{Fe}{2} $\lambda$2490 are presented for 22 and 67 sources, respectively. 

We also carried out  a photometric  calibration that allowed us to present a homogeneous B--band photometric data for 242 AGN of the MOJAVE/2cm sample. Using these data, \citet{arshakian_2_10,arshakian10} have discussed the relations between optical, radio, and $\gamma$-ray emission of 135 AGN from the flux-density-limited MOJAVE-1 sample.  \citet{torrealba11} used the photometric data of 233 AGN of the MOJAVE/2cm sample to confirm the relations between the optical and 15\,GHz emission found for MOJAVE-1.  In a forthcoming paper, we will  present the results of analysis of the emission line parameters and the properties of the pc-scale jets. Preliminary results about this study have been published in \citet{A05}, \citet{torrealba08}, \citet{torrealbaphd2010}, and \citet{hovatta10}.

\begin{acknowledgments}

The authors want to acknowledge an anonymous referee for very useful comments and suggestions, which helped to improve this work. We also are grateful to Dr. C.~R. Lawrence for kindly providing us his spectroscopic data. We thank Marianne Vestergaard for kindly providing us the Fe II template which was used in this work. Special thanks are given to the technical staff and night assistant of the OAN--SPM and OAGH. This work is supported by CONACyT basic research grants 48484-F, 54480, 151494 (M\'exico).

This research has made use of (1) data that were acquired at Observatorio Astron\'omico Nacional in  San Pedro M\'artir (OAN--SPM), M\'exico and at Observatorio Astron\'omico Guillermo Haro, in Cananea, Sonora (OAGH); (3) the USNO-B catalog (Monet et al. 2003); (4) the MAPS Catalog of POSS I supported by the University of Minnesota (the APS databases can be accessed at http://aps.umn.edu/); (5) the NASA/IPAC Extragalactic Database (NED) which is operated by the Jet Propulsion Laboratory, California Institute of Technology, under contract with the National Aeronautics and Space Administration;  (6) The \textit{SDSS}. Funding for the \textit{SDSS}\footnote{The \textit{SDSS} Web Site is http://www.sdss.org/.} and \textit{SDSS-II} has been provided by the Alfred P. Sloan Foundation, the Participating Institutions, the National Science Foundation, the U.S. Department of Energy, the National Aeronautics and Space Administration, the Japanese Monbukagakusho, the Max Planck Society, and the Higher Education Funding Council for England. The \textit{SDSS} is managed by the Astrophysical Research Consortium for the Participating Institutions. The Participating Institutions are the American Museum of Natural History, Astrophysical Institute Potsdam, University of Basel, University of Cambridge, Case Western Reserve University, University of Chicago, Drexel University, Fermilab, the Institute for Advanced Study, the Japan Participation Group, Johns Hopkins University, the Joint Institute for Nuclear Astrophysics, the Kavli Institute for Particle Astrophysics and Cosmology, the Korean Scientist Group, the Chinese Academy of Sciences (LAMOST), Los Alamos National Laboratory, the Max-Planck-Institute for Astronomy (MPIA), the Max-Planck-Institute for Astrophysics (MPA), New Mexico State University, Ohio State University, University of Pittsburgh, University of Portsmouth, Princeton University, the United States Naval Observatory, and the University of Washington. (7)  The Multimission Archive at the Space Telescope Science Institute (MAST). STScI is operated by the Association of Universities for Research in Astronomy, Inc., under NASA contract NAS5-26555. Support for MAST for non-HST data is provided by the NASA Office of Space Science via grant NNX09AF08G and by other grants and contracts.

\end{acknowledgments}

\newpage
%\appendix
\begin{appendices}

\section{MOJAVE/2cm AGN sample}
\label{secap:mojave}

%\begin{landscape}
\begin{scriptsize}
\tablecols{10}
\setlength{\tabnotewidth}{0.5\columnwidth}
\tabcaption{\bf MOJAVE/2cm AGN SAMPLE} \label{tabA0:sample}
\def\ColumnHeaders{ 
   \multicolumn{1}{c}{  {IAU}} &
   \multicolumn{1}{c}{  {R.A.}} &
   \multicolumn{1}{c}{  {Dec.}} &   
   \multicolumn{1}{c}{  {Optical}} &
   \multicolumn{1}{c}{  {}} &
   \multicolumn{1}{c}{  {}} &
   \multicolumn{1}{c}{  {}} &
   \multicolumn{1}{c}{  {}} &
   \multicolumn{1}{c}{  {MOJAVE}} &
   \multicolumn{1}{c}{  {Radio}}\\

   \multicolumn{1}{c}{  {Name}} &
   \multicolumn{1}{c}{  {(2000)}} &
   \multicolumn{1}{c}{  {(2000)}} &   
   \multicolumn{1}{c}{  {Class}} &
   \multicolumn{1}{c}{  {z}} &
   \multicolumn{1}{c}{  {$A_{B}$}} &
   \multicolumn{1}{c}{  {$B_{J}$}} &
   \multicolumn{1}{c}{  {$M_{B}$}} & 
    \multicolumn{1}{c}{ {Id.}} & 
   \multicolumn{1}{c}{  {Spectra}} \\
}
\begin{longtable}{llcccccccc}
 %% Primera Cabeza
  \toprule
  \ColumnHeaders\\ \midrule
  \endfirsthead
  
  %% Otras Cabezas
  \tabcaptioncontinued
  \toprule
  \ColumnHeaders\\ \midrule
  \endhead
  
  %% Pies normales
  \bottomrule
  \endfoot
  0003$-$066	&	00 06 13.89	&	$-$06 23 35.33	&	B	&	0.347	&	0.157	&	20.12	&	$-$20.74	&	M1	&	F	\\
0007$+$106	&	00 10 31.03	&	$+$10 58 29.70	&	G	    &	0.089	&	0.422	&	13.54	&	$-$24.37	&	M1	&	F	\\
0014$+$813	&	00 17 8.475 &	$+$81 35 08.14  &	Q	    &	3.387	&	0.810	&	17.22	&	$-$28.44	&	2cm	&	F	\\
0016$+$731	&	00 19 45.78	&	$+$73 27 30.02	&	L	&	1.781	&	1.376	&	18.88	&	$-$25.46	&	M1	&	F	\\
0026$+$346	&	00 29 14.24	&	$+$34 56 32.26	&	G	&	0.517	&	0.466	&	20.99	&	$-$20.73	&	2cm	&	F	\\
0035$+$413	&	00 38 24.84	&	$+$41 37 06.00	&	Q	&	1.353	&	0.268	&	21.34	&	$-$22.43	&	2cm	&	F	\\
0039$+$230	&	00 42 4.54  &	$+$23 20 01.06  &	U	&	\nodata	&	0.124	&	20.08	&	\nodata	 &	2cm	&	F	\\
0048$-$097	&	00 50 41.33	&	$-$09 29 05.20	&	B	&	\nodata	&	0.139	&	16.79	&	\nodata	 &	M1	&	F	\\
0055$+$300	&	00 57 48.93	&	$+$30 21 08.20	&	G	&	0.017	&	0.279	&	13.05	&	$-$21.26	&	M2	&	F	\\
0059$+$581	&	01 02 45.76	&	$+$58 24 11.11	&	Q	&	0.644	&	2.374	&	19.24	&	$-$22.94	&	M1	&	F	\\
0106$+$013	&	01 08 38.77	&	$+$01 35 00.32	&	H	&	2.099	&	0.105	&	18.75	&	$-$25.94	&	M1	&	F	\\
0108$+$388	&	01 11 37.39	&	$+$39 06 28.10	&	G	&	0.668	&	0.204	&	16.17	&	$-$26.10	&	M2	&	G2	\\
0109$+$224	&	01 12 05.82	&	$+$22 44 38.79	&	B	&	0.265	&	0.161	&	15.60	&	$-$24.67 &	M1	&	F	\\
0112$-$017	&	01 15 17.10	&	$-$01 27 04.58	&	Q	&	1.365	&	0.267	&	17.47	&	$-$26.32	&	2cm	&	F	\\
0113$-$118	&	01 16 12.52	&	$-$11 36 15.43	&	Q	&	0.672	&	0.125	&	18.09	&	$-$24.19	&	M2	&	F	\\
0119$+$041	&	01 21 56.86	&	$+$04 22 24.73	&	H	&	0.637	&	0.152	&	18.64	&	$-$23.52	&	2cm	&	F	\\
0119$+$115	&	01 21 41.67	&	$+$11 49 50.60	&	Q	&	0.570	&	0.138	&	19.64	&	$-$22.29	&	M1	&	F	\\
0122$-$003	&	01 25 28.84	&	$-$00 05 55.96  &	Q	&	1.077	&	0.138	&	16.41	&	$-$26.88	&	M2	&	F	\\
0133$+$476	&	01 36 58.59	&	$+$47 51 29.10	&	H	&	0.859	&	0.658	&	17.43	&	$-$25.37	&	M1	&	F	\\
0133$-$203	&	01 35 37.51	&	$-$20 08 45.89  &	Q	&	1.141	&	0.075	&	18.31	&	$-$25.10	&	2cm	&	F	\\
0138$-$097	&	01 41 25.83	&	$-$09 28 43.67  &	B	&	0.733	&	0.127	&	17.68	&	$-$24.79	&	2cm	&	F	\\
0146$+$056	&	01 49 22.37	&	$+$05 55 53.57  &	Q	&	2.345	&	0.265	&	20.32	&	$-$24.59	&	2cm	&	F	\\
0149$+$218	&	01 52 18.06	&	$+$22 07 07.70	&	Q	&	1.320	&	0.324	&	19.37	&	$-$24.35	&	M2	&	F	\\
0153$+$744	&	01 57 34.88	&	$+$74 42 43.00	&	Q	&	2.341	&	2.090	&	17.98	&	$-$26.93	&	2cm	&	F	\\
0201$+$113	&	02 03 46.72	&	$+$11 34 45.60	&	Q	&	3.610	&	0.617	&	21.85	&	$-$23.94	&	2cm	&	F	\\
0202$+$149	&	02 04 50.41	&	$+$15 14 11.04	&	H	&	0.405	&	0.230	&	22.40	&	$-$18.79	&	M1	&	F	\\
0202$+$319	&	02 05 04.92	&	$+$32 12 30.10	&	L	&	1.466	&	0.254	&	16.54	&	$-$27.40	&	M1	&	F	\\
0212$+$735	&	02 17 30.81	&	$+$73 49 32.62	&	H	&	2.367	&	3.212	&	20.30	&	$-$24.63	&	M1	&	F	\\
0215$+$015	&	02 17 48.95	&	$+$01 44 49.60	&	H	&	1.715	&	0.144	&	20.17	&	$-$24.09	&	M1	&	F	\\
0218$+$357	&	02 21 05.47	&	$+$35 56 13.72	&	Q	&	0.944	&	0.293	&	20.28	&	$-$22.73	&	2cm	&	F	\\
0221$+$067	&	02 24 28.49	&	$+$06 59 23.50	&	Q	&	0.511	&	0.321	&	20.65	&	$-$21.04	&	2cm	&	F	\\
0224$+$671	&	02 28 50.05	&	$+$67 21 03.00	&	Q	&	0.523	&	4.443	&	20.91	&	$-$20.83	&	M1	&	F	\\
0234$+$285	&	02 37 52.40	&	$+$28 48 08.99	&	H	&	1.207	&	0.677	&	19.08	&	$-$24.45	&	M1	&	F	\\
0235$+$164	&	02 38 38.93	&	$+$16 36 59.28	&	B	&	0.940	&	0.341	&	20.21	&	$-$22.79	&	M1	&	F	\\
0238$-$084	&	02 41 04.71	&	$-$08 15 21.50	&	G	&	0.005	&	0.114	&	12.90	&	$-$18.76	&	M1	&	F	\\
0248$+$430	&	02 51 34.54	&	$+$43 15 15.83	&	Q	&	1.310	&	0.463	&	14.77	&	$-$28.93	&	2cm	&	F	\\
0300$+$470	&	03 03 35.24	&	$+$47 16 16.20	&	B	&	\nodata	&	1.132	&	18.21	&	\nodata	     &	M1	&	F	\\
0310$+$013	&	03 12 43.60	&	$+$01 33 17.54	&	Q	&	0.664	&	0.658	&	18.39	&	$-$23.86	&	2cm	&	F	\\
0316$+$162	&	03 18 57.76	&	$+$16 28 32.34	&	Q	&	0.907	&	0.673	&	23.31	&	\nodata	        &	2cm	&	G1	\\
0316$+$413	&	03 19 48.16	&	$+$41 30 42.10	&	G	&	0.018	&	0.703	&	12.85	&	$-$21.59	&	M1	&	F	\\
0333$+$321	&	03 36 30.49	&	$+$32 18 28.70	&	L	&	1.259	&	3.244	&	18.18	&	$-$25.44	&	M1	&	F	\\
0336$-$019	&	03 39 30.93	&	$-$01 46 35.80	&	H	&	0.852	&	0.377	&	17.31	&	$-$25.48	&	M1	&	F	\\
0355$+$508	&	03 59 29.74	&	$+$50 57 50.16	&	Q	&	1.510	&	6.362	&	\nodata	&	\nodata	        &	M2	&	F	\\
0402$-$362	&	04 03 53.75 &	$-$36 05 01.91  &	Q	&	1.417	&	0.022	&	16.81	&	$-$27.05	&	2cm	&	G	\\
0403$-$132	&	04 05 34.00	&	$-$13 08 13.60	&	H	&	0.571	&	0.250	&	17.71	&	$-$24.22	&	M1	&	F	\\
0405$-$385	&	04 06 59.07	&	$-$38 26 27.80	&	Q	&	1.285	&	0.024	&	19.18	&	$-$24.48	&	2cm	&	F	\\
0415$+$379	&	04 18 21.27	&	$+$38 01 35.51	&	G	&	0.049	&	7.107	&	20.36	&	$-$16.25	&	M1	&	S	\\
0420$+$022	&	04 22 52.22	&	$+$02 19 26.94	&	Q	&	2.277	&	0.936	&	20.41	&	$-$24.44	&	2cm	&	F	\\
0420$-$014	&	04 23 15.80	&	$-$01 20 33.06	&	H	&	0.914	&	0.567	&	17.65	&	$-$25.29	&	M1	&	F	\\
0422$+$004	&	04 24 46.84	&	   00 36 06.30	&	B	&	\nodata	&	0.436	&	16.56	&	\nodata	&	M1	&	F	\\
0429$+$415	&	04 32 36.50	&	$+$41 38 28.43	&	Q	&	1.022	&	2.413	&	19.13	&	$-$24.04	&	M2	&	C	\\
0430$+$052	&	04 33 11.08	&	$+$05 21 15.90	&	G	&	0.033	&	1.283	&	16.01	&	$-$19.74	&	M1	&	F	\\
0438$-$436	&	04 40 17.18	&	$-$43 33 08.60	&	Q	&	2.852	&	0.060	&	19.58	&	$-$25.73	&	2cm	&	F	\\
0440$-$003	&	04 42 38.66	&	$+$00 17 43.47	&	Q	&	0.844	&	0.227	&	19.47	&	$-$23.30	&	M2	&	F	\\
0446$+$112	&	04 49 07.67	&	$+$11 21 28.50	&	U	&	\nodata	&	2.170	&	21.00	&	\nodata	        &	M1	&	F	\\
0454$+$844	&	05 08 42.52	&	$+$84 32 04.50	&	B	&	0.112	&	0.412	&	18.61	&	$-$19.80	&	M2	&	F	\\
0454$-$234	&	04 57 03.18 &	$-$23 24 52.02	&	Q	&	1.003	&	0.204	&	18.58	&	$-$24.55	&	M2	&	F	\\
0458$-$020	&	05 01 12.81	&	$-$01 59 14.26	&	H	&	2.286	&	0.327	&	19.05	&	$-$25.82	&	M1	&	F	\\
0521$-$365	&	05 22 57.99	&	$-$36 27 31.00	&	G	&	0.055	&	0.169	&	12.35	&	$-$24.51	&	2cm	&	S	\\
0524$+$034	&	05 27 32.70	&	$+$03 31 31.45  &	B	&	\nodata	&	0.681	&	18.99	&	\nodata	      &	2cm	&	F	\\
0528$+$134	&	05 30 56.41	&	$+$13 31 55.18	&	H	&	2.070	&	3.621	&	20.37	&	$-$24.29	&	M1	&	F	\\
0529$+$075	&	05 32 38.99	&	$+$07 32 43.30	&	Q	&	1.254	&	1.352	&	19.75	&	$-$23.86	&	M1	&	F	\\
0529$+$483	&	05 33 15.86	&	$+$48 22 52.80	&	Q	&	1.162	&	1.767	&	20.62	&	$-$22.83	&	M1	&	F	\\
0537$-$286	&	05 39 54.28	&	$-$28 39 55.95	&	Q	&	3.104	&	0.106	&	20.00	&	$-$25.49	&	2cm	&	F	\\
0552$+$398	&	05 55 30.80	&	$+$39 48 49.16	&	Q	&	2.363	&	1.862	&	18.09	&	$-$26.84	&	M1	&	F	\\
0602$+$673	&	06 07 52.67	&	$+$67 20 55.42	&	Q	&	1.970	&	0.676	&	21.05	&	$-$23.50	&	2cm	&	F	\\
0605$-$085	&	06 07 59.69	&	$-$08 34 49.98	&	Q	&	0.872	&	2.572	&	18.77	&	$-$24.07	&	M1	&	F	\\
0607$-$157	&	06 09 40.93	&	$-$15 42 40.70	&	Q	&	0.324	&	1.107	&	18.77	&	$-$21.94	&	M1	&	F	\\
0615$+$820	&	06 26 03.04	&	$+$82 02 25.50	&	Q	&	0.710	&	0.357	&	18.39	&	$-$24.01	&	2cm	&	F	\\
0642$+$449	&	06 46 32.02	&	$+$44 51 16.59	&	Q	&	3.396	&	0.483	&	19.63	&	$-$26.05	&	M1	&	F	\\
0648$-$165	&	06 50 24.58	&	$-$16 37 39.70	&	U	&	\nodata	&	2.456	&	\nodata	&	\nodata	      &	M1	&	F	\\
0707$+$476	&	07 10 46.10	&	$+$47 32 11.14	&	Q	&	1.292	&	0.342	&	14.33	&	$-$29.34	&	M2	&	F	\\
0710$+$439	&	07 13 38.17	&	$+$43 49 17.00	&	G	&	0.518	&	0.309	&	19.98	&	$-$21.74	&	M2	&	G2	\\
0711$+$356 	&	07 14 24.82	&	$+$35 34 39.80	&	Q	&	1.620	&	0.246	&	17.85	&	$-$26.30	&	2cm	&	G	\\
0716$+$714	&	07 21 53.46	&	$+$71 20 36.20	&	B	&	\nodata	&	0.132	&	14.69	&	\nodata	 &	M1	&	F	\\
0723$-$008	&	07 25 50.64	&	$-$00 54 56.30	&	B	&	0.127	&	0.324	&	17.66	&	$-$21.02	&	M2	&	F	\\
0727$-$115	&	07 30 19.11	&	$-$11 41 12.60	&	Q	&	1.591	&	1.271	&	\nodata	&	\nodata	        &	M1	&	F	\\
0730$+$504	&	07 33 52.52	&	$+$50 22 09.00	&	Q	&	0.720	&	0.273	&	19.47	&	$-$22.96	&	M1	&	F	\\
0735$+$178	&	07 38 07.42	&	$+$17 42 19.20	&	B	&	\nodata	&	0.151	&	16.47	&	\nodata	 &	M1	&	F	\\
0736$+$017	&	07 39 18.09	&	$+$01 37 03.17	&	H	&	0.191	&	0.549	&	17.25	&	$-$22.31	&	M1	&	F	\\
0738$+$313	&	07 41 10.73	&	$+$31 11 59.10	&	Q	&	0.631	&	0.202	&	16.25	&	$-$25.89	&	M1	&	F	\\
0742$+$103	&	07 45 33.06	&	$+$10 11 12.69	&	Q	&	2.624	&	0.111	&	\nodata	&	\nodata	        &	M1	&	G1	\\
0745$+$241	&	07 48 36.11	&	$+$24 00 24.15	&	H	&	0.409	&	0.244	&	19.41	&	$-$21.80	&	M2	&	F	\\
0748$+$126	&	07 50 52.04	&	$+$12 31 04.83	&	Q	&	0.889	&	0.133	&	17.49	&	$-$25.39	&	M1	&	F	\\
0754$+$100	&	07 57 06.67	&	$+$09 56 34.00	&	B	&	0.266	&	0.097	&	15.29	&	$-$24.99	&	M1	&	F	\\
0804$+$499	&	08 08 39.66	&	$+$49 50 36.50	&	H	&	1.436	&	0.228	&	14.90	&	$-$28.99	&	M1	&	F	\\
0805$-$077	&	08 08 15.53	&	$-$07 51 09.80	&	Q	&	1.837	&	0.613	&	17.96	&	$-$26.45	&	M1	&	F	\\
0808$+$019	&	08 11 26.71	&	$+$01 46 52.30	&	B	&	1.148	&	0.142	&	18.03	&	$-$25.39	&	M1	&	F	\\
0814$+$425	&	08 18 16.00	&	$+$42 22 45.41	&	B	&	0.245	&	0.274	&	20.05	&	$-$20.05	&	M1	&	F	\\
0821$+$394	&	08 24 55.48	&	$+$39 16 41.90	&	Q	&	1.216	&	0.197	&	18.19	&	$-$25.35	&	2cm	&	F	\\
0823$+$033	&	08 25 50.33	&	$+$03 09 24.52	&	B	&	0.506	&	0.197	&	15.53	&	$-$26.14	&	M1	&	F	\\
0827$+$243	&	08 30 52.08	&	$+$24 10 59.80	&	L	&	0.940	&	0.142	&	17.01	&	$-$25.99	&	M1	&	F	\\
0829$+$046	&	08 31 48.89	&	$+$04 29 38.70	&	B	&	0.174	&	0.141	&	14.64	&	$-$24.80	&	M1	&	F	\\
0831$+$557 	&	08 34 54.90	&	$+$55 34 21.07	&	U	&	0.240	&	0.189	&	19.51	&	$-$20.55	&	2cm	&	F	\\
0834$-$201	&	08 36 39.26	&	$-$20 16 59.70	&	Q	&	2.752	&	0.432	&	19.36	&	$-$25.88	&	M2	&	F	\\
0836$+$710	&	08 41 24.37	&	$+$70 53 42.10	&	L	&	2.218	&	0.132	&	16.78	&	$-$28.02	&	M1	&	F	\\
0838$+$133 	&	08 40 47.68	&	$+$13 12 23.88	&	Q	&	0.681	&	0.403	&	17.89	&	$-$24.43	&	M1	&	F	\\
0850$+$581	&	08 54 41.99	&	$+$57 57 29.95	&	L	&	1.322	&	0.232	&	18.24	&	$-$25.48	&	M2	&	F	\\
0851$+$202	&	08 54 48.88	&	$+$20 06 30.70	&	B	&	0.306	&	0.122	&	14.53	&	$-$26.05	&	M1	&	F	\\
0859$+$470	&	09 03 04.05	&	$+$46 51 04.10	&	L	&	1.462	&	0.085	&	18.61	&	$-$25.32	&	2cm	&	F	\\
0859$-$140	&	09 02 16.83	&	$-$14 15 30.80	&	Q	&	1.339	&	0.269	&	16.46	&	$-$27.29	&	M2	&	C	\\
0906$+$015	&	09 09 10.09	&	$+$01 21 35.80	&	H	&	1.024	&	0.148	&	16.76	&	$-$26.41	&	M1	&	F	\\
0917$+$449	&	09 20 58.45	&	$+$44 41 53.98	&	Q	&	2.180	&	0.089	&	18.82	&	$-$25.94	&	M2	&	F	\\
0917$+$624	&	09 21 36.23	&	$+$62 15 52.10	&	Q	&	1.446	&	0.203	&	19.63	&	$-$24.28	&	M1	&	F	\\
0919$-$260	&	09 21 29.35	&	$-$26 18 43.38	&	Q	&	2.300	&	0.574	&	18.90	&	$-$25.97	&	2cm	&	G	\\
0923$+$392	&	09 27 03.01	&	$+$39 02 20.85	&	L	&	0.695	&	0.062	&	16.61	&	$-$25.75	&	M1	&	F	\\
0945$+$408	&	09 48 55.33	&	$+$40 39 44.60	&	L	&	1.249	&	0.060	&	17.62	&	$-$25.98	&	M1	&	F	\\
0953$+$254	&	09 56 49.89	&	$+$25 15 16.00	&	L	&	0.712	&	0.158	&	16.73	&	$-$25.67	&	M2	&	F	\\
0954$+$658	&	09 58 47.24	&	$+$65 33 54.82	&	B	&	0.367	&	0.495	&	16.62	&	$-$24.36	&	M2	&	F	\\
0955$+$476	&	09 58 19.67	&	$+$47 25 07.80	&	Q	&	1.882	&	0.064	&	18.31	&	$-$26.14	&	M1	&	F	\\
1012$+$232	&	10 14 47.09	&	$+$23 01 16.80	&	Q	&	0.565	&	0.106	&	17.50	&	$-$24.41	&	2cm	&	F	\\
1015$+$359	&	10 18 10.98	&	$+$35 42 39.44	&	Q	&	1.226	&	0.050	&	17.94	&	$-$25.62	&	M2	&	F	\\
1032$-$199	&	10 35 02.21	&	$-$20 11 34.70	&	Q	&	2.198	&	0.192	&	18.00	&	$-$26.78	&	2cm	&	F	\\
1034$-$293	&	10 37 16.08 &	$-$29 34 02.81  &	Q	&	0.312	&	0.221	&	17.52	&	$-$23.11	&	M2	&	F	\\
1036$+$054	&	10 38 46.77	&	$+$05 12 29.00	&	Q	&	0.473	&	0.110	&	20.20	&	$-$21.32	&	M1	&	F	\\
1038$+$064	&	10 41 17.16	&	$+$06 10 16.90	&	Q	&	1.265	&	0.107	&	16.46	&	$-$27.17	&	M1	&	F	\\
1045$-$188	&	10 48 06.62	&	$-$19 09 35.70	&	Q	&	0.595	&	0.164	&	17.77	&	$-$24.25	&	M1	&	F	\\
1049$+$215	&	10 51 48.79	&	$+$21 19 52.35	&	Q	&	1.300	&	0.109	&	18.23	&	$-$25.45	&	2cm	&	F	\\
1055$+$018	&	10 58 29.60	&	$+$01 33 58.82	&	H	&	0.890	&	0.116	&	17.90	&	$-$24.98	&	M1	&	F	\\
1055$+$201	&	10 58 17.90	&	$+$19 51 51.40	&	Q	&	1.110	&	0.106	&	18.51	&	$-$24.84	&	2cm	&	F	\\
1101$+$384	&	11 04 27.34	&	$+$38 12 31.50	&	B	&	0.031	&	0.066	&	13.61	&	$-$22.01	&	M2	&	F	\\
1116$+$128	&	11 18 57.30	&	$+$12 34 41.72	&	Q	&	2.118	&	0.109	&	18.40	&	$-$26.30	&	2cm	&	F	\\
1124$-$186	&	11 27 04.46	&	$-$18 57 17.80	&	H	&	1.048	&	0.187	&	19.43	&	$-$23.80	&	M1	&	F	\\
1127$-$145	&	11 30 07.05	&	$-$14 49 27.00	&	Q	&	1.184	&	0.158	&	16.75	&	$-$26.74	&	M1	&	F	\\
1128$+$385	&	11 30 53.28	&	$+$38 15 18.55	&	Q	&	1.733	&	0.109	&	19.12	&	$-$25.17	&	M2	&	F	\\
1144$+$402	&	11 46 58.30	&	$+$39 58 34.30	&	Q	&	1.089	&	0.075	&	18.61	&	$-$24.70	&	2cm	&	F	\\
1145$-$071	&	11 47 51.62	&	$-$07 24 41.40	&	Q	&	1.342	&	0.180	&	18.43	&	$-$25.32	&	2cm	&	F	\\
1148$-$001	&	11 50 43.86	&	$-$00 23 54.38	&	Q	&	1.980	&	0.098	&	16.84	&	$-$27.72	&	M2	&	F	\\
1150$+$812	&	11 53 12.49	&	$+$80 58 29.10	&	Q	&	1.250	&	0.332	&	19.73	&	$-$23.87	&	M1	&	F	\\
1155$+$251	&	11 58 25.82	&	$+$24 50 18.00	&	Q	&	0.202	&	0.086	&	19.66	&	$-$20.03	&	2cm	&	F	\\
1156$+$295	&	11 59 31.83	&	$+$29 14 44.00	&	H	&	0.729	&	0.084	&	16.94	&	$-$25.51	&	M1	&	F	\\
1213$-$172	&	12 15 46.75	&	$-$17 31 45.40	&	U	&	\nodata	&	0.253	&	\nodata	&	\nodata	        &	M1	&	F	\\
1219$+$044	&	12 22 22.50	&	$+$04 13 16.00	&	H	&	0.965	&	0.085	&	17.27	&	$-$25.78	&	M1	&	F	\\
1219$+$285	&	12 21 31.70	&	$+$28 13 58.40	&	B	&	0.102	&	0.097	&	17.09	&	$-$21.11	&	M2	&	F	\\
1222$+$216	&	12 24 54.45	&	$+$21 22 46.30	&	Q	&	0.432	&	0.101	&	17.77	&	$-$23.57	&	M1	&	F	\\
1226$+$023	&	12 29 06.70	&	$+$02 03 08.60	&	L	&	0.158	&	0.089	&	16.52	&	$-$22.66	&	M1	&	F	\\
1228$+$126	&	12 30 49.42	&	$+$12 23 28.00	&	G	&	0.004	&	0.096	&	11.12	&	$-$20.05	&	M1	&	S	\\
1244$-$255	&	12 46 46.80	&	$-$25 47 49.29	&	H	&	0.633	&	0.375	&	17.31	&	$-$24.86	&	M2	&	F	\\
1253$-$055	&	12 56 11.16	&	$-$05 47 21.52	&	H	&	0.536	&	0.123	&	16.02	&	$-$25.78	&	M1	&	F	\\
1255$-$316	&	12 57 59.06	&	$-$31 55 16.85	&	Q	&	1.924	&	0.380	&	18.64	&	$-$25.86	&	2cm	&	F	\\
1302$-$102	&	13 05 33.01	&	$-$10 33 19.30	&	L	&	0.278	&	0.184	&	15.20	&	$-$25.18	&	M2	&	F	\\
1308$+$326	&	13 10 28.66	&	$+$32 20 43.78	&	H	&	0.997	&	0.060	&	19.83	&	$-$23.29	&	M1	&	F	\\
1313$-$333	&	13 16 07.99	&	$-$33 38 59.17	&	Q	&	1.210	&	0.265	&	18.22	&	$-$25.31	&	2cm	&	F	\\
1323$+$321	&	13 26 16.51	&	$+$31 54 09.52	&	G	&	0.370	&	0.065	&	21.21	&	$-$19.78	&	2cm	&	G2	\\
1324$+$224	&	13 27 00.86	&	$+$22 10 50.10	&	Q	&	1.400	&	0.072	&	17.95	&	$-$25.89	&	M1	&	F	\\
1328$+$307	&	13 31 08.31	&	$+$30 30 32.90	&	Q	&	0.846	&	0.050	&	17.12	&	$-$25.65	&	2cm	&	C	\\
1334$-$127	&	13 37 39.80	&	$-$12 57 24.70	&	H	&	0.539	&	0.323	&	18.73	&	$-$23.08	&	M1	&	F	\\
1345$+$125	&	13 47 33.49	&	$+$12 17 23.50	&	G	&	0.121	&	0.145	&	17.33	&	$-$21.24	&	M2	&	G2	\\
1354$+$195	&	13 57 04.45	&	$+$19 19 07.30	&	Q	&	0.719	&	0.260	&	16.03	&	$-$26.39	&	2cm	&	F	\\
1354$-$152	&	13 57 11.33	&	$-$15 27 29.00	&	Q	&	1.890	&	0.406	&	17.53	&	$-$26.94	&	2cm	&	F	\\
1402$+$044	&	14 05 01.12 &	$+$4 15 35.82	&	Q	&	3.211	&	0.115	&	20.62	&	$-$24.94	&	2cm	&	F	\\
1404$+$286	&	14 07 00.39	&	$+$28 27 14.00	&	G	&	0.077	&	0.079	&	16.84	&	$-$20.75	&	M2	&	G2	\\
1413$+$135	&	14 15 58.81	&	$+$13 20 23.71	&	B	&	0.247	&	0.106	&	21.41	&	$-$18.71	&	M1	&	F	\\
1417$+$385	&	14 19 46.61	&	$+$38 21 48.40	&	Q	&	1.831	&	0.036	&	20.35	&	$-$24.05	&	M1	&	F	\\
1418$+$546	&	14 19 46.60	&	$+$54 23 14.79	&	B	&	0.152	&	0.058	&	14.79	&	$-$24.28	&	M2	&	F	\\
1424$+$366	&	14 26 37.08	&	$+$36 25 09.59	&	Q	&	1.091	&	0.035	&	18.89	&	$-$24.42	&	2cm	&	F	\\
1458$+$718	&	14 59 07.61	&	$+$71 40 19.90	&	L	&	0.904	&	0.110	&	16.95	&	$-$25.96	&	M1	&	C	\\
1502$+$106	&	15 04 25.02	&	$+$10 29 39.20	&	H	&	1.839	&	0.138	&	19.60	&	$-$24.81	&	M1	&	F	\\
1504$+$377	&	15 06 09.61	&	$+$37 30 51.20	&	Q	&	0.674	&	0.055	&	21.49	&	$-$20.80	&	2cm	&	F	\\
1504$-$166	&	15 07 04.88	&	$-$16 52 30.50	&	H	&	0.876	&	0.410	&	20.19	&	$-$22.66	&	M1	&	F	\\
1508$-$055	&	15 10 53.59	&	$-$05 43 07.10	&	Q	&	1.191	&	0.367	&	17.18	&	$-$26.32	&	M2	&	C	\\
1510$-$089	&	15 12 50.53	&	$-$09 05 59.70	&	H	&	0.360	&	0.416	&	16.10	&	$-$24.84	&	M1	&	F	\\
1511$-$100	&	15 13 44.98	&	$+$10 12 00.40	&	Q	&	1.513	&	0.456	&	18.77	&	$-$25.23	&	2cm	&	F	\\
1514$+$004	&	15 16 40.22	&	$+$00 15 01.91  &	G	&	0.052	&	0.238	&	17.05	&	$-$19.69	&	M2	&	F	\\
1514$-$241	&	15 17 41.80	&	$-$24 22 19.60	&	B	&	0.049	&	0.595	&	16.91	&	$-$19.70	&	M2	&	F	\\
1519$-$273	&	15 22 37.77	&	$-$27 30 11.00	&	B	&	1.297	&	1.026	&	18.12	&	$-$25.56	&	2cm	&	F	\\
1532$+$016	&	15 34 52.45	&	$+$01 31 04.21	&	Q	&	1.420	&	0.218	&	19.67	&	$-$24.20	&	M2	&	F	\\
1538$+$149	&	15 40 49.51	&	$+$14 47 46.00	&	B	&	0.605	&	0.238	&	15.91	&	$-$26.14	&	M1	&	F	\\
1546$+$027	&	15 49 29.43	&	$+$02 37 01.16	&	H	&	0.414	&	0.495	&	17.95	&	$-$23.28	&	M1	&	F	\\
1548$+$056	&	15 50 35.27	&	$+$05 27 10.47	&	H	&	1.422	&	0.289	&	18.16	&	$-$25.71	&	M1	&	F	\\
1555$+$001	&	15 57 51.52	&	$-$00 01 50.50	&	Q	&	1.772	&	0.602	&	19.55	&	$-$24.78	&	2cm	&	F	\\
1606$+$106	&	16 08 46.20	&	$+$10 29 07.78	&	Q	&	1.226	&	0.270	&	18.10	&	$-$25.46	&	M1	&	F	\\
1607$+$268	&	16 09 13.32 &	$+$26 41 29.04	&	G	&	0.473	&	0.228	&	\nodata	&	\nodata	&	M2	&	G1	\\
1611$+$343	&	16 13 41.07	&	$+$34 12 48.10	&	L	&	1.397	&	0.077	&	17.67	&	$-$26.17	&	M1	&	F	\\
1622$-$253	&	16 25 46.89	&	$-$25 27 38.30	&	Q	&	0.786	&	4.960	&	20.89	&	$-$21.72	&	M2	&	F	\\
1622$-$297	&	16 26 06.02 &	$-$29 51 26.97	&	Q	&	0.815	&	1.861	&	18.06	&	$-$24.63	&	M2	&	F	\\
1624$+$416 	&	16 25 57.67 &	$+$41 34 40.63  &	Q	&	2.550	&	0.035	&	20.00	&	$-$25.09	&	2cm	&	F	\\
1633$+$382	&	16 35 15.49	&	$+$38 08 04.50	&	H	&	1.814	&	0.048	&	18.19	&	$-$26.18	&	M1	&	F	\\
1637$+$574	&	16 38 13.45	&	$+$57 20 23.90	&	L	&	0.751	&	0.054	&	16.68	&	$-$25.84	&	M1	&	F	\\
1638$+$398	&	16 40 29.63	&	$+$39 46 46.03	&	Q	&	1.666	&	0.044	&	18.49	&	$-$25.71	&	M1	&	F	\\
1641$+$399	&	16 42 58.81	&	$+$39 48 36.90	&	H	&	0.593	&	0.057	&	15.99	&	$-$26.02	&	M1	&	F	\\
1642$+$690	&	16 42 07.84	&	$+$68 56 39.76	&	H	&	0.751	&	0.165	&	20.02	&	$-$22.50	&	M2	&	F	\\
1652$+$398	&	16 53 52.26	&	$+$39 45 36.70	&	B	&	0.033	&	0.084	&	15.15	&	$-$20.60	&	M2	&	F	\\
1655$+$077	&	16 58 09.01	&	$+$07 41 27.54	&	H	&	0.621	&	0.661	&	20.13	&	$-$21.98	&	M1	&	F	\\
1656$+$053	&	16 58 33.48	&	$+$05 15 16.40	&	H	&	0.879	&	0.684	&	16.84	&	$-$26.01	&	2cm	&	F	\\
1656$+$477	&	16 58 02.77	&	$+$47 37 49.24	&	Q	&	1.622	&	0.091	&	17.15	&	$-$27.00	&	2cm	&	F	\\
1726$+$455	&	17 27 27.65	&	$+$45 30 39.70	&	Q	&	0.717	&	0.105	&	18.79	&	$-$23.62	&	M1	&	F	\\
1730$-$130	&	17 33 02.70	&	$-$13 04 49.55	&	H	&	0.902	&	2.203	&	21.56	&	$-$21.35	&	M1	&	F	\\
1739$+$522	&	17 40 36.07	&	$+$52 11 43.50	&	H	&	1.379	&	0.153	&	18.20	&	$-$25.61	&	M1	&	F	\\
1741$-$038	&	17 43 58.85	&	$-$03 50 04.62	&	H	&	1.054	&	2.457	&	19.76	&	$-$23.49	&	M1	&	F	\\
1749$+$096	&	17 51 32.82	&	$+$09 39 00.60	&	B	&	0.322	&	0.779	&	18.48	&	$-$22.20	&	M1	&	F	\\
1749$+$701	&	17 48 32.90	&	$+$70 05 50.70	&	B	&	0.770	&	0.133	&	16.41	&	$-$26.16	&	M2	&	F	\\
1751$+$288	&	17 53 42.47	&	$+$28 48 04.90	&	Q	&	1.118	&	0.250	&	20.77	&	$-$22.59	&	M1	&	F	\\
1758$+$388	&	18 00 24.76	&	$+$38 48 30.70	&	Q	&	2.092	&	0.116	&	17.76	&	$-$26.92	&	M1	&	F	\\
1800$+$440	&	18 01 32.42	&	$+$44 04 20.60	&	Q	&	0.663	&	0.264	&	16.98	&	$-$25.27	&	M1	&	F	\\
1803$+$784	&	18 00 45.70	&	$+$78 28 04.20	&	B	&	0.680	&	0.225	&	15.35	&	$-$26.95	&	M1	&	F	\\
1807$+$698	&	18 06 50.71	&	$+$69 49 28.20	&	B	&	0.051	&	0.155	&	17.00	&	$-$19.70	&	M1	&	F	\\
1821$+$107	&	18 24 02.86 &	$+$10 44 23.77	&	Q	&	1.364	&	0.818	&	18.28	&	$-$25.50	&	2cm	&	G	\\
1823$+$568	&	18 24 07.06	&	$+$56 51 01.49	&	B	&	0.664	&	0.264	&	18.60	&	$-$23.65	&	M1	&	F	\\
1828$+$487	&	18 29 31.78	&	$+$48 44 46.20	&	L	&	0.692	&	0.333	&	17.18	&	$-$25.16	&	M1	&	C	\\
1845$+$797	&	18 42 08.89	&	$+$79 46 16.70	&	G	&	0.056	&	0.308	&	13.76	&	$-$23.14	&	M2	&	S	\\
1849$+$670	&	18 49 16.07	&	$+$67 05 41.60	&	Q	&	0.657	&	0.243	&	16.18	&	$-$26.05	&	M1	&	F	\\
1901$+$319	&	19 02 55.94	&	$+$31 59 41.70	&	Q	&	0.635	&	0.523	&	18.26	&	$-$23.90	&	M2	&	C	\\
1908$-$201	&	19 11 09.65 &	$-$20 06 55.11  &	Q	&	1.119	&	0.696	&	18.33	&	$-$25.04	&	M2	&	F	\\
1921$-$293	&	19 24 51.04	&	$-$29 14 30.30	&	H	&	0.352	&	0.536	&	18.34	&	$-$22.55	&	M2	&	F	\\
1928$+$738	&	19 27 48.45	&	$+$73 58 01.80	&	L	&	0.302	&	0.574	&	15.71	&	$-$24.85	&	M1	&	F	\\
1936$-$155	&	19 39 26.65	&	$-$15 25 43.00	&	H	&	1.657	&	0.690	&	19.37	&	$-$24.82	&	M1	&	F	\\
1937$-$101	&	19 39 57.26	&	$-$10 02 41.52  &	Q	&	3.787	&	0.911	&	18.55	&	$-$27.34	&	2cm	&	F	\\
1954$+$513	&	19 55 42.83	&	$+$51 31 48.60	&	L	&	1.223	&	0.650	&	19.14	&	$-$24.41	&	M2	&	F	\\
1954$-$388	&	19 57 59.82	&	$-$38 45 06.36  &	Q	&	0.630	&	0.347	&	18.83	&	$-$23.31	&	2cm	&	F	\\
1957$+$405	&	19 59 28.34	&	$+$40 44 02.02	&	G	&	0.056	&	1.644	&	13.79	&	$-$23.11	&	M1	&	S	\\
1958$-$179	&	20 00 57.09	&	$-$17 48 57.60	&	H	&	0.650	&	0.573	&	17.86	&	$-$24.35	&	M1	&	F	\\
2000$-$330	&	20 03 24.12 &	$-$32 51 45.13	&	Q	&	3.783	&	0.562	&	19.48	&	$-$26.41	&	2cm	&	G	\\
2005$+$403	&	20 07 44.94	&	$+$40 29 48.61	&	Q	&	1.736	&	2.995	&	19.78	&	$-$24.51	&	M1	&	F	\\
2007$+$777	&	20 05 31.08	&	$+$77 52 43.30	&	B	&	0.342	&	0.696	&	17.07	&	$-$23.75	&	M2	&	F	\\
2008$-$159	&	20 11 15.71	&	$-$15 46 40.20	&	Q	&	1.180	&	0.613	&	17.29	&	$-$26.19	&	M1	&	F	\\
2010$+$463	&	20 12 05.64 &	$+$46 28 55.78	&	B	&	\nodata	&	2.953	&	17.47	&	\nodata	 &	2cm	&	F	\\
2021$+$317	&	20 23 19.01	&	$+$31 53 02.31	&	U	&	\nodata	&	4.565	&	\nodata	&	\nodata	        &	M1	&	F	\\
2021$+$614	&	20 22 06.68	&	$+$61 36 58.81	&	G	&	0.227	&	0.884	&	19.83	&	$-$20.11	&	M1	&	F	\\
2029$+$121	&	20 31 54.99	&	$+$12 19 41.34  &	Q	&	1.215	&	0.349	&	\nodata	&	\nodata	        &	M2	&	F	\\
2037$+$511	&	20 38 37.03	&	$+$51 19 12.60	&	Q	&	1.686	&	4.732	&	19.88	&	$-$24.35	&	M1	&	F	\\
2059$+$034	&	21 01 38.83 &	$+$03 41 31.32  &	Q	&	1.015	&	0.447	&	17.24	&	$-$25.92	&	2cm	&	F	\\
2113$+$293	&	21 15 29.41	&	$+$29 33 38.37	&	Q	&	1.514	&	0.564	&	20.05	&	$-$23.95	&	M2	&	F	\\
2121$+$053	&	21 23 44.61	&	$+$05 35 22.30	&	H	&	1.941	&	0.313	&	17.98	&	$-$26.54	&	M1	&	F	\\
2126$-$158	&	21 29 12.18	&	$-$15 38 41.04  &	Q	&	3.280	&	0.344	&	17.24	&	$-$28.36	&	M2	&	F	\\
2128$+$048	&	21 30 32.97	&	$+$05 02 17.70	&	G	&	0.990	&	0.257	&	23.71	&	$-$19.40	&	2cm	&	G2	\\
2128$-$123	&	21 31 35.35	&	$-$12 07 04.50	&	L	&	0.501	&	0.265	&	15.63	&	$-$26.02	&	M1	&	F	\\
2131$-$021	&	21 34 10.31	&	$-$01 53 17.24	&	B	&	1.285	&	0.238	&	18.74	&	$-$24.92	&	M1	&	F	\\
2134$+$004	&	21 36 38.58	&	$+$00 41 54.20	&	L	&	1.932	&	0.281	&	16.93	&	$-$27.58	&	M1	&	F	\\
2136$+$141	&	21 39 01.31	&	$+$14 23 35.99	&	Q	&	2.427	&	0.427	&	18.74	&	$-$26.24	&	M1	&	F	\\
2144$+$092	&	21 47 10.16	&	$+$09 29 46.68	&	Q	&	1.113	&	0.310	&	18.41	&	$-$24.95	&	M2	&	F	\\
2145$+$067	&	21 48 05.45	&	$+$06 57 38.61	&	L	&	0.990	&	0.345	&	16.50	&	$-$26.63	&	M1	&	F	\\
2155$-$152	&	21 58 06.37	&	$-$15 01 09.00	&	H	&	0.672	&	0.207	&	17.65	&	$-$24.63	&	M1	&	F	\\
2200$+$420	&	22 02 43.29	&	$+$42 16 39.98	&	B	&	0.069	&	1.420	&	15.62	&	$-$21.73	&	M1	&	F	\\
2201$+$171	&	22 03 26.89	&	$+$17 25 48.20	&	Q	&	1.076	&	0.240	&	19.50	&	$-$23.78	&	M1	&	F	\\
2201$+$315	&	22 03 14.97	&	$+$31 45 38.27	&	L	&	0.295	&	0.534	&	14.98	&	$-$25.55	&	M1	&	F	\\
2209$+$236	&	22 12 05.96	&	$+$23 55 40.59	&	Q	&	1.125	&	0.330	&	19.60	&	$-$23.78	&	M1	&	F	\\
2216$-$038	&	22 18 52.03	&	$-$03 35 36.80	&	Q	&	0.901	&	0.409	&	16.87	&	$-$26.04	&	M1	&	F	\\
2223$-$052	&	22 25 47.25	&	$-$04 57 01.39	&	H	&	1.404	&	0.325	&	18.03	&	$-$25.82	&	M1	&	F	\\
2227$-$088	&	22 29 40.17	&	$-$08 32 54.10	&	H	&	1.560	&	0.221	&	17.74	&	$-$26.33	&	M1	&	F	\\
2230$+$114	&	22 32 36.40	&	$+$11 43 50.89	&	H	&	1.037	&	0.312	&	17.08	&	$-$26.13	&	M1	&	F	\\
2234$+$282	&	22 36 22.47	&	$+$28 28 57.42	&	H	&	0.795	&	0.273	&	19.55	&	$-$23.09	&	M2	&	F	\\
2243$-$123	&	22 46 18.23	&	$-$12 06 51.28	&	H	&	0.632	&	0.215	&	16.26	&	$-$25.88	&	M1	&	F	\\
2251$+$158	&	22 53 57.74	&	$+$16 08 53.56	&	H	&	0.859	&	0.462	&	16.51	&	$-$26.29	&	M1	&	F	\\
2255$-$282	&	22 58 06.05	&	$-$27 58 20.90	&	Q	&	0.927	&	0.145	&	16.54	&	$-$26.43	&	M2	&	F	\\
2318$+$049	&	23 20 44.94	&	$+$05 13 50.20	&	Q	&	0.623	&	0.282	&	18.53	&	$-$23.59	&	2cm	&	F	\\
2329$-$162	&	23 31 38.65	&	$-$15 56 57.01	&	Q	&	1.153	&	0.111	&	20.06	&	$-$23.37	&	2cm	&	F	\\
2331$+$073	&	23 34 12.82	&	$+$07 36 27.50	&	Q	&	0.401	&	0.353	&	16.40	&	$-$24.77	&	M1	&	F	\\
2345$-$167	&	23 48 02.60	&	$-$16 31 12.02	&	H	&	0.576	&	0.112	&	18.50	&	$-$23.45	&	M1	&	F	\\
2351$+$456	&	23 54 21.68	&	$+$45 53 04.20	&	Q	&	1.986	&	0.527	&	20.50	&	$-$24.07	&	M1	&	F	\\
\end{longtable}

\end{scriptsize}

\renewcommand{\arraystretch}{1.0}

\clearpage\onecolumn

\newpage

\section{Spectroscopic Atlas}
\label{sec:atlas}
%%%%%%%%%%%%%%%%%%%%%%%%%
\begin{figure*}[htbp!]
\includegraphics[width=\columnwidth]{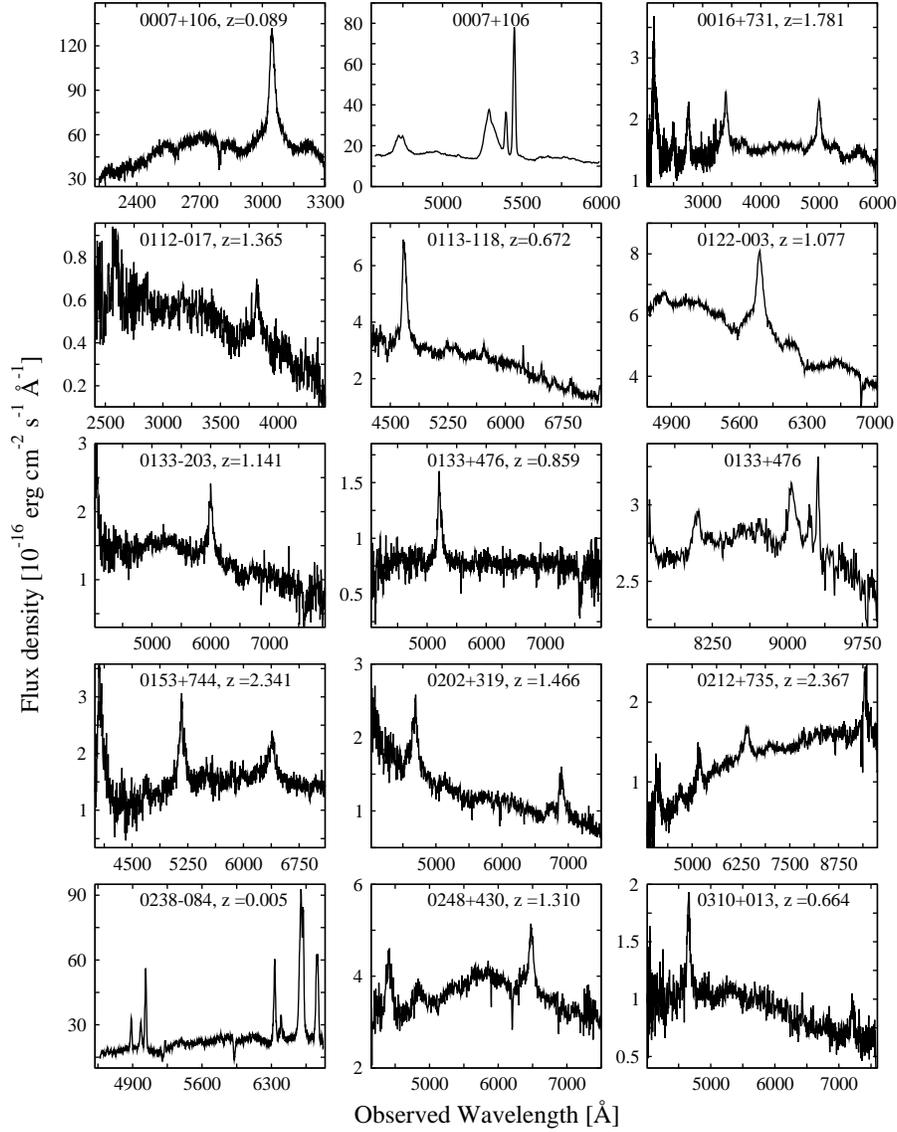}
\caption[]{\label{fig:atlas1}\small{Spectra for the 123 objects of the MOJAVE/2cm sample  that was available to study the H$\beta$ region,  and/or \ion{Mg}{2}\,$\lambda$2798/\ion{C}{4}\,$\lambda$1549; the abcissa is the observed wavelength in \AA\, and the ordinate is the flux density in units of $\rm 10^{-16}\, erg\, cm^{-2}\,s^{-1}\,$\AA$^{-1}$}, the name and the redshift of the source are shown in the top  of the corresponding frame.}
\end{figure*}

\begin{figure*}[!]
\setcounter{figure}{7}
\includegraphics[width=\columnwidth]{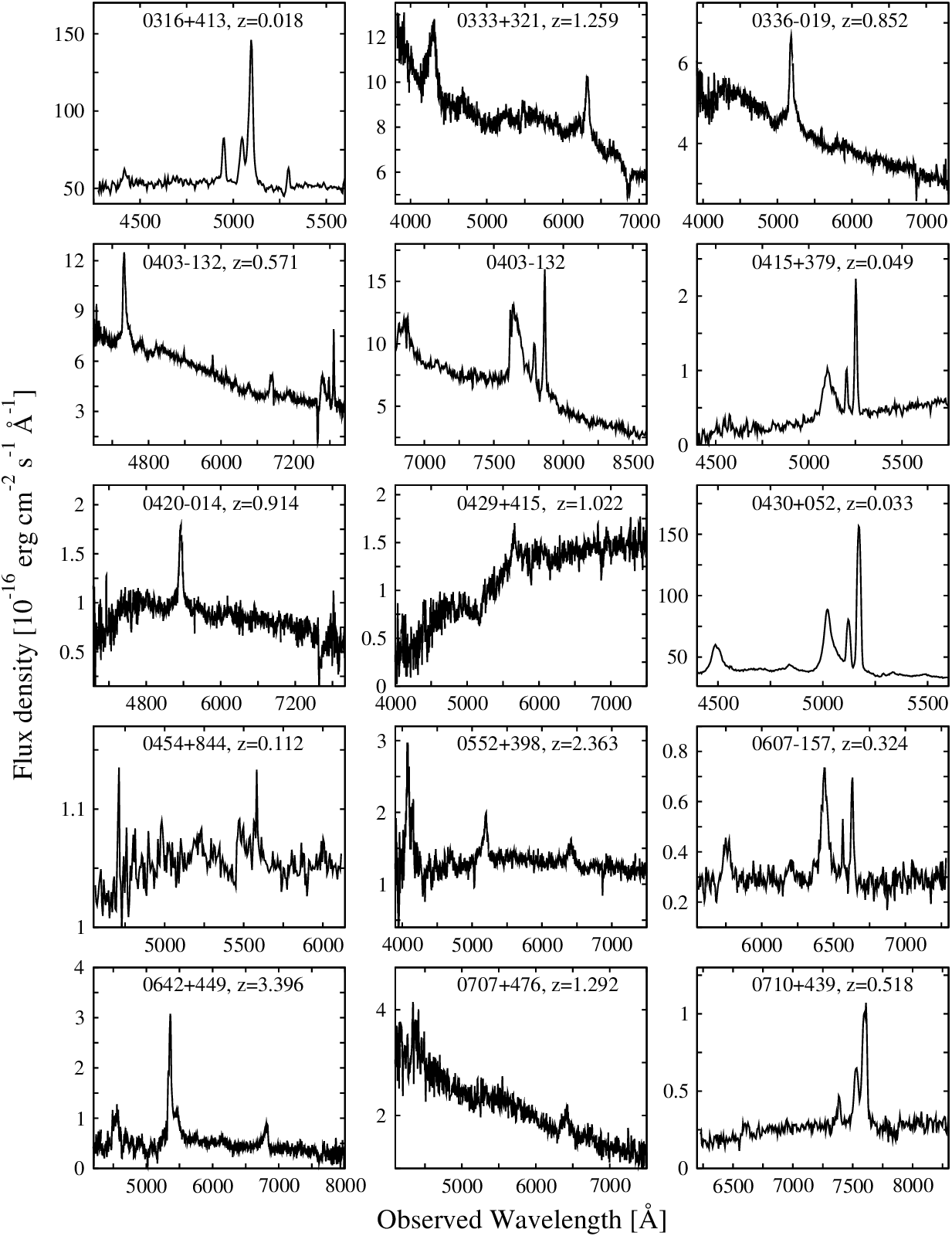}
\caption[]{\small{(Continued)}}
\end{figure*}

\begin{figure*}[!]
\setcounter{figure}{7}
\includegraphics[width=\columnwidth]{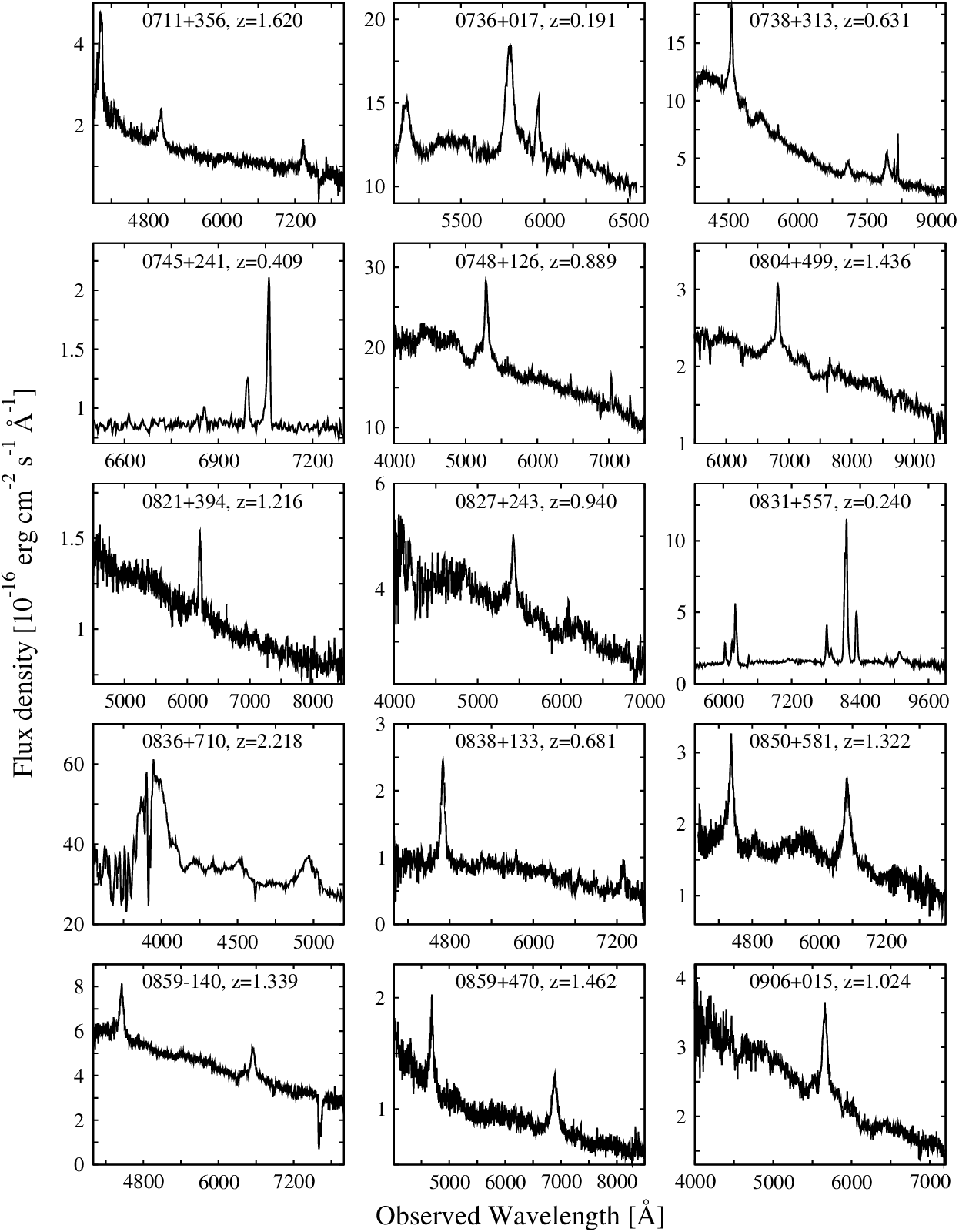}
\caption[]{\small{(Continued)}}
\end{figure*}

\begin{figure*}[!]
\setcounter{figure}{7}
\includegraphics[width=\columnwidth]{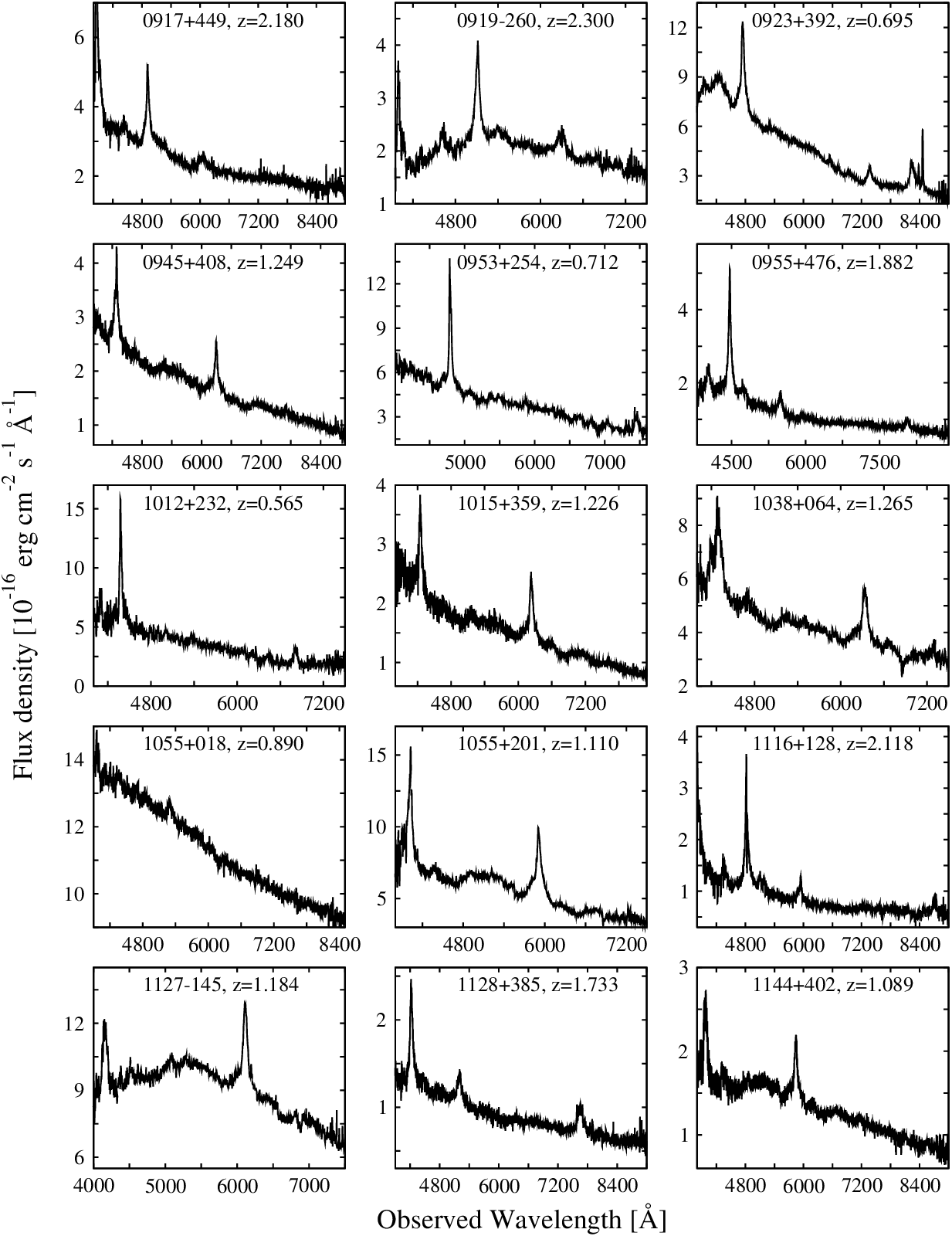}
\caption[]{\small{(Continued)}}
\end{figure*}

\begin{figure*}[!]
\setcounter{figure}{7}
\includegraphics[width=\columnwidth]{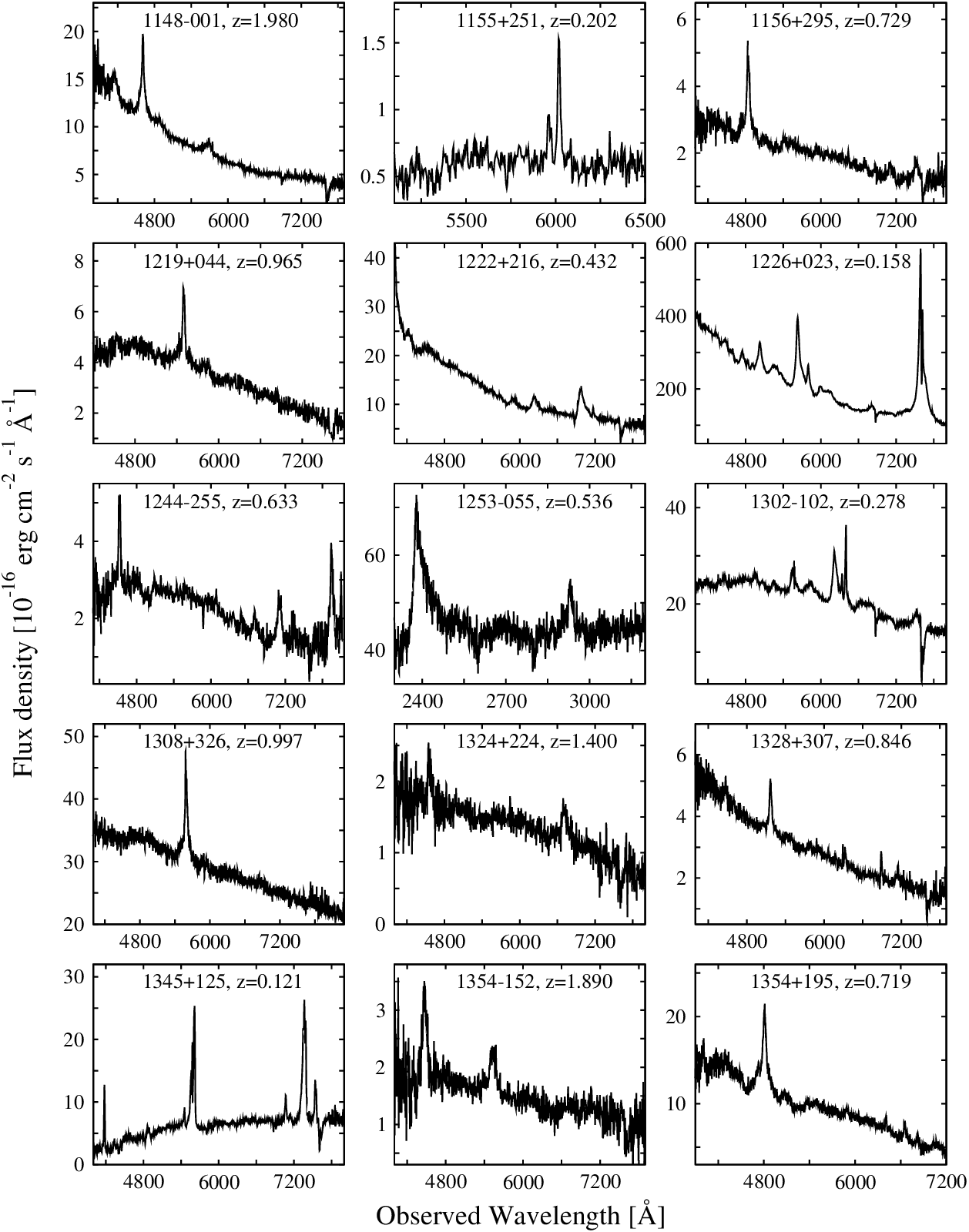}
\caption[]{\small{(Continued)}}
\end{figure*}

\begin{figure*}[!]
\setcounter{figure}{7}
\includegraphics[width=\columnwidth]{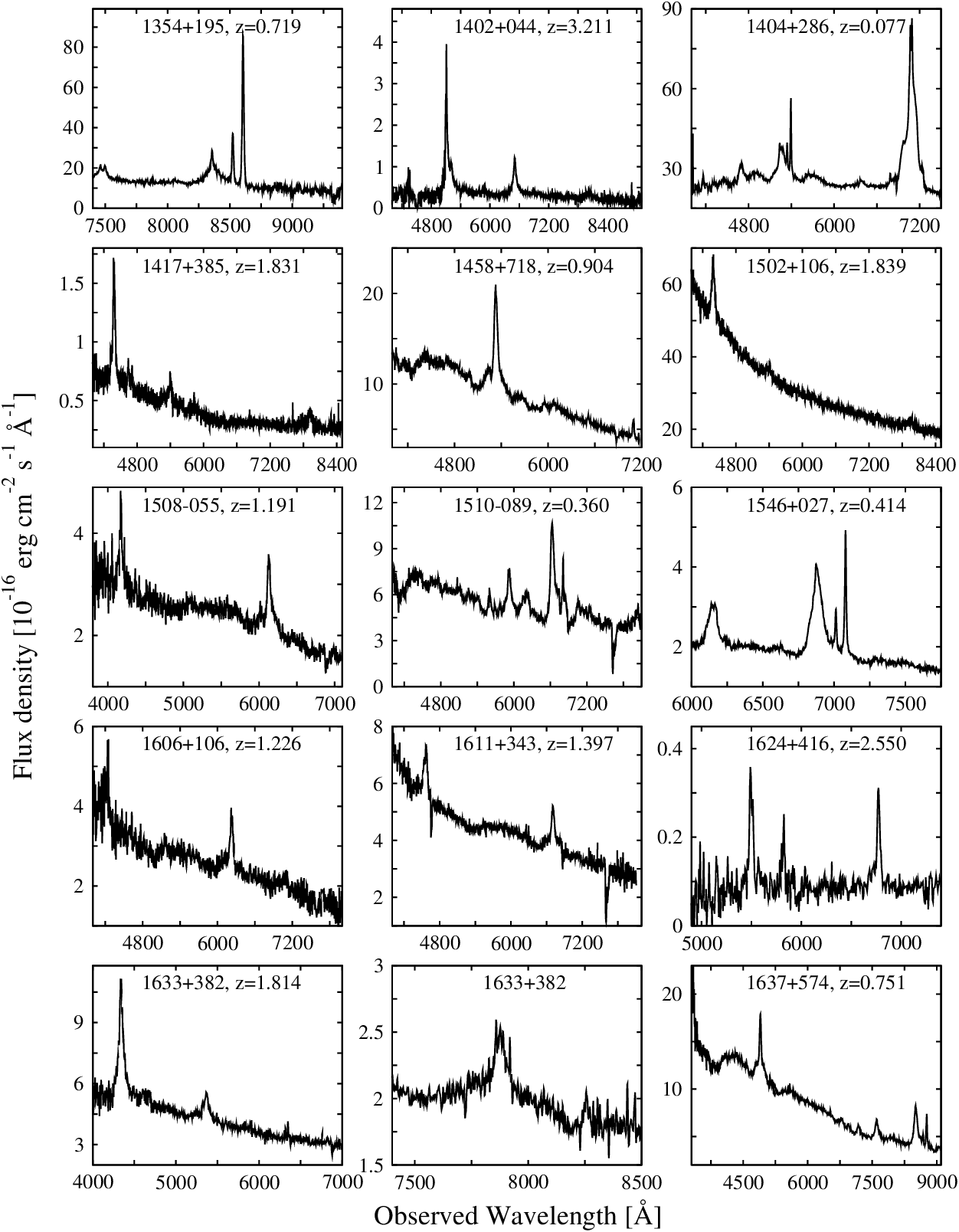}
\caption[]{\small{(Continued)}}
\end{figure*}

\begin{figure*}[!]
\setcounter{figure}{7}
\includegraphics[width=\columnwidth]{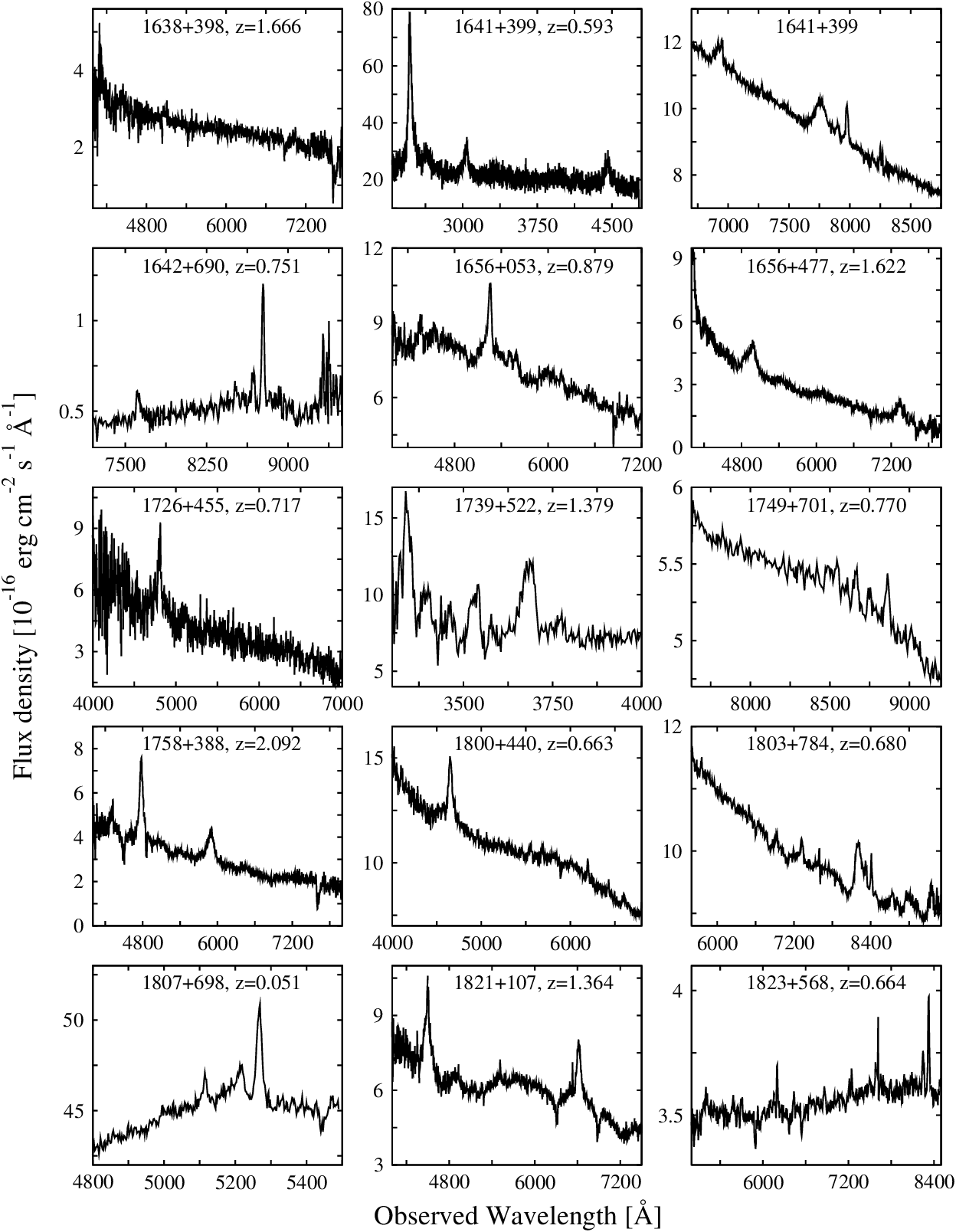}
\caption[]{\small{(Continued)}}
\end{figure*}

\begin{figure*}[!]
\setcounter{figure}{7}
\includegraphics[width=\columnwidth]{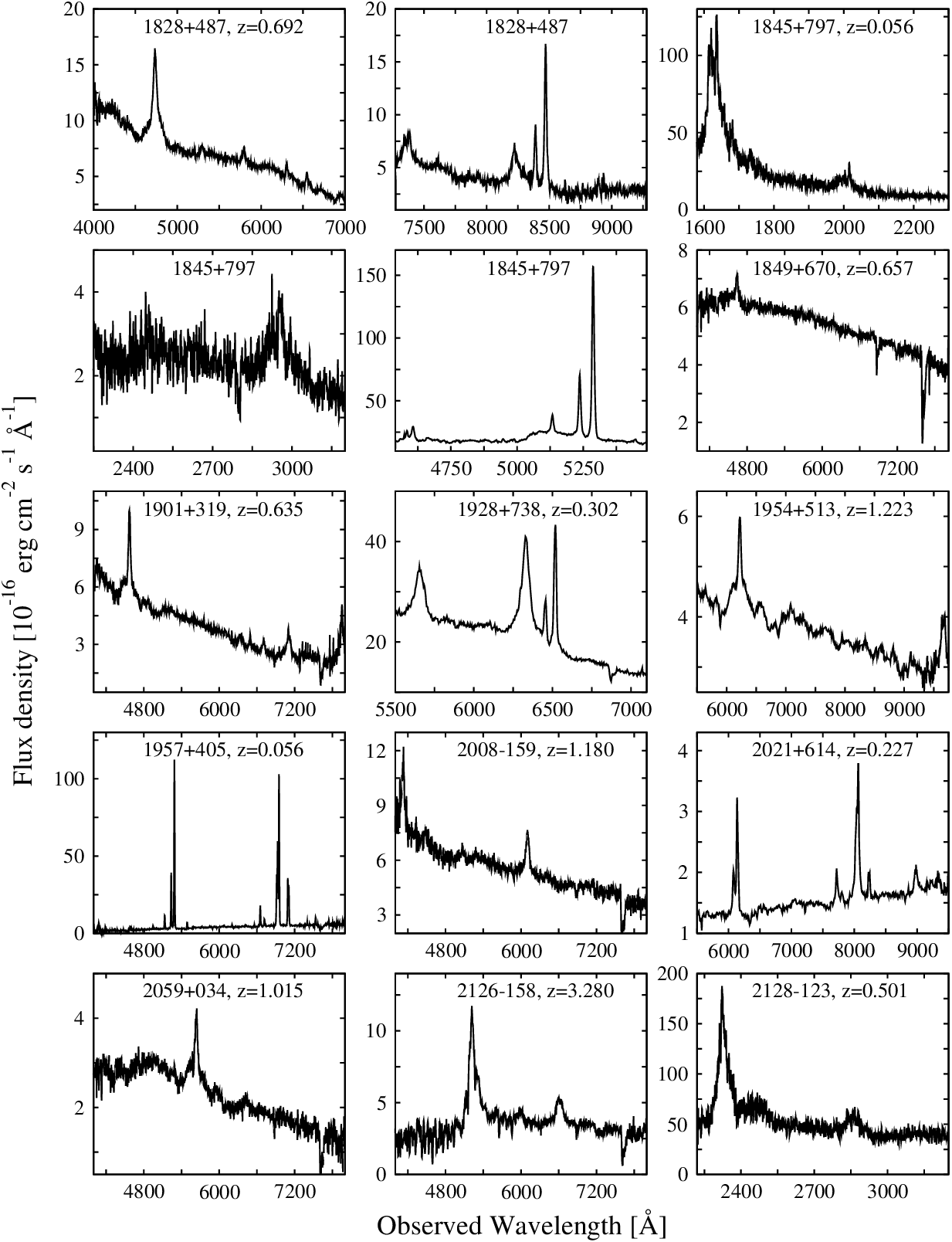}
\caption[]{\small{(Continued)}}
\end{figure*}

\begin{figure*}[!]
\setcounter{figure}{7}
\includegraphics[width=\columnwidth]{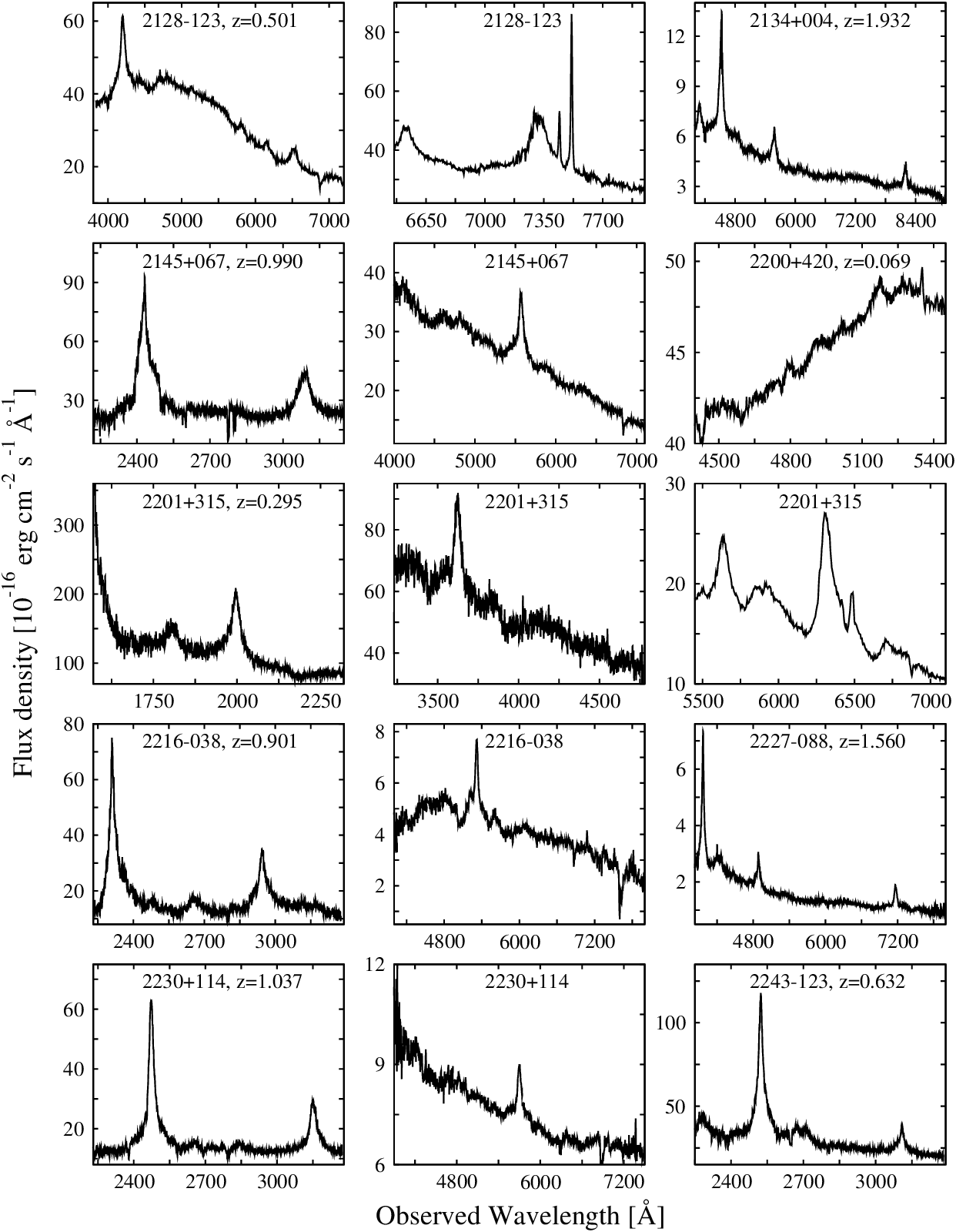}
\caption[]{\small{(Continued)}}
\end{figure*}

\begin{figure*}[!]
\setcounter{figure}{7}
\includegraphics[width=\columnwidth]{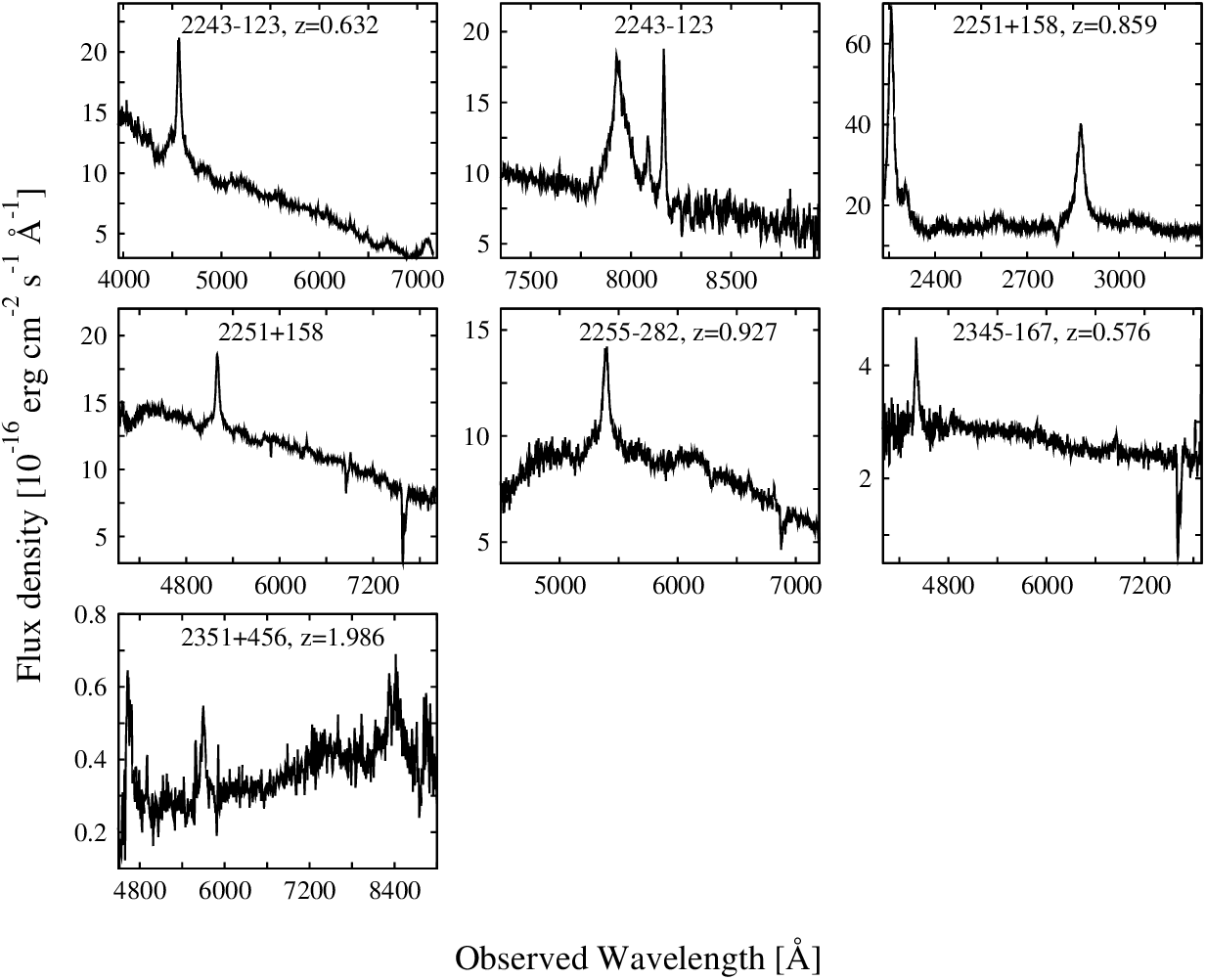}
\caption[]{\small{(Continued)}}
\end{figure*}

%%%%%%%%%%%%%%%%%%%%%%%%%

\newpage
\section{Spectroscopic emission lines  and continuum parameters}
\label{secap:parameters}
\begin{scriptsize}
\begin{landscape}
\tablecols{11}
\setlength{\tabnotewidth}{0.9\columnwidth}
\tabcaption{H$\beta$ and {[\ion{O}{III}]}\,$\lambda$5007 parameters for 41 AGN} \label{tabA1:hb1} 
\def\ColumnHeaders{ \multicolumn{1}{c}{{}} &
   \multicolumn{1}{c}{{}} &
   \multicolumn{4}{c}{{H$\beta$ (BC)}} &
   \multicolumn{1}{c}{{}} &
   \multicolumn{4}{c}{{[\ion{O}{III}] \,$\lambda$5007}} \\ %\hline \\%\\[0.5ex] \hline \\[-1.8ex]
   \cline{3-6}\cline{8-11} \\[-1.8ex]

   \multicolumn{1}{c}{{Source}} &
   \multicolumn{1}{c}{{}} &
   \multicolumn{1}{c}{{FWHM}} &
   \multicolumn{1}{c}{{EW}} &
\multicolumn{1}{c}{{Flux}\tablenotemark{c}} &
   \multicolumn{1}{c}{{$L_{\mathrm{H_{\beta}}}$}\tablenotemark{d}} &
\multicolumn{1}{c}{{}} &
   \multicolumn{1}{c}{{FWHM}} &
   \multicolumn{1}{c}{{EW}} &
\multicolumn{1}{c}{{Flux}\tablenotemark{c}} &
   \multicolumn{1}{c}{{$L_{\mathrm{[OIII]}}$}\tablenotemark{d}} \\

   \multicolumn{1}{c}{{Name}} &
   \multicolumn{1}{c}{{Ref.\tablenotemark{b}}} &
   \multicolumn{1}{c}{{(km\,s$^{-1}$)}} &
   \multicolumn{1}{c}{{(\AA)}} &
\multicolumn{1}{c}{} & %  
   \multicolumn{1}{c}{} & %
\multicolumn{1}{c}{{}} &
   \multicolumn{1}{c}{{(km\,s$^{-1}$)}} &
   \multicolumn{1}{c}{{(\AA)}} &
\multicolumn{1}{c}{} & %
   \multicolumn{1}{c}{} 
}
\begin{longtable}{lcccccccccc}
 %% Primera Cabeza
\toprule
  \ColumnHeaders\\ \midrule
  \endfirsthead
  
  %% Otras Cabezas
  \tabcaptioncontinued
  \toprule
  \ColumnHeaders\\ \midrule
  \endhead
  \bottomrule

%% Pies normales
\endfoot
%Now the data...
0007$+$106	&	1,4	&	4600$\,\pm\,$400	&	100.00$\,\pm\,$6.00	&	151.98$\,\pm\,$20.98	&	4.34	$\,\pm\,$0.60	&	&	620$\,\pm\,$170	&	75.00$\,\pm\,$20.00	&	105.00$\,\pm\,$7.80	&	2.96	$\,\pm\,$0.22	\\
0133$+$476	&	6	&	4223$\,\pm\,$124	&	13.34$\,\pm\,$0.39	&	3.42$\,\pm\,$0.10	&	20.97$\,\pm\,$0.61	&	&	985$\,\pm\,$29	&	4.52$\,\pm\,$0.13	&	1.15	$\,\pm\,$0.03	&	6.92	$\,\pm\,$0.20	\\
0238$-$084\tablenotemark{a}	&	5	&	1070$\,\pm\,$80	&	10.00$\,\pm\,$2.00	&	23.74	$\,\pm\,$2.00	&	0.001$\,\pm\,$0.000	&	&	1040$\,\pm\,$60	&	35.00$\,\pm\,$3.00	&	65.700$\,\pm\,$8.269	&	0.004$\,\pm\,$0.001	\\
0316$+$413\tablenotemark{a}	&	2,4	&	890$\,\pm\,$35	&	\nodata			&	98.50$\,\pm\,$58.70	&	0.13	$\,\pm\,$0.08	&	&	1350$\,\pm\,$20	&	80.00$\,\pm\,$50.00	&	359.99$\,\pm\,$179.99	&	0.46$\,\pm\,$0.23	\\
0403$-$132	&	4	&	3800$\,\pm\,$350	&	55.00$\,\pm\,$5.00	&	37.00$\,\pm\,$4.50	&	59.96$\,\pm\,$7.29	&	&	540$\,\pm\,$55	&	20.00$\,\pm\,$3.00	&	11.00$\,\pm\,$1.20	&	17.70$\,\pm\,$1.93	\\
0415$+$379	&	2	&	4100$\,\pm\,$400	&	125.00$\,\pm\,$15.00	&	4.10$\,\pm\,$0.60	&	7.85	$\,\pm\,$1.15	&	&	450	$\,\pm\,$35	&	60.00$\,\pm\,$10.00	&	2.60	$\,\pm\,$0.30	&	4.07	$\,\pm\,$0.47	\\
0430$+$052	&	1,4	&	2750$\,\pm\,$400	&	65.00$\,\pm\,$2.00	&	219.96$\,\pm\,$14.00	&	1.29	$\,\pm\,$0.08	&	&	450$\,\pm\,$150	&	70.00$\,\pm\,$8.00	&	235.00$\,\pm\,$49.83	&	1.33	$\,\pm\,$0.28	\\
0454$+$844	&	6	&	\nodata			&	\nodata			&	\nodata			&	\nodata			&	&	596$\,\pm\,$5	&	0.64	$\,\pm\,$0.01	&	0.068$\,\pm\,$0.001	&	0.003$\,\pm\,$0.000	\\
0607$-$157	&	2	&	3200$\,\pm\,$450	&	60.00$\,\pm\,$5.00	&	1.70$\,\pm\,$0.30	&	1.44	$\,\pm\,$0.25	&	&	640$\,\pm\,$130	&	25.00$\,\pm\,$2.00	&	0.65	$\,\pm\,$0.10	&	0.53	$\,\pm\,$0.08	\\
0710$+$439\tablenotemark{a}	&	6	&	906$\,\pm\,$219	&	21.29$\,\pm\,$3.13	&	0.50	$\,\pm\,$0.07	&	0.67	$\,\pm\,$0.10	&	&	1716$\,\pm\,$252	&	99.60$\,\pm\,$14.62	&	2.42	$\,\pm\,$0.36	&	3.22$\,\pm\,$0.47	\\
0736$+$017	&	1	&	2650$\,\pm\,$520	&	25.00$\,\pm\,$5.00	&	28.00$\,\pm\,$5.60	&	4.49	$\,\pm\,$0.90	&	&	720$\,\pm\,$160	&	7.00$\,\pm\,$2.00	&	8.00	$\,\pm\,$1.60	&	1.26	$\,\pm\,$0.25	\\
0738$+$313	&	3	&	5200$\,\pm\,$520	&	55.00$\,\pm\,$5.00	&	18.10$\,\pm\,$1.80	&	35.81$\,\pm\,$3.56	&	&	390	$\,\pm\,$50	&	20.00$\,\pm\,$2.00	&	3.50	$\,\pm\,$0.35	&	6.89	$\,\pm\,$0.69	\\
0745$+$241\tablenotemark{a}	&	3	&	370$\,\pm\,$20	&	2.00	$\,\pm\,$0.30	&	0.01	$\,\pm\,$0.00	&	0.01	$\,\pm\,$0.00	&	&	370$\,\pm\,$20	&	12.00$\,\pm\,$3.00	&	0.98	$\,\pm\,$0.05	&	0.70	$\,\pm\,$0.04	\\
0831$+$557\tablenotemark{a}	&	6	&	466$\,\pm\,$100	&	21.94$\,\pm\,$1.66	&	3.09	$\,\pm\,$0.23	&	0.64	$\,\pm\,$0.05	&	&	1585$\,\pm\,$120	&	93.74$\,\pm\,$7.07	&	12.41$\,\pm\,$0.94	&	2.54$\,\pm\,$0.19	\\
0923$+$392	&	3	&	4250$\,\pm\,$400	&	50.00$\,\pm\,$5.00	&	11.00$\,\pm\,$3.00	&	25.11$\,\pm\,$6.85	&	&	540$\,\pm\,$50	&	20.00$\,\pm\,$2.00	&	3.90	$\,\pm\,$0.40	&	8.89	$\,\pm\,$0.91	\\
1155$+$251\tablenotemark{a}	&	2	&	540	$\,\pm\,$60	&	8.00	$\,\pm\,$2.00	&	0.59	$\,\pm\,$0.06	&	0.07	$\,\pm\,$0.01	&	&	500$\,\pm\,$50	&	65.00$\,\pm\,$10.00	&	8.13	$\,\pm\,$0.80	&	1.02$\,\pm\,$0.10	\\
1222$+$216	&	2	&	5750$\,\pm\,$1000	&	60.00$\,\pm\,$15.00	&	37.30$\,\pm\,$7.50	&	27.88$\,\pm\,$5.61	&	&	360$\,\pm\,$70	&	10.00$\,\pm\,$3.00	&	2.16	$\,\pm\,$0.40	&	1.61	$\,\pm\,$0.30	\\
1226$+$023	&	1,2,4	&	3800$\,\pm\,$550	&	50.00	$\,\pm\,$7.00	&	1129.99$\,\pm\,$99.99	&	85.17$\,\pm\,$7.54	&	&	1300$\,\pm\,$150	&	10.00$\,\pm\,$1.00	&	191.00	$\,\pm\,$59.99	&	14.36$\,\pm\,$4.51	\\
1302$-$102	&	2,4	&	4300$\,\pm\,$1000	&	30.00$\,\pm\,$5.00	&	57.20	$\,\pm\,$3.00	&	16.08$\,\pm\,$0.84	&	&	800$\,\pm\,$15	&	10.00	$\,\pm\,$3.00	&	20.20$\,\pm\,$0.20	&	5.65	$\,\pm\,$0.06	\\
1345$+$125\tablenotemark{a}	&	2	&	1000	$\,\pm\,$70	&	10.00$\,\pm\,$2.00	&	5.76	$\,\pm\,$0.60	&	0.25	$\,\pm\,$0.03	&	&	1350$\,\pm\,$120	&	80.00$\,\pm\,$6.00	&	52.10$\,\pm\,$5.50	&	2.27	$\,\pm\,$0.24	\\
1354$+$195	&	4	&	3600$\,\pm\,$310	&	55.00$\,\pm\,$7.00	&	62.00$\,\pm\,$13.00	&	178.18$\,\pm\,$37.36	&	&	450	$\,\pm\,$60	&	60.00$\,\pm\,$5.00	&	67.00$\,\pm\,$6.00	&	191.13$\,\pm\,$17.12	\\
1404$+$286	&	1,2	&	8600$\,\pm\,$1600	&	70.00$\,\pm\,$6.00	&	148.00$\,\pm\,$30.00	&	2.30	$\,\pm\,$0.47	&	&	500$\,\pm\,$150	&	30.00	$\,\pm\,$2.00	&	43.90$\,\pm\,$7.30	&	0.68	$\,\pm\,$0.11	\\
1510$-$089	&	1,2	&	3250$\,\pm\,$180	&	80.00$\,\pm\,$10.00	&	35.50$\,\pm\,$5.10	&	21.94$\,\pm\,$3.15	&	&	550$\,\pm\,$45	&	15.00$\,\pm\,$2.00	&	4.90	$\,\pm\,$0.50	&	2.99	$\,\pm\,$0.31	\\
1546$+$027	&	2,3	&	5200$\,\pm\,$900	&	80.00$\,\pm\,$2.00	&	15.90$\,\pm\,$3.00	&	14.40$\,\pm\,$2.72	&	&	600$\,\pm\,$90	&	20.00$\,\pm\,$3.00	&	4.17	$\,\pm\,$0.80	&	3.72	$\,\pm\,$0.71	\\
1637$+$574	&	6	&	3226$\,\pm\,$172	&	52.42$\,\pm\,$2.80	&	20.82$\,\pm\,$1.11	&	56.06$\,\pm\,$2.99	&	&	1031$\,\pm\,$55	&	15.50$\,\pm\,$0.83	&	5.82	$\,\pm\,$0.31	&	15.66$\,\pm\,$0.83	\\
1641$+$399	&	3	&	3700$\,\pm\,$190	&	15.00$\,\pm\,$5.00	&	7.20$\,\pm\,$0.36	&	10.80$\,\pm\,$0.54	&	&	850	$\,\pm\,$40	&	5.00$\,\pm\,$2.00	&	2.60	$\,\pm\,$0.13	&	3.89	$\,\pm\,$0.19	\\
1642$+$690\tablenotemark{a}	&	6	&	565	$\,\pm\,$223	&	9.41$\,\pm\,$0.86	&	0.45	$\,\pm\,$0.04	&	1.32	$\,\pm\,$0.12	&	&	946$\,\pm\,$87	&	23.56$\,\pm\,$2.16	&	1.15	$\,\pm\,$0.11	&	3.37	$\,\pm\,$0.31	\\
1749$+$701	&	6	&	\nodata			&	\nodata			&	\nodata			&	\nodata			&	&	1022$\,\pm\,$23	&	1.38	$\,\pm\,$0.03	&	0.66	$\,\pm\,$0.01	&	2.01	$\,\pm\,$0.05	\\
1803$+$784	&	6	&	4320$\,\pm\,$24	&	6.94	$\,\pm\,$0.04	&	6.42	$\,\pm\,$0.04	&	15.59$\,\pm\,$0.09	&	&	1082$\,\pm\,$6	&	1.56$\,\pm\,$0.01	&	1.43	$\,\pm\,$0.01	&	3.45	$\,\pm\,$0.02	\\
1807$+$698	&	6	&	1487$\,\pm\,$16	&	0.88	$\,\pm\,$0.01	&	3.93$\,\pm\,$0.04	&	0.03	$\,\pm\,$0.00	&	&	856$\,\pm\,$9	&	1.86	$\,\pm\,$0.02	&	8.39	$\,\pm\,$0.09	&	0.06	$\,\pm\,$0.00	\\
1823$+$568	&	6	&	1173$\,\pm\,$18	&	0.46	$\,\pm\,$0.01	&	0.16$\,\pm\,$0.00	&	0.39	$\,\pm\,$0.01	&	&	559$\,\pm\,$9	&	1.66	$\,\pm\,$0.03	&	0.59	$\,\pm\,$0.01	&	1.39	$\,\pm\,$0.02	\\
1828$+$487	&	4	&	3050$\,\pm\,$340	&	45.00$\,\pm\,$3.00	&	15.20$\,\pm\,$2.00	&	42.07$\,\pm\,$5.54	&	&	640$\,\pm\,$60	&	50.00$\,\pm\,$3.00	&	16.00$\,\pm\,$2.00	&	43.86$\,\pm\,$5.48	\\
1845$+$797	&	4	&	12000$\,\pm\,$1200	&	35.00$\,\pm\,$6.00	&	23.00$\,\pm\,$1.20	&	0.26	$\,\pm\,$0.01	&	&	580$\,\pm\,$50	&	75.00$\,\pm\,$10.00	&	140.00$\,\pm\,$9.99	&	1.54	$\,\pm\,$0.11	\\
1901$+$319	&	2	&	1670$\,\pm\,$210	&	45.00$\,\pm\,$10.00	&	7.22	$\,\pm\,$0.90	&	18.91$\,\pm\,$2.36	&	&	500$\,\pm\,$30	&	60.00$\,\pm\,$15.00	&	11.70$\,\pm\,$1.50	&	30.19$\,\pm\,$3.87	\\
1928$+$738	&	1,4	&	3360$\,\pm\,$400	&	55.00$\,\pm\,$2.00	&	85.20$\,\pm\,$25.00	&	39.20$\,\pm\,$11.50	&	&	490$\,\pm\,$50	&	20.00$\,\pm\,$1.00	&	29.50$\,\pm\,$10.00	&	13.35$\,\pm\,$4.53	\\
1957$+$405\tablenotemark{a}	&	2	&	560	$\,\pm\,$60	&	34.00$\,\pm\,$3.00	&	9.78	$\,\pm\,$1.50	&	0.28	$\,\pm\,$0.04	&	&	550	$\,\pm\,$40	&	380.00$\,\pm\,$20.00	&	116.00	$\,\pm\,$20.00	&	3.15	$\,\pm\,$0.54	\\
2021$+$614	&	6	&	\nodata			&	\nodata			&	\nodata			&	\nodata			&	&	1394$\,\pm\,$30	&	39.19	$\,\pm\,$0.83	&	5.08	$\,\pm\,$0.11	&	1.55	$\,\pm\,$0.03	\\
2128$-$123	&	4	&	4900	$\,\pm\,$370	&	45.00$\,\pm\,$2.00	&	152.00$\,\pm\,$20.00	&	181.94$\,\pm\,$23.94	&	&	370$\,\pm\,$50	&	15.00$\,\pm\,$2.00	&	42.00$\,\pm\,$1.00	&	49.90$\,\pm\,$1.19	\\
2200$+$420	&	6	&	\nodata			&	\nodata			&	\nodata			&	\nodata			&	&	178	$\,\pm\,$2	&	0.24	$\,\pm\,$0.01	&	1.16	$\,\pm\,$0.02	&	0.04	$\,\pm\,$0.00	\\
2201$+$315	&	1	&	4230$\,\pm\,$300	&	60.00$\,\pm\,$15.00	&	8.80	$\,\pm\,$1.60	&	3.92	$\,\pm\,$0.71	&	&	1200$\,\pm\,$100	&	15.00$\,\pm\,$3.00	&	15.00$\,\pm\,$0.80	&	6.58	$\,\pm\,$0.35	\\
2243$-$123	&	4	&	3900$\,\pm\,$310	&	55.00$\,\pm\,$2.00	&	49.60$\,\pm\,$7.00	&	99.18$\,\pm\,$14.00	&	&	490	$\,\pm\,$40	&	13.00$\,\pm\,$1.00	&	10.50$\,\pm\,$2.00	&	20.87$\,\pm\,$3.98	\\
\midrule
\tabnotetext{a}  {Source with  FWHM H$\beta\,\lesssim$ 1000 km\, s$^{-1}$.}
\tabnotetext{b}{Spectrum reference: 1=OAGH; 2=OAN-SPM; 3=\textit{SDSS} (Sloan Digital Sky Survey); 4=\citet{M03}; 5=\textit{HST-FOS} (Hubble Space Telescope) and 6=\citet{L96}. Values shown are mean and standard deviation for multiple spectra.}
\tabnotetext{c}{Flux in units of 10$^{-15}$ erg\,s$^{-1}$\,cm$^{-2}$.}
\tabnotetext{d}{Luminosities in units of 10$^{42}$ erg\,s$^{-1}$.}
\end{longtable}

\end{landscape}
\end{scriptsize}

\begin{scriptsize}

\begin{landscape}
\renewcommand{\arraystretch}{1.0}
\tablecols{8}
\setlength{\tabnotewidth}{0.9\columnwidth}
\tabcaption{Parameters for the $\lambda$5100\,\AA\, continuum and \ion{Fe}{2}\, $\lambda$4570 for 41 AGN} \label{tabA2:hb2}

\def\ColumnHeaders{\multicolumn{1}{c}{{}} &
   \multicolumn{3}{c}{{5100 \AA\, Continuum}} &
   \multicolumn{1}{c}{{}} &
   \multicolumn{3}{c}{{\ion{Fe}{2}\, $\lambda$4570}} \\
   \cline{2-4}\cline{6-8}  \\[-1.8ex]

   \multicolumn{1}{c}{{Source}} &
   \multicolumn{1}{c}{{Flux density}} &
   \multicolumn{1}{c}{{$\lambda\,L_{5100}$}} &
   \multicolumn{1}{c}{Spectral} &
   \multicolumn{1}{c}{{}} &
   \multicolumn{1}{c}{{Flux}} &
   \multicolumn{1}{c}{{$L_{4570}$}} &
   \multicolumn{1}{c}{{EW}} \\
   
   \multicolumn{1}{c}{{}} &
   \multicolumn{1}{c}{{(10$^{-16}$ erg\,s$^{-1}$\,cm$^{-2}$\,\AA$^{-1}$)}} &
   \multicolumn{1}{c}{{(10$^{44}$ erg\,s$^{-1}$)}} &
   \multicolumn{1}{c}{{index\tablenotemark{a}}} &
   \multicolumn{1}{c}{{}} &
   \multicolumn{1}{c}{{(10$^{-16}$ erg\,s$^{-1}$\,cm$^{-2}$)}} &
   \multicolumn{1}{c}{{(10$^{41}$ erg\,s$^{-1}$)}} &
   \multicolumn{1}{c}{{(\AA)}} 
}
\begin{longtable}{lccccccc}
%% Primera Cabeza
  \toprule
  \ColumnHeaders\\ \midrule
  \endfirsthead
  
  %% Otras Cabezas
  \tabcaptioncontinued
  \toprule
  \ColumnHeaders\\ \midrule
  \endhead
  
  %% Pies normales
 \bottomrule

\endfoot
%Now the data...
0007$+$106	&	16.10$\,\pm\,$1.60	&	2.30	$\,\pm\,$0.23	&	$-$2.53	&	&	159.41$\,\pm\,$0.54	&	4.65	$\,\pm\,$0.01	&	10.50$\,\pm\,$0.67	\\
0133$+$476	&	2.56	$\,\pm\,$0.07	&	77.63$\,\pm\,$2.27	&	$-$4.44	&	&	42.90$\,\pm\,$0.03	&	272.95$\,\pm\,$0.20	&	16.71$\,\pm\,$0.30	\\
0238$-$084	&	17.80$\,\pm\,$1.00	&	0.01	$\,\pm\,$0.00	&	$-$3.11	&	&	\nodata			&	\nodata			&	\nodata			\\
0316$+$413	&	45.00$\,\pm\,$8.50	&	0.29	$\,\pm\,$0.05	&	$-$2.43	&	&	320.90$\,\pm\,$0.90	&	0.43	$\,\pm\,$0.00	&	6.64	$\,\pm\,$0.35	\\
0403$-$132	&	4.70	$\,\pm\,$0.40	&	38.40$\,\pm\,$3.27	&	$-$2.67	&	&	60.29$\,\pm\,$2.22	&	99.11$\,\pm\,$3.65	&	7.89	$\,\pm\,$0.40	\\
0415$+$379	&	0.45	$\,\pm\,$0.05	&	3.16	$\,\pm\,$0.35	&	$-$1.64	&	&	4.87	$\,\pm\,$0.02	&	13.94$\,\pm\,$0.07	&	33.12$\,\pm\,$10.59	\\
0430$+$052	&	31.90$\,\pm\,$3.80	&	0.90	$\,\pm\,$0.11	&	$-$2.83	&	&	273.49$\,\pm\,$1.23	&	1.72	$\,\pm\,$0.01	&	8.21	$\,\pm\,$0.40	\\
0454$+$844	&	1.04	$\,\pm\,$0.01	&	0.24	$\,\pm\,$0.00	&	$-$1.13	&	&	4.92	$\,\pm\,$0.00	&	0.23	$\,\pm\,$0.00	&	4.75	$\,\pm\,$0.07	\\
0607$-$157	&	0.27	$\,\pm\,$0.03	&	1.11	$\,\pm\,$0.12	&	$-$3.89	&	&	11.24$\,\pm\,$0.02	&	10.11	$\,\pm\,$0.02	&	50.94$\,\pm\,$7.35	\\
0710$+$439	&	0.25	$\,\pm\,$0.04	&	1.67	$\,\pm\,$0.24	&	$-$3.33	&	&	10.64$\,\pm\,$0.01	&	14.50$\,\pm\,$0.01	&	53.24$\,\pm\,$5.66	\\
0736$+$017	&	11.20$\,\pm\,$2.20	&	8.92	$\,\pm\,$1.75	&	$-$1.16	&	&	204.70$\,\pm\,$0.17	&	33.88$\,\pm\,$0.03	&	18.20$\,\pm\,$0.32	\\
0738$+$313	&	2.56	$\,\pm\,$0.30	&	25.59$\,\pm\,$3.00	&	$-$2.55	&	&	85.67$\,\pm\,$0.02	&	171.52$\,\pm\,$0.05	&	28.28$\,\pm\,$0.93	\\
0745$+$241	&	0.83	$\,\pm\,$0.03	&	3.04	$\,\pm\,$0.11	&	$-$0.86	&	&	3.03	$\,\pm\,$0.02	&	2.23	$\,\pm\,$0.01	&	3.54	$\,\pm\,$0.16	\\
0831$+$557	&	1.32	$\,\pm\,$0.10	&	1.37	$\,\pm\,$0.10	&	$-$3.26	&	&	16.23$\,\pm\,$0.10	&	3.37	$\,\pm\,$0.01	&	13.63$\,\pm\,$0.69	\\
0923$+$392	&	1.93	$\,\pm\,$0.20	&	22.41$\,\pm\,$2.32	&	$-$1.75	&	&	11.52$\,\pm\,$0.02	&	26.35$\,\pm\,$0.05	&	4.93	$\,\pm\,$0.20	\\
1155$+$251	&	1.27	$\,\pm\,$0.13	&	0.81	$\,\pm\,$0.08	&	$-$5.10	&	&	28.13$\,\pm\,$0.05	&	3.56	$\,\pm\,$0.01	&	64.36$\,\pm\,$9.92	\\
1222$+$216	&	6.84	$\,\pm\,$0.31	&	25.96$\,\pm\,$1.18	&	$-$0.83	&	&	\nodata			&	\nodata			&	\nodata			\\
1226$+$023	&	218.00$\,\pm\,$40.00	&	83.45$\,\pm\,$15.31	&	$-$2.08	&	&	4520.37$\,\pm\,$29.91	&	342.38$\,\pm\,$2.26	&	13.59$\,\pm\,$0.67	\\
1302$-$102	&	19.20$\,\pm\,$1.00	&	27.30	$\,\pm\,$1.42	&	$-$2.27	&	&	186.77$\,\pm\,$0.54	&	53.13$\,\pm\,$0.13	&	9.81	$\,\pm\,$0.25	\\
1345$+$125	&	5.72	$\,\pm\,$0.23	&	1.27	$\,\pm\,$0.05	&	$-$2.49	&	&	100.65$\,\pm\,$0.19	&	4.45	$\,\pm\,$0.01	&	27.46$\,\pm\,$1.40	\\
1354$+$195	&	9.80	$\,\pm\,$0.90	&	141.91$\,\pm\,$13.03	&	$-$2.61	&	&	60.12$\,\pm\,$0.54	&	175.28	$\,\pm\,$1.41	&	4.82	$\,\pm\,$0.26	\\
1404$+$286	&	21.80$\,\pm\,$2.40	&	1.72	$\,\pm\,$0.19	&	$-$2.59	&	&	449.82$\,\pm\,$0.54	&	7.02	$\,\pm\,$0.01	&	20.47$\,\pm\,$0.76	\\
1510$-$089	&	3.86	$\,\pm\,$0.25	&	11.94$\,\pm\,$0.77	&	$-$3.20	&	&	270.95$\,\pm\,$0.25	&	171.50$\,\pm\,$0.16	&	68.72$\,\pm\,$8.58	\\
1546$+$027	&	2.00	$\,\pm\,$0.50	&	9.01	$\,\pm\,$2.26	&	$-$1.52	&	&	28.36$\,\pm\,$0.08	&	26.45$\,\pm\,$0.08	&	15.50$\,\pm\,$2.26	\\
1637$+$574	&	3.56	$\,\pm\,$0.19	&	48.81$\,\pm\,$2.61	&	$-$0.60	&	&	47.53$\,\pm\,$0.07	&	128.41$\,\pm\,$0.18	&	10.69$\,\pm\,$0.31	\\
1641$+$399	&	11.40$\,\pm\,$0.20	&	86.95$\,\pm\,$1.53	&	$-$3.14	&	&	18.53$\,\pm\,$0.30	&	27.83$\,\pm\,$0.45	&	1.79	$\,\pm\,$0.04	\\
1642$+$690	&	0.53	$\,\pm\,$0.05	&	7.87	$\,\pm\,$0.72	&	$-$2.82	&	&	7.00	$\,\pm\,$0.00	&	20.77$\,\pm\,$0.01	&	15.72$\,\pm\,$1.03	\\
1749$+$701	&	4.92	$\,\pm\,$0.11	&	76.20$\,\pm\,$1.72	&	$-$1.30	&	&	17.88$\,\pm\,$0.02	&	55.11$\,\pm\,$0.05	&	3.29	$\,\pm\,$0.04	\\
1803$+$784	&	9.13	$\,\pm\,$0.05	&	111.91$\,\pm\,$0.61	&	$-$2.05	&	&	19.41$\,\pm\,$0.11	&	47.72$\,\pm\,$0.26	&	2.02	$\,\pm\,$0.03	\\
1807$+$698	&	44.82$\,\pm\,$0.47	&	1.59	$\,\pm\,$0.02	&	$-$2.23	&	&	362.20$\,\pm\,$0.37	&	2.55	$\,\pm\,$0.00	&	9.00	$\,\pm\,$0.16	\\
1823$+$568	&	3.56	$\,\pm\,$0.05	&	42.42$\,\pm\,$0.65	&	$-$0.15	&	&	17.73$\,\pm\,$0.00	&	42.50$\,\pm\,$0.01	&	5.07	$\,\pm\,$0.13	\\
1828$+$487	&	2.52	$\,\pm\,$0.30	&	35.02$\,\pm\,$4.17	&	$-$3.61	&	&	46.79$\,\pm\,$0.30	&	131.98$\,\pm\,$0.85	&	10.59$\,\pm\,$1.37	\\
1845$+$797	&	17.50$\,\pm\,$1.00	&	0.98	$\,\pm\,$0.06	&	$-$8.86	&	&	296.66$\,\pm\,$0.31	&	3.35	$\,\pm\,$0.00	&	18.51$\,\pm\,$0.96	\\
1901$+$319	&	2.18	$\,\pm\,$0.30	&	28.42$\,\pm\,$3.91	&	$-$2.25	&	&	77.17$\,\pm\,$0.36	&	208.24$\,\pm\,$0.98	&	38.84$\,\pm\,$3.58	\\
1928$+$738	&	12.80$\,\pm\,$5.30	&	29.25$\,\pm\,$12.11	&	$-$2.12	&	&	205.07$\,\pm\,$0.54	&	97.43$\,\pm\,$0.26	&	9.29	$\,\pm\,$0.46	\\
1957$+$405	&	1.80	$\,\pm\,$0.39	&	0.24	$\,\pm\,$0.05	&	$-$2.53	&	&	138.58$\,\pm\,$0.22	&	4.35	$\,\pm\,$0.01	&	85.74$\,\pm\,$12.31	\\
2021$+$614	&	1.28	$\,\pm\,$0.03	&	1.96	$\,\pm\,$0.04	&	$-$2.41	&	&	6.27	$\,\pm\,$0.02	&	2.07	$\,\pm\,$0.01	&	5.08	$\,\pm\,$0.33	\\
2128$-$123	&	30.00$\,\pm\,$1.00	&	180.90$\,\pm\,$6.03	&	$-$0.68	&	&	\nodata			&	\nodata			&	\nodata			\\
2200$+$420	&	47.21$\,\pm\,$0.63	&	8.28	$\,\pm\,$0.11	&	$-$2.70	&	&	205.61$\,\pm\,$0.20	&	8.20	$\,\pm\,$0.01	&	4.73	$\,\pm\,$0.07	\\
2201$+$315	&	12.00	$\,\pm\,$1.00	&	26.59$\,\pm\,$2.22	&	$-$1.07	&	&	232.43$\,\pm\,$0.25	&	106.49$\,\pm\,$0.12	&	13.29$\,\pm\,$0.31	\\
2243$-$123	&	7.27	$\,\pm\,$0.80	&	73.41$\,\pm\,$8.08	&	$-$1.96	&	&	\nodata			&	\nodata			&	\nodata	\\ 
\midrule

\tabnotetext{a} {Spectral index of the local continuum power-law.}

\end{longtable}
\end{landscape}
\end{scriptsize}

\renewcommand{\arraystretch}{1.0}

\clearpage\onecolumn

\begin{scriptsize}
\begin{landscape}%para que aparezca rotada
\renewcommand{\arraystretch}{0.9}
\tablecols{10}
\setlength{\tabnotewidth}{0.9\columnwidth}
\tabcaption{Parameters for the 3000 \AA\, continuum and  \ion{Mg}{2}\, $\lambda$2798 for 78 AGN} \label{tabA3:MgII_1}
\def\ColumnHeaders{\multicolumn{1}{c}{{}} &
   \multicolumn{1}{c}{{}} &
   \multicolumn{4}{c}{{\ion{Mg}{2} (BC)}} &
   \multicolumn{1}{c}{{}} &
   \multicolumn{3}{c}{{3000 \AA, Continuum }} \\
   \cline{3-6}\cline{8-10}  \\[-1.8ex]
   
   \multicolumn{1}{c}{{}} &
   \multicolumn{1}{c}{{}} &
   \multicolumn{1}{c}{{FWHM}} &
   \multicolumn{1}{c}{{EW}} &
   \multicolumn{1}{c}{{Flux}\tablenotemark{b}} &
   \multicolumn{1}{c}{{$L_{\mathrm{Mg\,II}}$}} &
\multicolumn{1}{c}{{}} &
     \multicolumn{1}{c}{{Flux}\tablenotemark{c}} &
     \multicolumn{1}{c}{{$\lambda\,L_{3000}$}} &
     \multicolumn{1}{c}{{Spectral}} \\

   \multicolumn{1}{c}{{Source}} &
   \multicolumn{1}{c}{{Ref\tablenotemark{a}}} &
   \multicolumn{1}{c}{{(km\,s$^{-1}$)}} &
   \multicolumn{1}{c}{{(\AA)}} &
   \multicolumn{1}{c}  {} & %{{(10$^{-15}$ erg\,s$^{-1}$\,cm$^{-2}$)}} &
   \multicolumn{1}{c}{{(10$^{42}$ erg\,s$^{-1}$)}} &
\multicolumn{1}{c}{{}}  &
   \multicolumn{1}{c} {}& %{{(10$^{-16}$ erg\,s$^{-1}$\,cm$^{-2}$\,\AA$^{-1}$)}} &
   \multicolumn{1}{c}{{(10$^{44}$ erg\,s$^{-1}$)}} &
   \multicolumn{1}{c}{{index\tablenotemark{d}}}
}
\begin{longtable}{lccccccccc}

%% Primera Cabeza
  \toprule
  \ColumnHeaders\\ \midrule
  \endfirsthead
  
  %% Otras Cabezas
  \tabcaptioncontinued
  \toprule
  \ColumnHeaders\\ \midrule
  \endhead

 \bottomrule

\endfoot
%Now the data...
 0007$+$106	&	5	&	6322    $\,\pm\,$635	&	57.69	$\,\pm\,$5.80	&	256.10	$\,\pm\,$25.74	&	9.06	$\,\pm\,$0.91	&	&	42.84	$\,\pm\,$4.31	&	4.35	$\,\pm\,$0.44	&	$-$0.90	\\
0016$+$731	&	6	&	4989	$\,\pm\,$132	&	20.60	$\,\pm\,$3.33	&	3.72	$\,\pm\,$0.10	&	498.72	$\,\pm\,$13.38	&	&	1.39	$\,\pm\,$0.06	&	484.93	$\,\pm\,$19.22	&	$-$0.09	\\
0112$-$017	&	2	&	5304	$\,\pm\,$777	&	40.53	$\,\pm\,$5.94	&	1.33	$\,\pm\,$0.19	&	21.16	$\,\pm\,$3.09	&	&	0.29	$\,\pm\,$0.04	&	13.47	$\,\pm\,$1.97	&	$-$0.30	\\
0113$-$118	&	2	&	4751	$\,\pm\,$42	&	70.57	$\,\pm\,$0.83	&	17.06	$\,\pm\,$1.63	&	39.35	$\,\pm\,$3.75	&	&	2.50	$\,\pm\,$0.36	&	17.04	$\,\pm\,$2.45	&	$-$0.23	\\
0122$-$003	&	2	&	5140	$\,\pm\,$491	&	37.30	$\,\pm\,$1.69	&	16.90	$\,\pm\,$0.58	&	127.30	$\,\pm\,$4.37	&	&	4.05	$\,\pm\,$0.26	&	90.22	$\,\pm\,$5.88	&	$-$0.85	\\
0133$-$203	&	2	&	5072	$\,\pm\,$648	&	74.43	$\,\pm\,$9.51	&	6.00	$\,\pm\,$0.77	&	48.28	$\,\pm\,$6.17	&	&	0.95	$\,\pm\,$0.12	&	22.76	$\,\pm\,$2.90	&	$-$1.09	\\
0133$+$476	&	2	&	5367	$\,\pm\,$453	&	46.26	$\,\pm\,$3.90	&	3.27	$\,\pm\,$0.28	&	28.10	$\,\pm\,$2.37	&	&	0.73	$\,\pm\,$0.06	&	17.68	$\,\pm\,$1.49	&	$-$2.62	\\
0202$+$319	&	2	&	4271	$\,\pm\,$467	&	34.86	$\,\pm\,$3.81	&	2.67	$\,\pm\,$0.29	&	50.61	$\,\pm\,$5.53	&	&	0.69	$\,\pm\,$0.08	&	38.39	$\,\pm\,$4.20	&	$-$2.73	\\
0248$+$430	&	2	&	5560	$\,\pm\,$442	&	37.77	$\,\pm\,$3.00	&	10.69	$\,\pm\,$0.85	&	201.10	$\,\pm\,$15.97	&	&	2.98	$\,\pm\,$0.24	&	160.17	$\,\pm\,$12.74	&	$-$9.65	\\
0310$+$013	&	2	&	4808	$\,\pm\,$396	&	79.49	$\,\pm\,$17.60	&	4.65	$\,\pm\,$0.27	&	21.18	$\,\pm\,$1.23	&	&	0.81	$\,\pm\,$0.15	&	10.29	$\,\pm\,$1.98	&	$-$7.88	\\
0333$+$321	&	1	&	3735	$\,\pm\,$272	&	14.60	$\,\pm\,$1.06	&	9.03	$\,\pm\,$0.66	&	6091.17	$\,\pm\,$443.18	&	&	5.67	$\,\pm\,$0.41	&	8205.02	$\,\pm\,$597.65	&	$-$0.61	\\
0336$-$019	&	2	&	4781	$\,\pm\,$958	&	22.85	$\,\pm\,$3.45	&	9.15	$\,\pm\,$1.16	&	52.68	$\,\pm\,$6.68	&	&	3.88	$\,\pm\,$0.06	&	64.45	$\,\pm\,$0.96	&	$-$0.15	\\
0403$-$132	&	2,5	&	4752	$\,\pm\,$138	&	41.36	$\,\pm\,$13.45	&	23.64	$\,\pm\,$5.15	&	43.52	$\,\pm\,$9.50	&	&	6.11	$\,\pm\,$0.26	&	32.85	$\,\pm\,$1.43	&	$-$1.35	\\
0420$-$014	&	2	&	4846	$\,\pm\,$65	&	73.72	$\,\pm\,$0.99	&	4.82	$\,\pm\,$0.06	&	43.37	$\,\pm\,$0.58	&	&	7.70	$\,\pm\,$0.10	&	196.00	$\,\pm\,$2.65	&	$-$6.70	\\
0429$+$415	&	2	&	5757	$\,\pm\,$19	&	33.46	$\,\pm\,$7.69	&	3.14	$\,\pm\,$0.59	&	421.56	$\,\pm\,$78.54	&	&	1.23	$\,\pm\,$0.06	&	386.04	$\,\pm\,$19.97	&	$-$3.47	\\
0707$+$476	&	2	&	6849	$\,\pm\,$522	&	32.02	$\,\pm\,$2.44	&	4.20	$\,\pm\,$0.32	&	64.75	$\,\pm\,$4.93	&	&	1.40	$\,\pm\,$0.11	&	62.50	$\,\pm\,$4.78	&	$-$0.76	\\
0711$+$356 	&	2	&	3908	$\,\pm\,$657	&	33.92	$\,\pm\,$5.70	&	2.56	$\,\pm\,$0.43	&	61.02	$\,\pm\,$10.27	&	&	0.68	$\,\pm\,$0.11	&	47.33	$\,\pm\,$7.95	&	$-$0.75	\\
0738$+$313	&	3	&	5357	$\,\pm\,$181	&	39.69	$\,\pm\,$1.34	&	36.43	$\,\pm\,$1.23	&	80.13	$\,\pm\,$2.71	&	&	8.38	$\,\pm\,$0.28	&	54.20	$\,\pm\,$1.83	&	$-$0.50	\\
0748$+$126	&	2	&	5297	$\,\pm\,$194	&	26.29	$\,\pm\,$0.96	&	41.90	$\,\pm\,$1.53	&	195.69	$\,\pm\,$7.15	&	&	15.87	$\,\pm\,$0.58	&	219.74	$\,\pm\,$8.02	&	$-$3.56	\\
0804$+$499  	&	6	&	7495	$\,\pm\,$323	&	11.79	$\,\pm\,$1.55	&	2.84	$\,\pm\,$0.11	&	48.75	$\,\pm\,$1.85	&	&	1.91	$\,\pm\,$0.06	&	96.07	$\,\pm\,$3.11	&	$-$0.12	\\
0821$+$394	&	3	&	2445	$\,\pm\,$82	&	10.73	$\,\pm\,$0.36	&	1.13	$\,\pm\,$0.04	&	12.43	$\,\pm\,$0.42	&	&	0.97	$\,\pm\,$0.03	&	31.23	$\,\pm\,$1.05	&	$-$0.97	\\
0827$+$243	&	2	&	5304	$\,\pm\,$209	&	22.67	$\,\pm\,$0.89	&	7.15	$\,\pm\,$0.28	&	38.78	$\,\pm\,$1.53	&	&	3.09	$\,\pm\,$0.12	&	49.54	$\,\pm\,$1.96	&	$-$0.11	\\
0838$+$133 	&	2	&	4683	$\,\pm\,$427	&	78.99	$\,\pm\,$7.20	&	6.35	$\,\pm\,$0.58	&	21.96	$\,\pm\,$2.00	&	&	0.79	$\,\pm\,$0.07	&	7.83	$\,\pm\,$0.71	&	$-$0.58	\\
0850$+$581	&	3	&	5359	$\,\pm\,$318	&	53.09	$\,\pm\,$3.15	&	6.79	$\,\pm\,$0.40	&	95.81	$\,\pm\,$5.69	&	&	1.17	$\,\pm\,$0.07	&	48.36	$\,\pm\,$2.88	&	$-$3.31	\\
0859$-$140	&	2,1	&	4996	$\,\pm\,$268	&	22.95	$\,\pm\,$8.65	&	7.81	$\,\pm\,$2.47	&	117.91	$\,\pm\,$37.29	&	&	3.18	$\,\pm\,$0.20	&	140.08	$\,\pm\,$8.85	&	$-$0.22	\\
0859$+$470	&	3	&	5252	$\,\pm\,$282	&	42.70	$\,\pm\,$2.29	&	2.99	$\,\pm\,$0.16	&	44.72	$\,\pm\,$2.39	&	&	0.68	$\,\pm\,$0.04	&	30.11	$\,\pm\,$1.61	&	$-$3.16	\\
0906$+$015	&	3,2	&	4616	$\,\pm\,$43	&	28.70	$\,\pm\,$5.97	&	7.18	$\,\pm\,$0.81	&	47.64	$\,\pm\,$5.40	&	&	2.39	$\,\pm\,$0.75	&	46.88	$\,\pm\,$14.81	&	$-$1.01	\\
0923$+$392	&	3	&	4927	$\,\pm\,$247	&	52.59	$\,\pm\,$2.63	&	30.95	$\,\pm\,$1.55	&	73.05	$\,\pm\,$3.65	&	&	5.58	$\,\pm\,$0.28	&	39.20	$\,\pm\,$1.96	&	$-$0.12	\\
0945$+$408	&	2,3	&	5116	$\,\pm\,$307	&	40.57	$\,\pm\,$6.24	&	4.82	$\,\pm\,$0.05	&	47.35	$\,\pm\,$0.53	&	&	1.13	$\,\pm\,$0.19	&	33.09	$\,\pm\,$5.62	&	$-$2.36	\\
0953$+$254	&	2	&	4006	$\,\pm\,$1434	&	42.31	$\,\pm\,$0.28	&	12.77	$\,\pm\,$8.80	&	35.45	$\,\pm\,$24.38	&	&	2.82	$\,\pm\,$1.86	&	23.05	$\,\pm\,$15.21	&	$-$1.77	\\
1012$+$232	&	2	&	3381	$\,\pm\,$314	&	70.34	$\,\pm\,$6.53	&	28.22	$\,\pm\,$2.62	&	42.72	$\,\pm\,$3.97	&	&	3.85	$\,\pm\,$0.36	&	17.31	$\,\pm\,$1.60	&	$-$0.15	\\
1015$+$359	&	3	&	4100	$\,\pm\,$196	&	36.07	$\,\pm\,$1.72	&	4.44	$\,\pm\,$0.21	&	41.36	$\,\pm\,$1.97	&	&	1.12	$\,\pm\,$0.05	&	31.14	$\,\pm\,$1.48	&	$-$1.13	\\
1038$+$064	&	2	&	4696	$\,\pm\,$317	&	42.39	$\,\pm\,$2.86	&	12.27	$\,\pm\,$0.83	&	132.42	$\,\pm\,$8.91	&	&	2.83	$\,\pm\,$0.19	&	90.40	$\,\pm\,$6.10	&	$-$2.58	\\
1055$+$018	&	3	&	6039	$\,\pm\,$130	&	5.83	$\,\pm\,$0.13	&	7.46	$\,\pm\,$0.16	&	34.07	$\,\pm\,$0.73	&	&	12.60	$\,\pm\,$0.27	&	170.56	$\,\pm\,$3.65	&	$-$1.64	\\
1055$+$201	&	2	&	6208	$\,\pm\,$404	&	53.79	$\,\pm\,$3.50	&	22.62	$\,\pm\,$1.47	&	175.83	$\,\pm\,$11.44	&	&	3.88	$\,\pm\,$0.25	&	89.57	$\,\pm\,$5.84	&	$-$0.72	\\
1116$+$128	&	3	&	4054	$\,\pm\,$272	&	33.29	$\,\pm\,$2.23	&	1.50	$\,\pm\,$0.10	&	58.32	$\,\pm\,$3.93	&	&	0.44	$\,\pm\,$0.03	&	50.75	$\,\pm\,$3.40	&	$-$2.95	\\
1127$-$145	&	2	&	5101	$\,\pm\,$190	&	26.76	$\,\pm\,$5.06	&	19.57	$\,\pm\,$4.52	&	194.05	$\,\pm\,$44.75	&	&	7.14	$\,\pm\,$0.52	&	208.63	$\,\pm\,$15.28	&	$-$0.12	\\
1128$+$385	&	3	&	5543	$\,\pm\,$432	&	35.84	$\,\pm\,$2.79	&	2.15	$\,\pm\,$0.17	&	50.26	$\,\pm\,$3.93	&	&	0.60	$\,\pm\,$0.05	&	41.95	$\,\pm\,$3.27	&	$-$1.16	\\
1144$+$402	&	3	&	4822	$\,\pm\,$237	&	32.13	$\,\pm\,$1.58	&	3.88	$\,\pm\,$0.19	&	27.70	$\,\pm\,$1.36	&	&	1.19	$\,\pm\,$0.06	&	25.29	$\,\pm\,$1.25	&	$-$0.69	\\
1156$+$295	&	2	&	4245	$\,\pm\,$225	&	50.36	$\,\pm\,$2.66	&	10.74	$\,\pm\,$0.57	&	28.77	$\,\pm\,$1.53	&	&	2.05	$\,\pm\,$0.11	&	16.39	$\,\pm\,$0.87	&	$-$1.16	\\
1219$+$044	&	2	&	5268	$\,\pm\,$335	&	16.56	$\,\pm\,$4.37	&	7.36	$\,\pm\,$0.52	&	39.40	$\,\pm\,$2.77	&	&	3.17	$\,\pm\,$0.14	&	50.46	$\,\pm\,$2.26	&	$-$0.46	\\
1244$-$255	&	2	&	2565	$\,\pm\,$713	&	8.30	$\,\pm\,$3.55	&	2.61	$\,\pm\,$0.69	&	7.28	$\,\pm\,$1.92	&	&	2.52	$\,\pm\,$0.14	&	20.28	$\,\pm\,$1.10	&	$-$0.09	\\
1308$+$326  	&	3	&	5267	$\,\pm\,$360	&	11.12	$\,\pm\,$1.73	&	3.81	$\,\pm\,$0.22	&	21.33	$\,\pm\,$1.25	&	&	2.81	$\,\pm\,$0.09	&	46.91	$\,\pm\,$1.43	&	0.00	\\
1324$+$224	&	2	&	5129	$\,\pm\,$317	&	36.83	$\,\pm\,$10.38	&	4.38	$\,\pm\,$0.24	&	57.85	$\,\pm\,$3.21	&	&	0.79	$\,\pm\,$0.15	&	31.23	$\,\pm\,$6.06	&	$-$1.18	\\
1328$+$307	&	2	&	3378	$\,\pm\,$119	&	16.56	$\,\pm\,$0.58	&	5.32	$\,\pm\,$0.19	&	19.88	$\,\pm\,$0.70	&	&	2.99	$\,\pm\,$0.11	&	33.34	$\,\pm\,$1.18	&	$-$0.68	\\
1354$+$195	&	1	&	5321	$\,\pm\,$245	&	47.39	$\,\pm\,$2.18	&	44.12	$\,\pm\,$2.03	&	144.17	$\,\pm\,$6.64	&	&	9.07	$\,\pm\,$0.42	&	86.59	$\,\pm\,$3.99	&	$-$0.08	\\
1417$+$385	&	3	&	4287	$\,\pm\,$739	&	31.93	$\,\pm\,$5.50	&	0.71	$\,\pm\,$0.12	&	17.23	$\,\pm\,$2.98	&	&	0.25	$\,\pm\,$0.04	&	17.89	$\,\pm\,$3.08	&	$-$3.48	\\
1458$+$718	&	1	&	5131	$\,\pm\,$77	&	48.26	$\,\pm\,$0.83	&	31.94	$\,\pm\,$6.84	&	152.67	$\,\pm\,$32.74	&	&	6.15	$\,\pm\,$1.49	&	87.30	$\,\pm\,$21.15	&	$-$1.89	\\
1508$-$055	&	1,2	&	6216	$\,\pm\,$159	&	40.71	$\,\pm\,$3.01	&	8.87	$\,\pm\,$3.54	&	113.37	$\,\pm\,$45.25	&	&	2.24	$\,\pm\,$0.85	&	82.69	$\,\pm\,$31.56	&	$-$2.23	\\
1606$+$106	&	2	&	5012	$\,\pm\,$307	&	18.85	$\,\pm\,$4.20	&	5.04	$\,\pm\,$0.31	&	62.29	$\,\pm\,$3.83	&	&	2.01	$\,\pm\,$0.13	&	72.47	$\,\pm\,$4.69	&	$-$2.93	\\
1611$+$343	&	2	&	5561	$\,\pm\,$271	&	23.74	$\,\pm\,$1.16	&	7.48	$\,\pm\,$0.36	&	99.46	$\,\pm\,$4.84	&	&	3.09	$\,\pm\,$0.15	&	122.28	$\,\pm\,$5.94	&	$-$2.57	\\
1633$+$382	&	3	&	5583	$\,\pm\,$323	&	18.77	$\,\pm\,$1.09	&	3.25	$\,\pm\,$0.19	&	78.33	$\,\pm\,$4.53	&	&	1.70	$\,\pm\,$0.10	&	122.31	$\,\pm\,$7.06	&	$-$2.57	\\
1637$+$574	&	6	&	3899	$\,\pm\,$136	&	29.69	$\,\pm\,$1.03	&	28.72	$\,\pm\,$1.00	&	79.50	$\,\pm\,$2.77	&	&	8.57	$\,\pm\,$0.30	&	70.82	$\,\pm\,$2.46	&	$-$1.48	\\
1641$+$399	&	1,5	&	5520	$\,\pm\,$1135	&	46.44	$\,\pm\,$7.20	&	66.16	$\,\pm\,$6.61	&	102.18	$\,\pm\,$10.20	&	&	15.23	$\,\pm\,$0.94	&	69.97	$\,\pm\,$4.34	&	$-$1.75	\\
1656$+$053	&	2	&	4748	$\,\pm\,$133	&	18.41	$\,\pm\,$0.52	&	12.58	$\,\pm\,$0.35	&	121.93	$\,\pm\,$3.42	&	&	6.73	$\,\pm\,$0.19	&	182.04	$\,\pm\,$5.11	&	$-$0.60	\\
1656$+$477	&	2	&	6824	$\,\pm\,$1203	&	77.86	$\,\pm\,$39.91	&	11.39	$\,\pm\,$4.33	&	222.05	$\,\pm\,$84.34	&	&	0.64	$\,\pm\,$0.25	&	36.76	$\,\pm\,$14.65	&	$-$3.35	\\
1726$+$455	&	1	&	5771	$\,\pm\,$934	&	176.80	$\,\pm\,$28.63	&	32.42	$\,\pm\,$5.25	&	84.83	$\,\pm\,$13.75	&	&	3.37	$\,\pm\,$0.54	&	26.18	$\,\pm\,$4.23	&	$-$5.60	\\
1800$+$440	&	1	&	4659	$\,\pm\,$837	&	13.05	$\,\pm\,$2.66	&	15.31	$\,\pm\,$3.65	&	40.79	$\,\pm\,$9.73	&	&	11.32	$\,\pm\,$4.77	&	87.94	$\,\pm\,$6.51	&	$-$0.94	\\
1821$+$107	&	2	&	5265	$\,\pm\,$152	&	22.55	$\,\pm\,$4.36	&	13.30	$\,\pm\,$0.38	&	438.47	$\,\pm\,$12.59	&	&	4.19	$\,\pm\,$0.84	&	380.80	$\,\pm\,$15.90	&	$-$1.16	\\
1828$+$487	&	2	&	5427	$\,\pm\,$143	&	27.42	$\,\pm\,$0.72	&	29.04	$\,\pm\,$0.77	&	95.07	$\,\pm\,$2.52	&	&	9.97	$\,\pm\,$0.26	&	94.74	$\,\pm\,$2.50	&	$-$2.48	\\
1845$+$797	&	5	&	6874	$\,\pm\,$275	&	44.00	$\,\pm\,$13.62	&	11.08	$\,\pm\,$0.44	&	0.14	$\,\pm\,$0.01	&	&	1.26	$\,\pm\,$0.31	&	0.05	$\,\pm\,$0.01	&	$-$4.02	\\
1849$+$670	&	2	&	5868	$\,\pm\,$156	&	15.00	$\,\pm\,$0.40	&	8.13	$\,\pm\,$0.22	&	20.85	$\,\pm\,$0.55	&	&	5.83	$\,\pm\,$0.15	&	43.73	$\,\pm\,$1.16	&	$-$0.53	\\
1901$+$319	&	1,2	&	5099	$\,\pm\,$491	&	30.56	$\,\pm\,$6.53	&	12.90	$\,\pm\,$0.15	&	44.93	$\,\pm\,$0.52	&	&	4.04	$\,\pm\,$0.85	&	39.99	$\,\pm\,$8.46	&	$-$2.88	\\
1954$+$513  	&	6	&	8752	$\,\pm\,$354	&	15.14	$\,\pm\,$1.86	&	7.28	$\,\pm\,$0.30	&	147.80	$\,\pm\,$6.15	&	&	3.84	$\,\pm\,$0.11	&	218.69	$\,\pm\,$6.44	&	$-$0.15	\\
2008$-$159	&	2	&	4955	$\,\pm\,$219	&	15.72	$\,\pm\,$2.65	&	9.10	$\,\pm\,$0.41	&	161.02	$\,\pm\,$7.31	&	&	4.48	$\,\pm\,$0.14	&	223.21	$\,\pm\,$7.12	&	$-$3.18	\\
2059$+$034	&	2	&	4731	$\,\pm\,$260	&	37.13	$\,\pm\,$2.04	&	7.54	$\,\pm\,$0.41	&	73.44	$\,\pm\,$4.04	&	&	1.93	$\,\pm\,$0.11	&	53.85	$\,\pm\,$2.96	&	$-$0.87	\\
2128$-$123	&	1	&	5282	$\,\pm\,$131	&	27.79	$\,\pm\,$0.69	&	109.50	$\,\pm\,$2.72	&	150.67	$\,\pm\,$3.73	&	&	40.32	$\,\pm\,$1.00	&	161.12	$\,\pm\,$4.00	&	$-$2.44	\\
2134$+$004	&	3	&	5194	$\,\pm\,$351	&	24.91	$\,\pm\,$1.68	&	6.13	$\,\pm\,$0.41	&	235.34	$\,\pm\,$15.89	&	&	2.33	$\,\pm\,$0.16	&	260.67	$\,\pm\,$17.56	&	$-$2.42	\\
2145$+$067	&	1	&	5517	$\,\pm\,$152	&	25.20	$\,\pm\,$0.70	&	56.88	$\,\pm\,$1.57	&	457.53	$\,\pm\,$12.62	&	&	21.34	$\,\pm\,$0.59	&	495.82	$\,\pm\,$13.71	&	$-$0.47	\\
2201$+$315	&	5	&	5248	$\,\pm\,$308	&	36.96	$\,\pm\,$2.17	&	184.70	$\,\pm\,$10.85	&	108.11	$\,\pm\,$6.37	&	&	47.88	$\,\pm\,$2.81	&	79.47	$\,\pm\,$4.66	&	$-$2.37	\\
2216$-$038	&	2	&	6198	$\,\pm\,$287	&	25.23	$\,\pm\,$1.17	&	10.14	$\,\pm\,$0.47	&	70.04	$\,\pm\,$3.26	&	&	3.94	$\,\pm\,$0.18	&	78.57	$\,\pm\,$3.63	&	$-$0.31	\\
2227$-$088	&	3	&	5896	$\,\pm\,$322	&	18.75	$\,\pm\,$1.03	&	1.81	$\,\pm\,$0.10	&	38.15	$\,\pm\,$2.09	&	&	0.92	$\,\pm\,$0.05	&	56.98	$\,\pm\,$3.11	&	$-$3.71	\\
2230$+$114	&	2	&	4583	$\,\pm\,$77	&	11.80	$\,\pm\,$0.20	&	8.19	$\,\pm\,$0.14	&	71.21	$\,\pm\,$1.19	&	&	6.65	$\,\pm\,$0.11	&	167.96	$\,\pm\,$2.80	&	$-$3.53	\\
2243$-$123	&	1	&	5838	$\,\pm\,$411	&	31.84	$\,\pm\,$0.63	&	29.74	$\,\pm\,$2.74	&	66.25	$\,\pm\,$6.11	&	&	8.75	$\,\pm\,$0.66	&	57.27	$\,\pm\,$4.34	&	$-$2.53	\\
2251$+$158	&	1,2	&	5162	$\,\pm\,$473	&	14.94	$\,\pm\,$1.08	&	19.01	$\,\pm\,$2.28	&	125.98	$\,\pm\,$15.12	&	&	12.50	$\,\pm\,$1.77	&	237.05	$\,\pm\,$33.57	&	$-$3.84	\\
2255$-$282	&	1	&	5171	$\,\pm\,$190	&	26.32	$\,\pm\,$0.96	&	21.64	$\,\pm\,$0.79	&	113.03	$\,\pm\,$4.15	&	&	8.39	$\,\pm\,$0.31	&	129.76	$\,\pm\,$4.76	&	$-$2.82	\\
2345$-$167	&	2	&	4678	$\,\pm\,$222	&	26.07	$\,\pm\,$1.24	&	6.73	$\,\pm\,$0.32	&	10.72	$\,\pm\,$0.51	&	&	2.76	$\,\pm\,$0.13	&	13.04	$\,\pm\,$0.62	&	$-$4.83	\\
2351$+$456	&	6	&	6266	$\,\pm\,$656	&	36.88	$\,\pm\,$10.50	&	1.74	$\,\pm\,$0.34	&	99.19	$\,\pm\,$19.44	&	&	0.33	$\,\pm\,$0.06	&	54.25	$\,\pm\,$9.10	&	\,0.64	 \\
\midrule
\tabnotetext{a}  {Spectrum reference: 1=OAGH; 2=OAN-SPM; 3=\textit{SDSS} (Sloan Digital Sky Survey); 4=\citet{M03}; 5=\textit{HST-FOS} (Hubble Space Telescope) y 6=\citet{L96}. Values shown are the mean and standard deviation if multiple spectra were available.}
\tabnotetext{b} {Flux in units of 10$^{-15}$ erg\,s$^{-1}$\,cm$^{-2}$}
\tabnotetext{c} {Flux density in units of 10$^{-16}$ erg\,s$^{-1}$\,cm$^{-2}$\,\AA$^{-1}$}
\tabnotetext{d} {Spectral index of local continuum power-law.}
\end{longtable}
\end{landscape}
\end{scriptsize}

\renewcommand{\arraystretch}{1.0}

\clearpage

\begin{scriptsize}
\begin{landscape}%para que aparezca rotada
\renewcommand{\arraystretch}{0.9}
\tablecols{11}
\setlength{\tabnotewidth}{0.9\columnwidth}
\tabcaption{\ion{Fe}{2}\, $\lambda$2490 parameters for  78 AGN} \label{tabA4:MgII_2} 

\def\ColumnHeaders{\multicolumn{1}{c}{{}} &
   \multicolumn{4}{c}{{\ion{Fe}{2}\, $\lambda$2490\tablenotemark{a}}} &
   \multicolumn{1}{c}{{}} &
   \multicolumn{1}{c}{{}} &
   \multicolumn{4}{c}{{\ion{Fe}{2}\, $\lambda$2490\tablenotemark{a}}} \\
   \cline{2-5}\cline{8-11} \\[-0.5ex]
   
   \multicolumn{1}{c}{{}} &
   \multicolumn{1}{c}{{Flux}\tablenotemark{b}} &
   \multicolumn{1}{c}{{$L_{2490}$}\tablenotemark{c}} &
   \multicolumn{1}{c}{{EW}} &
   \multicolumn{1}{c}{{}} &
   \multicolumn{1}{c}{{}} &
   \multicolumn{1}{c}{{}} &
   \multicolumn{1}{c}{{Flux}\tablenotemark{b}} &
   \multicolumn{1}{c}{{$L_{2490}$}\tablenotemark{c}} &
   \multicolumn{1}{c}{{EW}} &
   \multicolumn{1}{c}{{}} \\

   \multicolumn{1}{c}{{Source}} &
   \multicolumn{1}{c}{{}} &
   \multicolumn{1}{c}{{}} &
   \multicolumn{1}{c}{{(\AA)}} &
   \multicolumn{1}{c}{{$\lambda_{Fe}$\tablenotemark{d}}} &
   \multicolumn{1}{c}{{}} &
   \multicolumn{1}{c}{{Source}} &
   \multicolumn{1}{c}{{}} &
   \multicolumn{1}{c}{{}} &
   \multicolumn{1}{c}{{(\AA)}} &
   \multicolumn{1}{c}{{$\lambda_{Fe}$\tablenotemark{d}}} 
}
\begin{longtable}{ccccccccccc}
%\begin{longtable}{lll lll lll lc}
%% Primera Cabeza
  \toprule
  \ColumnHeaders\\ \midrule
  \endfirsthead
  
  %% Otras Cabezas
  \tabcaptioncontinued
  \toprule
  \ColumnHeaders\\ \midrule
  \endhead
  
  %% Pies normales
 \bottomrule
%  \tabnotetext{a}{Spectral index of the local continuum power-law.}
\endfoot
%Now the data...
0007$+$106	&	405.42$\pm$1.55	&	16.09$\pm$0.06	&	67.05$\pm$6.74	&		&	&	1156$+$295	&	18.70$\pm$0.05	&	51.45$\pm$0.14	&	59.54$\pm$3.15	&	2300	\\
0016$+$731	&	8.10	$\pm$0.00	&	1584.21$\pm$0.72	&	50.30$\pm$8.13	&		&	&	1219$+$044	&	38.21$\pm$0.03	&	209.30	$\pm$0.16	&	79.30$\pm$20.91	&		\\
0112$-$017	&	4.12	$\pm$0.01	&	70.54$\pm$0.22	&	75.50$\pm$11.06	&		&	&	1244$-$255	&	23.57$\pm$0.02	&	72.94$\pm$0.05	&	70.27$\pm$30.05	&	2436	\\
0113$-$118	&	\nodata			&	\nodata			&	\nodata			&		&	&	1308$+$326  	&	10.14$\pm$0.02	&	57.49$\pm$0.09	&	30.17$\pm$4.68	&		\\
0122$-$003	&	36.24$\pm$0.03	&	283.20$\pm$0.25	&	56.66$\pm$2.57	&		&	&	1324$+$224	&	17.03$\pm$0.00	&	229.03$\pm$0.04	&	110.16$\pm$31.04	&		\\
0133$-$203	&	19.75$\pm$0.05	&	161.82$\pm$0.39	&	120.59$\pm$15.40	&		&	&	1328$+$307	&	\nodata			&	\nodata			&	\nodata			&		\\
0133$+$476	&	3.84	$\pm$0.01	&	39.52$\pm$0.09	&	46.02$\pm$3.88	&		&	&	1354$+$195	&	106.32$\pm$0.24	&	372.16$\pm$0.83	&	76.87$\pm$3.54	&		\\
0202$+$319	&	7.81	$\pm$0.02	&	158.72$\pm$0.31	&	64.38$\pm$7.04	&		&	&	1417$+$385	&	1.86	$\pm$0.01	&	45.90$\pm$0.12	&	54.91$\pm$9.46	&		\\
0248$+$430	&	29.82$\pm$0.06	&	635.95$\pm$1.36	&	78.69$\pm$6.26	&		&	&	1458$+$718	&	68.40$\pm$0.18	&	337.40	$\pm$0.88	&	67.07$\pm$1.15	&		\\
0310$+$013	&	\nodata			&	\nodata			&	\nodata			&		&	&	1508$-$055	&	25.65$\pm$0.07	&	361.88$\pm$0.92	&	76.59$\pm$5.67	&		\\
0333$+$321	&	53.76$\pm$0.11	&	88402$\pm$176	&	62.54$\pm$4.55	&		&	&	1606$+$106	&	15.67$\pm$0.01	&	208.97$\pm$0.06	&	55.75$\pm$12.43	&		\\
0336$-$019	&	15.35$\pm$0.03	&	97.69$\pm$0.18	&	30.68$\pm$4.63	&		&	&	1611$+$343	&	20.85$\pm$0.07	&	283.83$\pm$0.92	&	47.51$\pm$2.32	&		\\
0403$-$132	&	60.01$\pm$0.09	&	118.27$\pm$0.18	&	75.31$\pm$24.49	&		&	&	1633$+$382	&	7.56	$\pm$0.02	&	184.62$\pm$0.45	&	34.26$\pm$1.98	&		\\
0420$-$014	&	4.83	$\pm$0.01	&	50.77$\pm$0.10	&	6.20	$\pm$0.08	&		&	&	1637$+$574	&	56.15$\pm$0.21	&	157.73	$\pm$0.60	&	19.61$\pm$0.68	&		\\
0429$+$415	&	10.59$\pm$0.02	&	2759.66$\pm$4.87	&	97.16$\pm$22.32	&		&	&	1641$+$399	&	124.33$\pm$0.32	&	194.42$\pm$0.50	&	58.40$\pm$9.05	&		\\
0707$+$476	&	5.25$\pm$0.02	&	88.91$\pm$0.29	&	24.53$\pm$1.87	&		&	&	1656$+$053	&	38.62$\pm$0.09	&	450.67$\pm$1.05	&	45.80$\pm$1.29	&		\\
0711$+$356 	&	5.52$\pm$0.01	&	140.76$\pm$0.33	&	47.09$\pm$7.91	&		&	&	1656$+$477	&	19.49$\pm$0.01	&	389.42$\pm$0.14	&	91.67$\pm$46.99	&		\\
0738$+$313	&	35.62$\pm$0.12	&	82.83$\pm$0.28	&	28.05$\pm$0.95	&	2340	&	&	1726$+$455	&	74.74$\pm$0.24	&	201.30$\pm$0.64	&	111.20$\pm$18.00	&	2190	\\
0748$+$126	&	89.89$\pm$0.25	&	435.49$\pm$1.21	&	44.52$\pm$1.63	&		&	&	1800$+$440	&	33.17$\pm$0.11	&	95.15$\pm$0.31	&	21.80$\pm$4.44	&		\\
0804$+$499  	&	9.37$\pm$0.00	&	171.24$\pm$0.07	&	40.11$\pm$5.26	&	2245	&	&	1821$+$107	&	44.50$\pm$0.01	&	1836.31$\pm$0.58	&	70.62$\pm$13.65	&		\\
0821$+$394	&	2.64$\pm$0.01	&	30.64$\pm$0.07	&	21.19$\pm$0.71	&		&	&	1828$+$487	&	27.91$\pm$0.15	&	100.22$\pm$0.54	&	18.48$\pm$0.49	&	2360	\\
0827$+$243	&	20.29$\pm$0.05	&	114.47$\pm$0.28	&	48.26$\pm$1.90	&		&	&	1845$+$797	&	29.54$\pm$0.07	&	0.42$\pm$0.00	&	115.13$\pm$35.65	&		\\
0838$+$133 	&	\nodata			&	\nodata			&	\nodata			&	2300	&	&	1849$+$670	&	\nodata			&	\nodata			&	\nodata			&	2400	\\
0850$+$581	&	12.81$\pm$0.03	&	192.49$\pm$0.44	&	73.75$\pm$4.38	&		&	&	1901$+$319	&	22.03$\pm$0.08	&	88.45$\pm$0.33	&	40.87$\pm$8.74	&		\\
0859$-$140	&	15.40$\pm$0.04	&	250.31$\pm$0.61	&	34.21$\pm$12.90	&		&	&	1954$+$513  	&	11.49$\pm$0.01	&	279.08$\pm$0.30	&	26.10$\pm$3.20	&	2475	\\
0859$+$470	&	4.18$\pm$0.01	&	63.99$\pm$0.16	&	44.29$\pm$2.37	&		&	&	2008$-$159	&	21.64$\pm$0.01	&	452.22$\pm$0.25	&	36.13$\pm$6.09	&		\\
0906$+$015	&	13.94$\pm$0.04	&	96.12$\pm$0.26	&	40.70$\pm$8.46	&		&	&	2059$+$034	&	24.31$\pm$0.06	&	267.56$\pm$0.66	&	82.07$\pm$4.52	&		\\
0923$+$392	&	53.89$\pm$0.14	&	129.20$\pm$0.35	&	63.23$\pm$3.16	&	2236	&	&	2128$-$123	&	\nodata			&	\nodata			&	\nodata			&	2500	\\
0945$+$408	&	10.22$\pm$0.03	&	101.86$\pm$0.27	&	56.87$\pm$8.74	&		&	&	2134$+$004	&	15.86$\pm$0.05	&	659.36$\pm$1.95	&	45.64$\pm$3.09	&		\\
0953$+$254	&	12.47$\pm$0.05	&	36.16$\pm$0.15	&	36.00$\pm$0.23	&		&	&	2145$+$067	&	116.37$\pm$0.31	&	1025.39$\pm$2.77	&	37.93$\pm$1.05	&		\\
1012$+$232	&	\nodata			&	\nodata			&	\nodata			&	2558	&	&	2201$+$315	&	\nodata			&	\nodata			&	\nodata			&	2490	\\
1015$+$359	&	8.93$\pm$0.03	&	84.33$\pm$0.31	&	52.18$\pm$2.49	&		&	&	2216$-$038	&	37.04$\pm$0.10	&	287.05$\pm$0.74	&	71.84$\pm$3.33	&		\\
1038$+$064	&	29.15$\pm$0.07	&	322.63$\pm$0.82	&	67.30$\pm$4.54	&		&	&	2227$-$088	&	4.29$\pm$0.01	&	96.08$\pm$0.32	&	33.39$\pm$1.83	&		\\
1055$+$018	&	18.34$\pm$0.06	&	86.27$\pm$0.28	&	12.35$\pm$0.27	&		&	&	2230$+$114	&	14.26$\pm$0.05	&	135.46$\pm$0.51	&	17.61$\pm$0.30	&		\\
1055$+$201	&	38.59$\pm$0.09	&	309.18$\pm$0.75	&	61.79$\pm$4.02	&		&	&	2243$-$123	&	36.74$\pm$0.14	&	86.85$\pm$0.32	&	30.31$\pm$0.60	&		\\
1116$+$128	&	4.16$\pm$0.01	&	166.66$\pm$0.41	&	60.91$\pm$4.08	&		&	&	2251$+$158	&	36.58$\pm$0.11	&	275.48$\pm$0.83	&	24.98$\pm$1.80	&		\\
1127$-$145	&	50.37$\pm$0.11	&	521.09$\pm$1.13	&	52.36$\pm$9.91	&		&	&	2255$-$282	&	59.92$\pm$0.12	&	326.19$\pm$0.63	&	66.88$\pm$2.45	&		\\
1128$+$385	&	4.36$\pm$0.01	&	105.02$\pm$0.26	&	50.39$\pm$3.92	&		&	&	2345$-$167	&	\nodata			&	\nodata			&	\nodata			&	2450	\\
1144$+$402	&	10.39$\pm$0.04	&	75.79$\pm$0.27	&	62.73$\pm$3.08	&		&	&	2351$+$456	&	5.30$\pm$0.00	&	349.14$\pm$0.12	&	127.57$\pm$36.31	&		\\

\midrule
\tabnotetext{a}  {\ion{Fe}{2} $\lambda2490$ corresponds to the  \ion{Fe}{2} emission in the 2180-2800 \AA\, spectral region.}
\tabnotetext{b}  {Flux in units of 10$^{-15}$ erg\,s$^{-1}$\,cm$^{-2}$.}
\tabnotetext{c}  {Luminosity in units of 10$^{42}$ erg\,s$^{-1}$}
\tabnotetext{d}  {Lower value of wavelength (\AA)  at which spectra allowed the measurement of \ion{Fe}{2} $\lambda2490$ emission. If omitted, the value is 2180\,\AA.}
\end{longtable}
\end{landscape}
\end{scriptsize}

\begin{scriptsize}
\begin{landscape}%para que aparezca rotada
\renewcommand{\arraystretch}{0.9}
\tablecols{10}
\setlength{\tabnotewidth}{0.9\columnwidth}
\tabcaption{\ion{C}{4}\, $\lambda$1549 and continuum at 1350\,\AA\, for 35 AGN} \label{tabA5:CIV_1}
\def\ColumnHeaders{\multicolumn{1}{c}{{}} &
   \multicolumn{1}{c}{{}} &
   \multicolumn{4}{c}{{\ion{C}{4} (BC)}} &
   \multicolumn{1}{c}{{}} &
   \multicolumn{3}{c}{{Continuum 1350 \AA}}\\
   \cline{3-6}\cline{8-10}\\[-0.5ex]

   \multicolumn{1}{c}{{}} &
   \multicolumn{1}{c}{{}} &
   \multicolumn{1}{c}{{FWHM}} &
   \multicolumn{1}{c}{{EW}} &
   \multicolumn{1}{c}{{Flux}\tablenotemark{b}} &
   \multicolumn{1}{c}{{L$_{\mathrm{C\,IV}}$}} &
   \multicolumn{1}{c}{{}} &
   \multicolumn{1}{c}{{Flux}\tablenotemark{c}} &
   \multicolumn{1}{c}{{$\lambda\,L_{1350}$}} &
   \multicolumn{1}{c}{{Spectral}} \\

   \multicolumn{1}{c}{{Name}} &
   \multicolumn{1}{c}{{Ref\tablenotemark{a}}} &
   \multicolumn{1}{c}{{(km\,s$^{-1}$)}} &
   \multicolumn{1}{c}{{(\AA)}} &
   \multicolumn{1}{c}{{}} &
   \multicolumn{1}{c}{{(10$^{42}$ erg\,s$^{-1}$)}} &
   \multicolumn{1}{c}{{}} &
   \multicolumn{1}{c}{{}} &
   \multicolumn{1}{c}{{(10$^{44}$ erg\,s$^{-1}$)}} &
   \multicolumn{1}{c}{{index\tablenotemark{d}}}
}
\begin{longtable}{lccccccccc}
%% Primera Cabeza
  \toprule
  \ColumnHeaders\\ \midrule
  \endfirsthead
  
  %% Otras Cabezas
  \tabcaptioncontinued
  \toprule
  \ColumnHeaders\\ \midrule
  \endhead
  
  %% Pies normales
 \bottomrule
%  \tabnotetext{a}{Spectral index of the local continuum power-law.}
\endfoot
%Now the data...
0016$+$731	&	6	&	7906$\pm$952	&	22.78$\pm$2.74	&	2.91$\pm$0.35	&	705.03$\pm$85.04	&	&	1.45$\pm$0.12	&	594.71$\pm$49.63	&	\nodata	\\
0153$+$744	&	1	&	6220$\pm$655	&	22.36$\pm$2.35	&	3.14$\pm$0.33	&	5188.08$\pm$546.90	&	&	1.10$\pm$0.18	&	3460.14$\pm$556.77	&	$-$6.30	\\
0212$+$735	&	6	&	5579$\pm$408	&	10.48$\pm$0.77	&	0.94$\pm$0.07	&	11345.32$\pm$829.08	&	&	0.70$\pm$0.11	&	19565.88$\pm$2997.33	&	$-$1.85	\\
0552$+$398	&	2	&	4352$\pm$334	&	10.42$\pm$0.80	&	1.29$\pm$0.10	&	1462.35$\pm$112.00	&	&	1.22$\pm$0.11	&	2536.05$\pm$234.9	&	$-$8.12	\\
0642$+$449	&	2	&	4025$\pm$1652	&	24.91$\pm$5.44	&	1.06$\pm$0.45	&	258.23$\pm$109.04	&	&	0.65$\pm$0.18	&	232.9$\pm$63.24	&	$-$1.60	\\
0711$+$356 	&	2	&	4941$\pm$257	&	27.64$\pm$1.44	&	6.03$\pm$0.31	&	159.76$\pm$8.32	&	&	2.69$\pm$0.14	&	100.19$\pm$5.21	&	$-$2.34	\\
0836$+$710	&	1,6	&	7657$\pm$130	&	13.27$\pm$0.33	&	23.71$\pm$16.67	&	1111.15$\pm$782.39	&	&	23.00$\pm$15.90	&	1487.7$\pm$1024.52	&	$-$1.17	\\
0917$+$449	&	3	&	6935$\pm$436	&	17.61$\pm$1.11	&	4.96$\pm$0.31	&	206.76$\pm$13.01	&	&	3.35$\pm$0.17	&	191.31$\pm$9.48	&	$-$0.95	\\
0919$-$260	&	2	&	4895$\pm$390	&	15.77$\pm$1.26	&	3.42$\pm$0.27	&	380.03$\pm$30.22	&	&	1.90$\pm$0.22	&	313.24$\pm$36.93	&	$-$15.68	\\
0955$+$476	&	3	&	5673$\pm$649	&	36.28$\pm$4.15	&	5.88$\pm$0.67	&	161.82$\pm$18.52	&	&	1.84$\pm$0.26	&	69.08$\pm$9.65	&	$-$1.06	\\
1116$+$128	&	3	&	6514$\pm$1217	&	37.14$\pm$6.94	&	3.86$\pm$0.72	&	155.31$\pm$29.01	&	&	1.25$\pm$0.17	&	69.13$\pm$9.46	&	$-$3.82	\\
1128$+$385	&	3	&	7343$\pm$821	&	22.46$\pm$2.51	&	2.67$\pm$0.30	&	65.69$\pm$7.36	&	&	1.57$\pm$0.11	&	53.09$\pm$3.55	&	$-$0.40	\\
1148$-$001	&	2	&	6198$\pm$43	&	13.79$\pm$2.74	&	12.30$\pm$1.44	&	409.67$\pm$47.10	&	&	11.48$\pm$4.60	&	523.65$\pm$209.37	&	$-$1.84	\\
1253$-$055	&	5	&	8613$\pm$350	&	25.68$\pm$6.54	&	18.88$\pm$1.87	&	26.61$\pm$2.63	&	&	38.44$\pm$9.79	&	74.48$\pm$18.99	&	7.20	\\
1354$-$152	&	2	&	8505$\pm$2066	&	35.13$\pm$8.53	&	5.47$\pm$1.33	&	280.23$\pm$68.14	&	&	1.65$\pm$0.40	&	122$\pm$29.65	&	0.00	\\
1402$+$044	&	3	&	7864$\pm$1631	&	57.34$\pm$11.89	&	1.60$\pm$0.33	&	178.2$\pm$36.86	&	&	0.39$\pm$0.08	&	60.37$\pm$12.38	&	$-$0.08	\\
1417$+$385	&	3	&	5622$\pm$879	&	17.15$\pm$2.68	&	1.12$\pm$0.18	&	27.8$\pm$4.37	&	&	0.74$\pm$0.14	&	25.01$\pm$4.55	&	$-$1.07	\\
1502$+$106	&	3	&	5481$\pm$128	&	19.02$\pm$0.96	&	3.15$\pm$0.17	&	94.35$\pm$4.94	&	&	6.90$\pm$0.39	&	285.42$\pm$16.3	&	$-$2.50	\\
1624$+$416 	&	6	&	2818$\pm$1507	&	66.71$\pm$35.68	&	0.36$\pm$0.19	&	19.82$\pm$10.61	&	&	0.06$\pm$0.03	&	4.88$\pm$2.61	&	$-$4.12	\\
1633$+$382	&	2,6	&	6499$\pm$29	&	25.63$\pm$3.79	&	12.97$\pm$1.73	&	318.55$\pm$41.58	&	&	5.29$\pm$1.06	&	176.38$\pm$3.32	&	$-$2.55	\\
1638$+$398	&	2	&	9150$\pm$995	&	26.56$\pm$2.89	&	6.68$\pm$0.73	&	133.1$\pm$14.47	&	&	3.17$\pm$0.34	&	85.89$\pm$9.35	&	$-$0.01	\\
1641$+$399	&	5	&	6710$\pm$1064	&	49.47$\pm$7.84	&	120.30$\pm$19.07	&	192.96$\pm$30.71	&	&	24.78$\pm$3.93	&	54.34$\pm$8.61	&	$-$0.07	\\
1739$+$522	&	6	&	3445$\pm$942	&	12.53$\pm$3.43	&	8.52$\pm$2.33	&	128.87$\pm$35.24	&	&	10.55$\pm$2.39	&	219.87$\pm$50.05	&	$-$2.87	\\
1758$+$388	&	2	&	6744$\pm$733	&	27.36$\pm$2.97	&	9.33$\pm$1.01	&	368.74$\pm$39.92	&	&	4.51$\pm$0.42	&	245.26$\pm$22.79	&	$-$0.02	\\
1845$+$797	&	5	&	9135$\pm$210	&	92.48$\pm$18.70	&	284.65$\pm$37.12	&	3.65$\pm$0.48	&	&	69.00$\pm$2.19	&	1.26$\pm$0.04	&	$-$6.95	\\
2126$-$158	&	2	&	5959$\pm$603	&	17.62$\pm$1.78	&	5.84$\pm$0.59	&	1022.33$\pm$103.28	&	&	3.84$\pm$0.42	&	960.32$\pm$104.78	&	$-$4.34	\\
2128$-$123	&	5	&	8191$\pm$977	&	56.57$\pm$6.75	&	303.50$\pm$36.21	&	466.65$\pm$55.57	&	&	65.86$\pm$7.86	&	142.65$\pm$17.01	&	0.00	\\
2134$+$004	&	1	&	7418$\pm$850	&	24.19$\pm$2.77	&	18.26$\pm$2.09	&	794.82$\pm$90.77	&	&	9.83$\pm$0.75	&	603.64$\pm$46.18	&	$-$0.89	\\
2145$+$067	&	5	&	6423$\pm$638	&	32.44$\pm$3.22	&	69.59$\pm$6.91	&	663.85$\pm$65.91	&	&	24.01$\pm$2.68	&	327.08$\pm$36.52	&	4.83	\\
2201$+$315	&	5	&	7939$\pm$790	&	21.92$\pm$2.18	&	229.40$\pm$22.83	&	165.8$\pm$16.51	&	&	132.43$\pm$10.19	&	140.86$\pm$10.88	&	$-$0.01	\\
2216$-$038	&	5	&	8209$\pm$646	&	34.05$\pm$1.55	&	42.96$\pm$1.94	&	355.1$\pm$15.78	&	&	13.27$\pm$0.67	&	158.18$\pm$7.6	&	$-$3.18	\\
2230$+$114	&	5	&	5990$\pm$403	&	20.96$\pm$1.41	&	25.94$\pm$1.74	&	255.71$\pm$17.18	&	&	13.40$\pm$1.11	&	188.01$\pm$15.57	&	$-$0.06	\\
2243$-$123	&	5	&	7231$\pm$899	&	45.46$\pm$5.65	&	135.80$\pm$16.89	&	332.78$\pm$41.35	&	&	35.66$\pm$4.43	&	122.17$\pm$15.16	&	0.00	\\
2251$+$158	&	5	&	7875$\pm$182	&	17.56$\pm$6.39	&	31.01$\pm$2.82	&	250.01$\pm$22.81	&	&	18.53$\pm$5.66	&	217.32$\pm$66.45	&	0.00	\\
2351$+$456	&	6	&	5996$\pm$706	&	86.61$\pm$10.20	&	1.25$\pm$0.15	&	89.38$\pm$10.51	&	&	0.22$\pm$0.03	&	23.78$\pm$2.80	&	$-$4.49	\\ 
\midrule
\tabnotetext{a} {Spectrum reference: 1=OAGH; 2=OAN-SPM; 3=\textit{SDSS} (Sloan Digital Sky Survey); 4=\citet{M03}; 5=\textit{HST-FOS} (Hubble Space Telescope) y 6=\citet{L96}. If more spectra were available, the values are averaged and the error is the standard deviation. }
\tabnotetext{b} {Flux in units of 10$^{-15}$ erg\,s$^{-1}$\,cm$^{-2}$.}
\tabnotetext{c} {Flux density in units of 10$^{-16}$ erg\,s$^{-1}$\,cm$^{-2}$\,\AA$^{-1}$.}
\tabnotetext{d} {Spectral index for the local continuum.}
\end{longtable}
\end{landscape}
\end{scriptsize}
\renewcommand{\arraystretch}{1.00}

\end{appendices}

%%%%%%%%%%%%%%%viene la bibliografia

\bibliographystyle{rmaa} % style rmaa.bst
\bibliography{biblio_atlas.bib} % your references Yourfile.bib

\end{document}